\documentclass[reprint,aps,showpacs,prl,nofootinbib]{revtex4-2}
\usepackage{color}
\usepackage{newlfont}
\usepackage{graphicx}
\usepackage{amssymb}
\usepackage{amsmath}
\usepackage[caption=false]{subfig}
  
\usepackage{bm}
\usepackage{verbatim}
\usepackage[latin1]{inputenc}
\usepackage{bbm}
\usepackage{multirow}
\usepackage[thinlines]{easytable}
\usepackage{braket}
\usepackage{amsthm}
\usepackage{bbold}
\usepackage{subfig}
\usepackage{mathrsfs}
\usepackage[varg]{txfonts} 
\usepackage{amsfonts}

\newcommand{\be}{\begin{equation}}
\newcommand{\ee}{\end{equation}}
\newcommand{\beq}{\begin{eqnarray}}
\newcommand{\eeq}{\end{eqnarray}}
\newcommand{\ba}{\begin{array}}
\newcommand{\ea}{\end{array}}

\newcommand{\mf}{\mathfrak}

\DeclareMathOperator{\Tr}{Tr}

\newtheorem{result}{Result}

\begin{document}

\title{Quantum work: Reconciling quantum mechanics and thermodynamics}
\author{Thales A. B. Pinto Silva}
\email{pinto\_silva@campus.technion.ac.il}
\author{David Gelbwaser-Klimovsky}
\affiliation{Schulich Faculty of Chemistry and Helen Diller Quantum Center, Technion-Israel Institute of Technology, Haifa 3200003, Israel}

\begin{abstract} 
It has been claimed that no protocol for measuring quantum work can satisfy standard physical principles, casting doubts on the compatibility between quantum mechanics, thermodynamics, and the classical limit. In this Letter, we present a solution for this incompatibility. We demonstrate that the standard formulation of these principles fails to address the classical limit properly. By proposing changes in this direction, we prove that all the essential principles can be satisfied when work is defined as a quantum observable, reconciling quantum work statistics and thermodynamics.
\end{abstract}

\maketitle

The concept of work lies at the heart of thermodynamics, being crucial to determine the processes permissible under the first and second laws \cite{Callen1985,Fermi1936,Reif2009,Sekimoto2010}. Moreover, work's statistical properties have paved the way for groundbreaking results, among which, the Jarzynski equality \cite{Jarzynski1997} and Crook's fluctuation relation \cite{Crooks1999} certainly stand out. These so-called fluctuation theorems have unveiled the tight connection of work with equilibrium free energy, the reversibility of thermodynamic processes, and information theory \cite{Campisi2011,Campisi2009,Jarzynski2004,Alhambra2016,Aberg2018,Brandao2013,Huber2008,liphardt2002,Park2020, Goold2016,Maruyana2009,Sagawa2009,Esposito2011,Manzano2018,Alipour2016}.

In the quantum realm, the pursuit of a universal definition of work resulted in multiple definitions \cite{Bochkov1977,Alicki1979,Gemmer2005,Campisi2011,Hanggi2017,Millen2016,Talkner2016,Deffner2019,Strasberg2021,Gelbwaser2013,Gelbwaser2015,Gelbwaser2014,Binder2018,Rossnagel2016,Dann2023,Gallego2016,Esposito2010,Barra2015,Kewming2022,Gnezdilov2023,Das2023,Lacerda2023,Ohanesjan2023,Safranek2023}, allowing the thermodynamic analysis of various quantum systems \cite{Rossnagel2016,Auffeves2022,Levy2020,Lorch2018}.  Nevertheless, the statistical characterization of work continues to be the subject of intense debate \cite{Silva2021,Silva2023,Jacob2023,Janovitch2022, Lostaglio2018,Jarzynski2007,Horowitz2007,Vilar2008,Vilar2008a,Vilar2008b,Horowitz2008,Peliti2008,Miller2017,Deffner2016,Beyer2020,Allahverdyan2014,Binder2015,Talkner2007,Beyer2022,Baumer2018,Perarnau2017,Francica2022a,Francica2022b,Roncaglia2014,Sone2020}. For instance, in scenarios without heat flux, work average is expected to equal the variation of mean energy, a feature not generally fulfilled by broadly used work measurement schemes, such as the two-point-measurement (TPM) \cite{Talkner2016,Deffner2016,Allahverdyan2014,Perarnau2017}, the Gaussian \cite{Talkner2016,Roncaglia2014}, and post-selection \cite{Lostaglio2018} schemes. Other significant discussions in the literature include the definition of work in scenarios encompassing the energetic effects of measurements and quantum resources \cite{Campaioli2018,Gyhm2022,Bresque2021,Rodrigues2019,Popovic2023,Jarzynski2015,Elouard2017,Maruyana2009,Sagawa2009,Mohammady2019,Lostaglio2018}. Successfully overcoming these challenges can propel advancements in quantum thermodynamics and the design of efficient quantum devices \cite{Auffeves2022,Levy2020,Lorch2018,Andolina2019,Deffner2021,Faist2015,Bennett1982}.

To elucidate quantum work, it has been generally considered that its statistics should comply with the following \emph{criteria}, provided by the basic structure of quantum mechanics and thermodynamics \cite{Perarnau2017,Baumer2018}: work statistics should (i) be described by a positive operator-valued measure (POVM) \cite{Nielsen2010,Heinosaari2012} independent of the system's initial state, (ii) yield an average work equal to the variation of energy over time for externally controlled systems not interacting with a heat bath, and (iii) reproduce the classical limit. Surprisingly, it was demonstrated that no protocol for raising work statistics would comply simultaneously with criteria (i) and (ii) and a \emph{condition} inspired by (iii) \cite{Perarnau2017}. These findings have prompted the quest for new measurement protocols and a better understanding of the statistics of work \cite{Baumer2018,Lostaglio2018, Sone2020,Silva2021,Silva2023,Beyer2020,Beyer2022,Francica2022a,Francica2022b,Jacob2023,Janovitch2022}. Remarkably, there is still no statistical framework for describing work that aligns with criteria (i)-(iii), suggesting an incompatibility between quantum mechanics, its convergence to the classical limit, and thermodynamics.

In this Letter, we propose a solution to this challenge. We demonstrate that there is at least one protocol for raising work statistics in consonance with the basic criteria (i)-(iii). To demonstrate our results, we adopt criteria (i) and (ii) in a manner analogous to Ref. \cite{Perarnau2017} while imposing necessary conditions for the classical limit as stipulated by criterion (iii). Specifically, we require that in the classical limit, the average of general quantum observables approximates their corresponding classical counterpart and that the commutation of quantum observables is sufficiently small. Guided by these conditions, we demonstrate that treating work as a two-time quantum observable \cite{Allahverdyan2005,Gelin2008,Lindblad1983},  criteria (i)-(iii) are satisfied. Furthermore, we shed light on previous forms of imposing criterion (iii), highlighting their reliance on the two-point-measurement (TPM) methodology \cite{Campisi2011,Talkner2007,Roncaglia2014}, which we show neither guarantees the recovery of the classical limit nor represents the unique approach capable of achieving it. We expect that our study showcases the feasibility of achieving a satisfactory reconciliation between quantum and classical work statistics and contributes to the ongoing quest for a unified framework for quantum work statistics.

\section{Requirements for consistent work statistics}

We first present criteria (i)-(iii) in detail. Our analysis consists of a closed system whose initial state is described by $\rho$ governed by an externally controlled time-dependent Hamiltonian $H(t)$ with eigenbasis $\{\ket{e_{j}(t)}\}$. The system's dynamics are dictated by the evolution operator $U_{t}$, which satisfies the Schr{\" o}dinger equation $i\hbar \partial_{t}U_t=H(t)U_t$. Since the system is isolated, no heat is exchanged, and work equals the system energy variation. We thus posit that there exists a general protocol for measuring work, such that for any process defined by operators $H(t)$ and $U_t$, this protocol can describe the probability $P(w)$ of obtaining a value of work $w$. Following the quantum mechanics paradigm \cite{Nielsen2010,Heinosaari2012}, we require that $P(w)=\Tr[M(w)\,\rho]$, where $M(w)$ are the elements of a POVM $\{M(w)\}$, i.e., a set of non-negative Hermitian operators $M(w)$ satisfying the relation $\sum_{w}M(w)=1$. Considering this formalism, a  work measurement scheme implemented by $\{M(w)\}$ should satisfy the following criteria:

(i) \emph{The operators $M(w)$ may depend on $H(t)$ and $U_t$, but are independent of $\rho$}. With this criterion, we anticipate that the measurement protocol must depend on the process described by $U_t$ and $H(t)$ but is independent of the initial state, as one should not expect to adjust the measurement apparatus to the system's initial state \cite{Perarnau2017}. Importantly, an equivalent formulation of criterion (i) allows for state-dependent POVM, as long as $P(w)$ exhibits linear dependence on the state \cite{Perarnau2017,Baumer2018}.

(ii) \emph{For any state $\rho$, the work average equals the variation in the average energy over time, i.e. $\braket{W}=\sum_w w\,P(w) =\Tr[H(t)\,U_t\rho U_t^\dagger]-\Tr[H(0)\,\rho]=\braket{H(t)}-\braket{H(0)}$.} Considering the convention in which work is deemed positive when the energy $H(t)$ increases, this statement aligns with the first law of thermodynamics, as there is no heat flux for this non-autonomous and closed quantum scenario.

(iii) \emph{In the classical limit, the statistics raised via the POVM $M(w)$ must approach the classical results}. More precisely, it is anticipated that as the system approaches the classical limit, the quantum distribution $P(w)$ obtained through $M(w)$ should exhibit the same statistics as the classical distribution associated with the classical limit. This criterion put forward the expectation that the classical world can be reproduced within quantum mechanics in semi-classical regimes \cite{Jarzynski2015,Deffner2008,Deffner2010,Chen2021,Zhu2016,Garcia-Mata2017}.

Criteria (i)-(iii) collectively provide a comprehensive framework that aligns with the foundations of quantum mechanics and thermodynamics and the convergence to classical behavior. In fact, these have served as the basis for many recent works (see \cite{Perarnau2017,Baumer2018,Beyer2022,Lostaglio2018,Francica2022a,Francica2022b} and references therein).

We can straightforwardly examine how criteria (i) and (ii) can constrain work POVMs considering  two notable examples that shall be treated frequently in this Letter. On the one hand, we  consider the two-point-measurement (TPM) POVM, which is succinctly described as follows \cite{Campisi2011,Talkner2007,Roncaglia2014}. At an initial time, the system is submitted to a projective measurement of energy, collapsing to an $H(0)$ eigenstate $\ket{e_n(0)}$ with probability $\mf{p}_n=\braket{e_n(0)|\rho|e_n(0)}$. The system then evolves unitarily until the instant $\tau$, when a second measurement is performed and a random eigenvalue $e_m(\tau)$ of $H(\tau)$ is obtained with probability $\mf{p}_{m|n}=|\bra{e_m(\tau)} U_{t}\ket{e_n(0)}|^2$. The TPM work probability is thus computed as \cite{Perarnau2017,Roncaglia2014} $P_{\text{\tiny TPM}}(w)=\sum_{mn}\mf{p}_{m|n}\,\mf{p}_{n}\,\delta[w-(e_m(\tau)-e_n(0))]=\Tr[M_{\text{\tiny TPM}}(w)\,\rho]$, where $M_{\text{\tiny TPM}}(w)=\sum_{mn}\delta[w-(e_{m}(\tau)-e_{n}(0))]\mf{p}_{m|n}\ket{e_{n}(0)}\bra{e_{n}(0)}$ are the elements of the TPM POVM and $\delta$ is the Dirac's delta. Although TPM complies with criterion (i), it generally does not satisfy criterion (ii), as the first measurement destroys any coherence the initial state may possess \cite{Perarnau2017,Baumer2018,Beyer2022}.

On the other hand, the observable (OBS) POVM is connected with the two-time work observable \cite{Allahverdyan2005,Gelin2008,Lindblad1983,Silva2021,Silva2023}
\begin{equation}
W(\tau,0)=H_{h}(\tau)-H(0),\label{Wq}
\end{equation}
where $H_{h}(t)=U_t^\dagger H(t)U_t$ represents the Hamiltonian in the Heisenberg picture at time $t$. Notably, $W(\tau,0)$ is a Hermitian operator possessing a set of eigenvalues $w_{j}(\tau,0)$ and corresponding eigenvectors $\{\ket{w_{j}(\tau,0)}\}$, such that $W(\tau,0)=\sum_{j}w_{j}(\tau,0)\ket{w_{j}(\tau,0)}\bra{w_{j}(\tau,0)}$. Accordingly, the probability of finding a value $w$ of work is defined as $P_{\text{\tiny OBS}}(w)=\Tr[M_{\text{\tiny OBS}}(w)\rho]$, where $M_{\text{\tiny OBS}}(w)=\sum_{j}\delta(w-w_{j}(\tau,0))\ket{w_{j}(\tau,0)}\bra{w_{j}(\tau,0)}$. As can be checked from the discussion in \cite{Silva2021}, there always exists a corresponding Schr{\"o}dinger operator $S$ diagonalized by the same basis as $\{\ket{w_{j}(\tau,0)}\}$ for each arbitrary interval $[0,\tau]$. Therefore, by measuring $S$ at time $0$ and collapsing the state to an eigenstate $\ket{w_{j}(\tau,0)}$, one is thus ``preparing'' a state whose amount of work done on the system is known to be $w_{j}(\tau,0)$ during $[0,\tau]$. By making several measurements of $S$ at time $0$, one thus raises $P_{\text{\tiny OBS}}(w)$.

We remark that criteria (i) and (ii) are fulfilled if and only if $M(w)$ are state-independent operators that satisfy the relation $\sum_{w}w M(w) =H_{h}(\tau)-H(0)$ \cite{Perarnau2017,Baumer2018}. Comparing this equation with Eq. \eqref{Wq} and considering the explicit form of $M_{\text{\tiny OBS}}(w)$, we readily confirm that the OBS POVM satisfies criteria (i) and (ii). 

\section{TPM and the classicality criterion}

Unlike criteria (i) and (ii), verifying criterion (iii) for a specific POVM is challenging due to the elusive nature of the classical limit in the quantum realm--a topic that has been intensely debated since the inception of quantum mechanics \cite{EPR1935,Bell1964,Zurek2009,Bilobran2015,Dieguez2018,Dieguez2022}. Still, there is no unique form of characterizing the classical limit in the quantum context \cite{Mari2011,Kwon2019,Kenfack2004,Chabaud2021,Lemos2018}. Notwithstanding these difficulties, Perarnau-Llobet \emph{et al.} \cite{Perarnau2017} boldly proposed what we refer to as the classical stochastic (CS) hypothesis, aiming to implement criterion (iii): \emph{for initial states with no quantum coherences in the energy basis, the results of classical stochastic thermodynamics should be recovered}. More specifically, they suggested that for incoherent initial states where $[\rho,H(0)]=0$, the work probability should be equal to the TPM probability, i.e., $[\rho,H(0)]=0\implies P(w)=P_{\text{\tiny TPM}}(w)$. The CS hypothesis was justified on the basis that the TPM scheme used for incoherent states enabled classical stochastic thermodynamics relations to be reproduced similarly in the quantum regime \cite{Campisi2011,Hanggi2017,Binder2018,Aberg2018,Alhambra2016,Jarzynski2004,Campisi2009,Campisi2011a} and because of its good convergence in the classical limit \cite{Jarzynski2015,Deffner2008,Deffner2010,Chen2021,Zhu2016,Garcia-Mata2017}. Remarkably, the authors demonstrated that no POVM could simultaneously satisfy criteria (i) and (ii) along with the CS hypothesis. However, even though the CS hypothesis has been considered in many works aiming to recover the classical limit \cite{Baumer2018,Beyer2022,Lostaglio2018}, we present throughout this section three arguments showing that demanding CS as a way of implementing criterion (iii) may not be the most general or accurate approach.

1) \emph{There is no fundamental principle dictating that quantum processes beginning with initial incoherent states should recover classical stochastic thermodynamic results} \cite{Jarzynski2015,Allahverdyan2014,Beyer2020,Silva2021}. Because the Hamiltonian is time-dependent, an initial incoherent state will not always time-evolve to an incoherent state with respect to the instantaneous Hamiltonian energy bases. The emergence of coherences throughout the process may result in a different amount of work exchange relative to what is classically expected. Moreover, for incoherent states quantum corrections can still display significant thermodynamics effects \cite{Hartle2018,Gelbwaser2018}, as has been recognized ever since Wigner's seminal paper \cite{Wigner1932}.

2) \emph{TPM is not the unique approach successful in reproducing classical outcomes  when coherences are absent}. Indeed, the agreement of TPM with the classical results in the semi-classical regime \cite{Jarzynski2015,Deffner2008,Deffner2010,Chen2021,Zhu2016,Garcia-Mata2017} has served as a justification for its selection in the CS hypothesis \cite{Perarnau2017,Baumer2018}. However, TPM is not necessarily unique in this sense. To prove this statement, consider a process described by $H(t)$ and $U_t$ and a initial state $\rho$ taking place within the time interval $[0,\tau]$. Consider also that the average of energy is bounded, i.e., $\braket{ H(t)}<\infty$, for any $t\in[0,\tau]$. Let $C(\rho(t))$ be the 1-norm measure of coherences \cite{Baumgratz2014} for the evolved state $\rho(t)=U_t\rho U_{t}^{\dagger}$:
    \begin{equation}
        C(\rho(t))=\|\rho(t)-\Phi_{H(t)}(\rho(t)) \|_1=\sum_{j\neq k}|\bra{e_j(t)}\rho(t)\ket{e_k(t)}|,\label{Crhot}
    \end{equation}
where $\Phi_{H(t)}(\rho(t))=\sum_{j}\ket{e_j(t)}\bra{e_j(t)}\rho(t)\ket{e_j(t)}\bra{e_j(t)}$ is the dephasing map at time $t$, accounting for the incoherent part of $\rho(t)$. Within this scenario, we prove the first result of this Letter:
\begin{result}
   If there is a real non-negative scalar $\epsilon_{1}$ in which for any initial incoherent state $\rho$,  $C(\rho(t))\leq \epsilon_{1}$ during the entire interval $[0,\tau]$, then $\int_{-\infty}^{\infty}dw\,|P_{\text{\tiny TPM}}(w)-P_{\text{\tiny OBS}}(w)|\leq C_1\epsilon_1$, where $C_1$ is a \emph{bounded} real positive scalar independent of $\epsilon_1$. Therefore, $\lim_{\epsilon_1\to 0}\int_{-\infty}^{\infty}dw\,|P_{\text{\tiny TPM}}(w)-P_{\text{\tiny OBS}}(w)|=0$.
\label{res1}
\end{result}
Result 1 arises from the fact that processes with a small amount of coherence require the Hamiltonians at different times to \emph{almost} commute. Otherwise, a large amount of coherences will emerge. The difference between TPM and OBS, as well as the creation of coherences, stem from the lack of Hamiltonian commutation and will be reduced as the Hamiltonians approach to a perfect commutation. Therefore, the TPM protocol will approach the OBS scheme for processes close to the classical limit where coherences are \emph{all the time} small. Consequently, justifying the preference for TPM over OBS  based on its convergence to classical results can be misleading. We confirm this physical intuition by rigorously showing that $C_1$ is bounded and independent of $\epsilon_1$ (see the Supplemental Material (SM) \cite{SMref}). The case of zero coherence, $\epsilon_1=0$ can be directly proved considering the results presented in Ref. \cite{Talkner2007}.

3) \emph{The CS hypothesis lacks the necessary generality to encompass a broader range of processes in the classical limit}. Consider, for instance, an oscillator with a time-dependent frequency described by the Hamiltonian operator \cite{Deffner2008,Deffner2010}
    \begin{equation}
        H(t)=\frac{P^{2}}{2m}+\frac{m\omega^{2}(t)}{2}X^2,\label{QH}
    \end{equation}
where $\omega^2(t)=\omega_0^2+(\omega_1^2-\omega_0^2)\frac{t}{\tau}$, and $\omega_{0,1}$ and $\tau$ are fixed parameters. When initialized in a coherent state $\rho=\ket{\alpha}\bra{\alpha}$, defined as
\begin{equation}
    \ket{\alpha}=\mathrm{e}^{-|\alpha|^2/2}\sum_{n}\frac{\alpha^{n}}{\sqrt{n!}}\ket{e_{n}(0)},\label{alpha}
\end{equation}
then, for a sufficiently large $|\alpha|$, the center of the wave packet exhibits classical particle-like behavior \cite{Cohen2020}, following dynamics governed by a classical Hamiltonian analogous to Eq. \eqref{QH}. On the one hand, following criterion (iii), if we consider this scenario a classical limit, the quantum work statistics should converge to its classical counterpart. Under this perspective, this example extends beyond the scope of the CS hypothesis, exposing its limited coverage of the classical limit by associating it solely with incoherence in the energy basis. On the other hand, despite the oscillator's classical behavior, one might expect that a large amount of coherence could lead to work statistics diverging from the classical counterpart. This example thus highlights the need to clarify the classical limit beyond the CS hypothesis and when we expect the quantum statistics to converge to it. We address this issue in the next section and in \cite{SMref}, where we explicitly calculate the work statistic for a coherent state, revealing that the OBS protocol appropriately converges in the classical limit while the TPM does not.

\section{OBS protocol and the classical limit}

As exemplified above, formally characterizing criterion (iii) is daunting and the CS hypothesis is not able to implement it. In order to circumvent these difficulties and still obtain valuable results from criterion (iii), we propose an alternative strategy. Here, we first identify some necessary conditions for a quantum process to approach the classical limit. Then, we show that these are enough to obtain general results for quantum work statistics.

We consider again a process described by $U_t$ and $H(t)$, but now the Hamiltonian $H(t)$ depends on a Schr{\"o}dinger coordinate observable $X$ and its conjugate momentum $P$ ($[X,P]=i\hbar\,\mathbbm{1}$), and time $t$ within the interval $[0,\tau]$. Consequently, $U_t$ and the Heisenberg version of the Hamiltonian $H_h(t)$ are continuous functions of $X$, $P$, and $t$. We can identify $H_h(t) = G(X, P, t)$, where $G$ is a \emph{smooth} function that can be written as powers of $X$, $P$, and $t$. 

In order to establish the necessary conditions for the convergence to the classical limit of an initial preparation $\rho$ and the quantum process described by $U_t$, $H_h(t)$, it is essential to introduce the following  quantities: the classical probability distribution $\varrho(\Gamma_0)$,  representing the likelihood of finding a classical system at the phase points $\Gamma_0=(x_0,p_0)$ at time $0$; and a classical Hamiltonian functions  $H_{\text{\tiny CL}}^{t}(\Gamma_t,t)$, characterizing the energy of the system at time $t$ with respect to the phase point $\Gamma_t=(x_{t},p_t)$. Importantly, within the Hamiltonian formalism, the evolution of $\Gamma_t(\Gamma_0,t)=(x_{t}(\Gamma_0,t),p_t(\Gamma_0,t))$ can be expressed in terms of $\Gamma_0$ and $t$, allowing the Hamiltonian function to also be written as a function of $\Gamma_0$ and $t$, so that $H_{\text{\tiny CL}}^{t}(\Gamma_t,t)=H_{\text{\tiny CL}}(\Gamma_0,t)$. Here, $H_{\text{\tiny CL}}$ corresponds to the energy at time $t$ related to the system that was at $\Gamma_0$ at time $0$. With these definitions in place, the following  conditions are required for the convergence to the classical limit of the process characterized by $\{X,P,\rho,H_h(t)\}$:

(A) \emph{There exists at least one classical distribution $\varrho(\Gamma_0)$ along with a Hamiltonian function $H_{\text{\tiny CL}}(\Gamma_0,t)$, such that}
    \begin{equation}
        \left|\Tr\left[P^{m}X^{n}\rho \right]-\int d\Gamma_0\, p_0^{m}x_{0}^n \,\varrho(\Gamma_0)\right|\leq \epsilon_{A}\left|\Tr\left[P^{m}X^{n}\rho \right]\right|\label{eq6}
    \end{equation}
    \emph{for any integers $m$ and $n$, and}
    \begin{equation}
        H_{\text{\tiny CL}}(\Gamma_0,t')\equiv H_{\text{\tiny CL}}(x_0,p_0,t)=G(x_0, p_0, t'),\label{eq7}
    \end{equation}
\emph{where $G(x_0, p_0, t)$ has the same functional form as $G(X, P, t)$, previously defined, $t'$ can be either $0$ or $\tau$, and $\epsilon_{A}\ll 1$}. Equation \eqref{eq6} asserts that there must be at least one classical scenario whose averages approach the quantum ones at $t=0$.. Furthermore,  Eq. \eqref{eq7} states that the quantum energy in the classical limit should be described in terms of the same quantities that define its classical version at $t'=0,\tau$.
    
(B) \emph{For any functions $g_{l}\equiv g_l(X,P)$ and $g_r\equiv g_r(X,P)$ describing product of powers of $X$ and $P$, it follows that}
\begin{equation}
    |\Tr[g_l\,[X,P]\,g_r\,\rho]|\leq \epsilon_{B}|\Tr[g_l\, XP\, g_r\,\rho]|,
\end{equation}
\emph{and}
\begin{equation}
    |\Tr[g_l\,[X,P]\,g_r\,\rho]|\leq \epsilon_{B}|\Tr[g_l\, PX\, g_r\,\rho]|,
\end{equation}
\emph{where $\epsilon_{B}\ll 1$}. This requirement resembles the usual classical limit assumption $\hbar \to 0$. In our formulation, however, we allow the non-commutation to have a non-null yet significantly small value compared to the system's inherent scales. 
     
It is important to emphasize that conditions (A) and (B) alone are not sufficient to certify the classical limit. Specifically, they do not rule out the effects of quantum correlations, measurement's disturbance, or any other quantum phenomena. In our framework, processes influenced by these quantum effects may still adhere to conditions (A) and (B) without truly approximating the classical limit. Conversely, these conditions are \emph{necessary} for the classical limit. Therefore, we expect that any quantum scenario genuinely approaching the classical limit rigorously satisfies conditions (A) and (B). In these circumstances, the set $\{x_0,p_0,\varrho,H_{\text{\tiny CL}}\}$ describe the classical scenario to which the process characterized by $\{X,P,\rho,H_h(t)\}$ converges. Interestingly, we demonstrated in \cite{SMref} that the system described in Eqs. \eqref{QH} and \eqref{alpha} satisfies conditions (A) and (B) for $|\alpha|\to \infty$.

Now, consider a scenario fulfilling conditions (A) and (B) converging to the classical limit.. Within this context, the classical work is described as $W_{\text{\tiny CL}}(\Gamma_0,\tau,0)=H_{\text{\tiny CL}}(\Gamma_0,\tau)-H_{\text{\tiny CL}}(\Gamma_0,0)$, with the associated classical distribution defined as
\begin{equation}
P_{\text{\tiny CL}}(w)=\int_{\text{\tiny $\Gamma_0$}}\,d\Gamma_0\, \varrho(\Gamma_0) \,\delta[w-W_{\text{\tiny CL}}(\Gamma_0,\tau,0)].\label{pcl}
\end{equation}
According to criterion (iii), a consistent approach to describing quantum work statistics should converge to the results provided by $P_{\text{\tiny CL}}(w)$ in the classical limit. Following this reasoning, we derived the central result of this Letter.

\begin{result}
If  conditions (A) and (B) are satisfied, then
\begin{equation}
        \lim_{\epsilon_{\max}\to 0}\int_{-\infty}^{\infty}dw|P_{\text{\tiny OBS}}(w)- P_{\text{\tiny CL}}(w)|= 0
    \end{equation}
and 
\begin{equation}
    \int_{-\infty}^{\infty}|P_{\text{\tiny OBS}}(w)-P_{\text{\tiny CL}}(w)|dw\leq K\epsilon_{\max},
\end{equation}
where $P_{\text{\tiny OBS}}(w)$ is the probability related to the OBS protocol, $P_{\text{\tiny CL}}(w)$ is defined in Eq. \eqref{pcl}, $\epsilon_{\max}$ is the maximal value of $\epsilon_{A}$ and $\epsilon_{B}$ as defined in conditions (A) and (B), and $K$ is a bounded real positive scalar independent of $\epsilon_{A}$ and $\epsilon_{B}$.
\end{result}

The OBS protocol, derived from classical work using standard quantization rules, reveals differences between classical and quantum observables in scenarios with significant lack of commutation. This suggests that the difference in quantum and classical probabilities is tied to the lack of commutation, approaching zero when commutation is negligible [condition (B)], and the quantum scenario exhibits similar averages as a classical counterpart [condition (A)]. This intuition supports result 2, which we rigorously proved in \cite{SMref}, employing the Cauchy-Hadamard formula \cite{Suslov2005,Ahlfors1979} to demonstrate the boundedness of $K$.

According to result 2, the OBS protocol will converge to the description via Eq. \eqref{pcl} for any given quantum state $\rho$ and process $\{U_t,H(t)\}$ that satisfies condition (A) and (B). Since these conditions are necessary for the classical limit, then we can deduce that the statistics of work obtained through the OBS protocol invariably replicate the classical work statistics in the classical limit, thereby satisfying criterion (iii).  Furthermore, from the review conducted in Ref. \cite{Baumer2018}, the OBS protocol is the only well-known protocol that generally adheres to (i) and (ii). As a consequence of the result 2 and given our current understanding, we can also assert that OBS is the only well-known protocol capable of meeting criteria (i)--(iii) simultaneously.

    \section{Conclusions}

Quantum work is a fundamental concept for extending thermodynamics to general quantum scenarios. Its statistical description, however, still presents intriguing challenges. Therefore, testing its consistency with criteria (i)-(iii), expressing general features that any quantum work protocol should possess, is a paramount goal. By adopting a novel strategy to consider the classical limit, we demonstrated that OBS is the only well-known scheme for quantum work measurement that complies with all these criteria simultaneously. 

Our results are entirely general concerning the internal structure of the closed system. The latter may comprise multiple subsystems, including scenarios where heat is transferred among them. However, we did not address open quantum scenarios with \emph{external} sources transferring heat to the system of interest. Although we showed that the OBS POVM is the only well-known scheme capable of satisfying simultaneously all the criteria (i)--(iii) in the closed-case scenario, it is not necessarily expected that an observable precisely as it is defined in Eq. (1) should be the correct observable describing thermodynamics work in the open quantum case. This leads to an intriguing question: Is there a broader, more encompassing quantum work observable that not only adheres to criteria (i)-(iii) in open quantum scenarios but also aligns with the observable in Eq. (1) in the specific closed case? We believe our current work lays the groundwork for addressing this question, hopefully inspiring future research to expand and refine our understanding of the concept of work.

Interestingly, our results pave the way for many research opportunities exploring superposition, entanglement, and other correlations regarding OBS statistics of work and other two-time quantities \cite{Silva2021}. Conditions (A) and (B) introduced here could be used  to study  the classical limit in diverse fields like quantum chaos \cite{Fortes2019}, quantum optics \cite{Kenfack2004,Chabaud2021,Lemos2018,Schleich2001}, and the semiclassical formalism. A promising avenue in these lines involves conducting a comparative study between the results obtained in \cite{Jarzynski2015,Deffner2008,Deffner2010,Chen2021,Zhu2016,Garcia-Mata2017} and result 2.  Moreover, our findings support the viewpoint that conventional fluctuation theorems aimed at recovering classical expressions necessitate a departure from their conventional forms \cite{Upadhyaya2023}, revealing new connections between equilibrium properties and information resources with quantum work statistics. Furthermore, we hope our approach stimulates the analysis of the relation between work and other quantum thermodynamic facets, such as the lack of detailed balance at equilibrium (nonreciprocity) and persistent quantum currents \cite{Alicki2023,Gelbwaser2021}. These lines of research, we expect, will propel the field of quantum thermodynamics to new frontiers.

\begin{acknowledgments}
We thank Renato Angelo and Saar Rahav for their thoughtful feedback on an earlier version of this manuscript. We also thank Felipe Barra, Karen Hovhannisyan, M. Hamed Mohammady, and Thiago de Oliveira for fruitful conversations. T.A.B.P.S. was supported by the Helen Diller Quantum Center -- Technion under Grant No. 86632417 and by the ISRAEL SCIENCE FOUNDATION (grant No. 2247/22).  D.G.K. is supported by the ISRAEL SCIENCE FOUNDATION (grant No. 2247/22) and by the Council for Higher Education Support Program for Hiring Outstanding Faculty Members in Quantum Science and Technology in Research Universities.
\end{acknowledgments}



\begin{thebibliography}{80}

\bibitem{Callen1985} H. B. Callen, {\it Thermodynamics and an Introduction to Thermostatistics}, 2nd ed. (Wiley, New York, 1985).

\bibitem{Fermi1936} E. Fermi, {\it Thermodynamics} (Dover, New York, 1956).

\bibitem{Reif2009} F. Reif, {\it Fundamentals of Statistical and Thermal Physics} (Waveland Press, Long Grove, 2009).

\bibitem{Sekimoto2010} K. Sekimoto, {\it Stochastic Energetics} (Springer, Berlin, 2010).

\bibitem{Jarzynski1997} C. Jarzynski, Nonequilibrium equality for free energy differences, Phys. Rev. Lett. {\bf 78}, 2690 (1997).

\bibitem{Crooks1999} G. E. Crooks, Entropy production fluctuation theorem and the nonequilibrium work relation for free energy differences,  Phys. Rev. E {\bf 60}, 2721 (1999).

\bibitem{liphardt2002} J. Liphardt, S. Dumont, S. B. Smith, I. Tinoco Jr., and C. Bustamante, Equilibrium information from nonequilibrium measurements in an experimental test of Jarzynski's equality, Science {\bf 296}, 1832 (2002).

\bibitem{Huber2008} G. Huber, F. Schmidt-Kaler, S. Deffner, and E. Lutz, Employing trapped cold ions to verify the quantum Jarzynski equality, Phys. Rev. Lett. {\bf 101}, 070403 (2008).

\bibitem{Brandao2013} F. G. S. L. Brand{\~a}o, M. Horodecki, J. Oppenheim, J. M. Renes, and R. W. Spekkens, Resource theory of quantum states out of thermal equilibrium, Phys. Rev. Lett. {\bf 111}, 250404 (2013).

\bibitem{Aberg2018} J. \AA berg, Fully quantum fluctuation theorems, Phys. Rev. X {\bf 8}, 011019 (2018).

\bibitem{Alhambra2016} A. M. Alhambra, L. Masanes, J. Oppenheim, and C. Perry, Fluctuating work: from quantum thermodynamical identities to a second law equality, Phys. Rev. X {\bf 6}, 041017 (2016).

\bibitem{Jarzynski2004} C. Jarzynski, Nonequilibrium work theorem for a system strongly coupled to a thermal environment, J. Stat. Mech., P09005 (2004).

\bibitem{Campisi2009} M. Campisi, P. Talkner, and P. H{\"a}nggi, Fluctuation theorem for arbitrary open quantum systems, Phys. Rev. Lett. {\bf 102}, 210401 (2009).

\bibitem{Maruyana2009}  K. Maruyama, F. Nori, and V. Vedral, \emph{Colloquium}: The physics of Maxwell's demon and information, Rev. Mod. Phys. {\bf 81}, 1 (2009).

\bibitem{Sagawa2009} T. Sagawa and M. Ueda, Minimal energy cost for thermodynamic information processing: Measurement and information erasure, Phys. Rev. Lett. {\bf 102}, 250602 (2009).

\bibitem{Esposito2011} M. Esposito and C. Van den Broeck, Second law and Landauer principle far from equilibrium, Europhys. Lett. {\bf 95}, 40004 (2011).

\bibitem{Manzano2018} G. Manzano, F. Plastina, and R. Zambrini, Optimal work extraction and thermodynamics of quantum measurements and correlations, Phys. Rev. Lett. {\bf 121}, 120602 (2018).

\bibitem{Alipour2016} S. Alipour, F. Benatti, F. Bakhshinezhad, M. Afsary, S. Marcantoni, and A. T. Rezakhani, Correlations in quantum thermodynamics: Heat, work, and entropy production, Sci. Rep. {\bf 6}, 1 (2016).

\bibitem{Park2020} J. J. Park, H. Nha, S. W. Kim, and V. Vedral, Information fluctuation theorem for an open quantum bipartite system, Phys. Rev. E {\bf 101}, 052128 (2020).

\bibitem{Goold2016} J. Goold, M. Huber, A. Riera, L. del Rio, and P. Skrzypczyk, The role of quantum information in thermodynamics--a topical review, J. Phys. A: Math. Theor.{\bf 49}, 143001 (2016).

\bibitem{Campisi2011} M. Campisi, P. H{\"a}nggi, and P. Talkner, \emph{Colloquium}: Quantum fluctuation relations: Foundations and applications, Rev. Mod. Phys. {\bf 83}, 771 (2011).

\bibitem{Rossnagel2016}
J. Ro{\ss}nagel, S. T. Dawkins, K. N. Tolazzi, O. Abah, E. Lutz, F. Schmidt-Kaler, K. Singer, A single-atom heat engine, Science {\bf 352}, 325 (2016).

\bibitem{Bochkov1977} G. N. Bochkov and Y. E. Kuzovlev, General theory of thermal fluctuations in nonlinear systems, Zh. Eksp. Teor. Fiz. {\bf 72}, 238 (1977) [Sov. Phys. JETP {\bf 45}, 125 (1977)]. 

\bibitem{Alicki1979} R. Alicki, The quantum open system as a model of the heat engine, J. Phys. A {\bf 12}, L103 (1979).


\bibitem{Gemmer2005} J. Gemmer, M. Michel, and G. Mahler, {\it Quantum Thermodynamics} (Springer, Berlin, 2005).

\bibitem{Hanggi2017} P. H{\"a}nggi and P. Talkner, The other QFT, Nat. Phys. {\bf 11}, 108 (2015).

\bibitem{Millen2016} J. Millen and A. Xuereb, Perspective on quantum thermodynamics, New J. Phys. {\bf 18}, 011002 (2016).

\bibitem{Talkner2016} P. Talkner and P. H{\"a}nggi, Aspects of quantum work, Phys. Rev. E {\bf 93}, 022131 (2016).

\bibitem{Deffner2019} S. Deffner and S. Campbell {\it Quantum Thermodynamics: An Introduction to The Thermodynamics of Quantum Information} (Morgan \& Claypool, San Rafael, 2019).

\bibitem{Binder2018} F. Binder, L. A. Correa, C. Gogolin, J. Anders, and G. Adesso, {\it Thermodynamics in the Quantum Regime: Fundamental Aspects and New Directions} Fundamental Theories of Physics, Vol. 195 (Springer International Publishing, Cham, 2018)

\bibitem{Strasberg2021} P. Strasberg and A. Winter, First and second law of quantum thermodynamics: A consistent derivation based on a microscopic definition of entropy, PRX Quantum {\bf 2}, 030202 (2021).

\bibitem{Gelbwaser2013} D. Gelbwaser-Klimovsky, R. Alicki, and G. Kurizki, Work and energy gain of heat-pumped quantized amplifiers, Europhys. Lett. {\bf 103}, 60005 (2013).

\bibitem{Gelbwaser2015} D. Gelbwaser-Klimovsky and G. Kurizki, Work extraction from heat-powered quantized optomechanical setups, Sci. Rep. {\bf 5}, 7809 (2015).

\bibitem{Gelbwaser2014} D. Gelbwaser-Klimovsky and G. Kurizki, Heat-machine control by quantum-state preparation: From quantum engines to refrigerators, Phys. Rev. E {\bf 90}, 022102 (2014).

\bibitem{Gallego2016} R. Gallego, J. Eisert, and H. Wilming, Thermodynamic work from operational principles, New J. Phys. {\bf 18}, 103017 (2016).

\bibitem{Dann2023} R. Dann and R. Kosloff, Unification of the first law of quantum thermodynamics, New J. Phys. {\bf 25}, 043019 (2023).

\bibitem{Esposito2010} M. Esposito, K. Lindenberg, and C. Van den Broeck, Entropy production as correlation between system and reservoir, New J. Phys. {\bf 12}, 013013 (2010).

\bibitem{Barra2015} F. Barra, The thermodynamic cost of driving quantum systems by their boundaries, Sci. Rep. {\bf 5}, 14873 (2015).

\bibitem{Kewming2022} M. J. Kewming and S. Shrapnel, Entropy production and fluctuation theorems in a continuously monitored optical cavity at zero temperature, Quantum {\bf 6}, 685 (2022).

\bibitem{Gnezdilov2023} N. V. Gnezdilov, A. I. Pavlov, V. Ohanesjan, Y. Cheipesh, and K. Schalm, Ultrafast dynamics of cold Fermi gas after a local quench, Phys. Rev. A {\bf 107}, L031301 (2023).

\bibitem{Das2023} D. Das, G. Thomas, and A. N. Jordan, Quantum Stirling heat engine operating in finite time, Phys. Rev. A {\bf 108}, 012220 (2023).

\bibitem{Lacerda2023} A. M. Lacerda, A. Purkayastha, M. Kewming, G. T. Landi, and J. Goold, Quantum thermodynamics with fast driving and strong coupling via the mesoscopic leads approach, Phys. Rev. B {\bf 107}, 195117 (2023).

\bibitem{Ohanesjan2023} V. Ohanesjan, Y. Cheipesh, N. V. Gnezdilov, A. I. Pavlov, K. Schalm, Energy dynamics, information and heat flow in quenched cooling and the crossover from quantum to classical thermodynamics, JHEP {\bf 2023}, 1 (2023).

\bibitem{Safranek2023} D. {\v S}afr{\' a}nek, D. Rosa, and F. C. Binder, Work extraction from unknown quantum Ssurces, Phys. Rev. Lett. {\bf 130}, 210401 (2023).

\bibitem{Lorch2018} N. L\"orch, C. Bruder, N. Brunner and P. P. Hofer, Optimal work extraction from quantum states by photo-assisted Cooper pair tunneling, Quantum Sci. Technol. {\bf 3}, 035014 (2018).

\bibitem{Levy2020}
A. Levy, M. G{\"o}b, B. Deng, K. Singer, E. Torrontegui, and D. Wang, Single-atom heat engine as a sensitive thermal probe, New J. Phys. {\bf 22}, 093020 (2020).

\bibitem{Auffeves2022} A. Auff{\` e}ves, Quantum technologies need a quantum energy initiative, PRX Quantum {\bf 3}, 020101 (2022).

\bibitem{Jacob2023} S. L. Jacob, G. T. Landi, M. Esposito, and F. Barra, Two-point measurement energy statistics from particle scattering, Phys. Rev. Res. {\bf 5}, 043160 (2023).

\bibitem{Silva2021} T. A. B. Pinto Silva
and R. M. Angelo, Quantum mechanical work, Phys. Rev. A {\bf 104}, 042215 (2021).

\bibitem{Silva2023} T. A. B. Pinto Silva
and R. M. Angelo, Fluctuation theorems for genuine quantum mechanical regimes, Phys. Rev. A {\bf 107}, 052211 (2023).

\bibitem{Sone2020} A. Sone, Y.X. Liu, and P. Cappellaro, Quantum Jarzynski equality in ppen quantum systems from the one-time measurement scheme, Phys. Rev. Lett. {\bf 125}, 060602 (2020).

\bibitem{Janovitch2022} M. Janovitch and G. T. Landi, Quantum mean-square predictors and thermodynamics, Phys. Rev. A {\bf 105}, 022217 (2022).

\bibitem{Lostaglio2018} M. Lostaglio, Quantum fluctuation theorems, contextuality, and work quasiprobabilities, Phys. Rev. Lett. {\bf 120}, 040602 (2018).

\bibitem{Jarzynski2007} C. Jarzynski, Comparison of far-from-equilibrium work relations, CR Phys. {\bf 8}, 495 (2007).

\bibitem{Horowitz2007} J. Horowitz and C. Jarzynski, Comparison of work fluctuation relations, J. Stat. Phys.: Theor. Expt., P11002 (2007).

\bibitem{Vilar2008} J. M. G. Vilar and J. M. Rubi, Failure of the work-hamiltonian connection for free-energy calculations, Phys. Rev. Lett. {\bf 100}, 020601 (2008).

\bibitem{Vilar2008a} J. M. G. Vilar and J. M. Rubi, Failure of the work-hamiltonian connection for free-energy calculations, Phys. Rev. Lett. {\bf 101}, 098902 (2008).

\bibitem{Vilar2008b} J. M. G. Vilar and J. M. Rubi, Failure of the work-hamiltonian connection for free-energy calculations, Phys. Rev. Lett. {\bf 101}, 098904 (2008).

\bibitem{Horowitz2008} J. Horowitz and C. Jarzynski, Failure of the work-hamiltonian connection for free-energy calculations, Phys. Rev. Lett. {\bf 101}, 098901 (2008).

\bibitem{Peliti2008} L. Peliti, Phys. Rev. Lett. {\bf 101}, Failure of the work-hamiltonian connection for free-energy calculations, 098903 (2008).

\bibitem{Miller2017}  H. J. D. Miller and J. Anders, Time-reversal symmetric work distributions for closed quantum dynamics in the histories framework, New J. Phys. {\bf 19}, 062001 (2017).

\bibitem{Deffner2016} S. Deffner, J. P. Paz and W. H. Zurek, Quantum work and the thermodynamic cost of quantum measurements, Phys. Rev. E {\bf 94}, 010103(R) (2016).

\bibitem{Beyer2020} K. Beyer, K. Luoma and W. T. Strunz, Work as an external quantum observable and an operational quantum work fluctuation theorem, Phys. Rev. Res. {\bf 2}, 033508 (2020).

\bibitem{Beyer2022} K. Beyer, R. Uola, K. Luoma, and W. T. Strunz, Joint measurability in nonequilibrium quantum thermodynamics, Phys. Rev. E {\bf 106}, L022101 (2022).

\bibitem{Allahverdyan2014} A. E. Allahverdyan, Nonequilibrium quantum fluctuations of work, Phys. Rev. E {\bf 90}, 032137 (2014).

\bibitem{Binder2015} F. Binder, S. Vinjanampathy, K. Modi and J. Goold, Quantum thermodynamics of general quantum processes, Phys. Rev. E {\bf 91}, 032119 (2015).

\bibitem{Talkner2007} P. Talkner, E. Lutz, and P. H{\"a}nggi, Fluctuation theorems: Work is not an observable, Phys. Rev. E {\bf 75}, 050102(R) (2007).

\bibitem{Perarnau2017} M. Perarnau-Llobet, E. B{\"a}umer, K. V. Hovhannisyan, M. Huber, and A. Acin, No-go theorem for the characterization of work fluctuations in coherent quantum systems, Phys. Rev. Lett. {\bf 118}, 070601 (2017).

\bibitem{Baumer2018} E. B{\"a}umer, M. Lostaglio, M. Perarnau-Llobet, and R. Sampaio, in {\it Thermodynamics in the Quantum Regime: Fundamental Aspects and New Directions}, edited by F. Binder, L. A. Correa, C. Gogolin, J. Anders, and G. Adesso (Springer International, Cham, 2018), pp. 275-300.

\bibitem{Francica2022a} G. Francica, Class of quasiprobability distributions of work with initial quantum coherence, Phys. Rev. E {\bf 105}, 014101 (2022).

\bibitem{Francica2022b} G. Francica, Most general class of quasiprobability distributions of work, Phys. Rev. E {\bf 106}, 054129 (2022).

\bibitem{Roncaglia2014} A. J. Roncaglia, F. Cerisola, and J. P. Paz, Work measurement as a generalized quantum measurement, Phys. Rev. Lett. {\bf 113}, 250601 (2014).

\bibitem{Campaioli2018} F. Campaioli, F. A. Pollock, and S. Vinjanampathy, in {\it Thermodynamics in the Quantum Regime: Fundamental Aspects and New Directions}, edited by F. Binder, L. A. Correa, C. Gogolin, J. Anders, and G. Adesso (Springer International, Cham, 2018), pp. 207-225.

\bibitem{Gyhm2022} J.-Y. Gyhm, D. {\v S}afr{\' a}nek, and D. Rosa, Quantum charging advantage cannot be extensive without global operations, Phys. Rev. Lett. {\bf 128}, 140501 (2022).

\bibitem{Bresque2021} L. Bresque, P. A. Camati, S. Rogers, K. Murch, A. N. Jordan, and A. Auff{\`e}ves, Two-qubit engine fueled by entanglement and local measurements, Phys. Rev. Lett. {\bf 126}, 120605 (2021).

\bibitem{Rodrigues2019} F. L. S. Rodrigues, G. De Chiara, M. Paternostro, and G. T. Landi, Thermodynamics of weakly coherent collisional models, Phys. Rev. Lett. {\bf 123}, 140601 (2019).

\bibitem{Popovic2023} M. Popovic, M. T. Mitchison, J. Goold, Thermodynamics of decoherence, Proc. R. Soc. A {\bf 479}, 20230040 (2023).

\bibitem{Jarzynski2015} C. Jarzynski,
H. T. Quan, and S. Rahav, Quantum-classical correspondence principle for work distributions, Phys. Rev. X {\bf 5}, 031038 (2015).

\bibitem{Elouard2017} C. Elouard, D. A. Herrera-Mart{\'i}, M. Clusel and A. Auff{\'e}ves, The role of quantum measurement in stochastic thermodynamics, npj Quantum Inf. {\bf 3}, 9 (2017).

\bibitem{Mohammady2019} M. H. Mohammady and A. Romito, Conditional work statistics of quantum measurements, Quantum {\bf 3}, 175 (2019).

\bibitem{Faist2015} P. Faist, F. Dupuis, J. Oppenheim, and R. Renner, The minimal work cost of information processing, Nat. Comm. {\bf 6}, 7669 (2015).

\bibitem{Andolina2019} G. M. Andolina, M. Keck, A. Mari, M. Campisi, V. Giovannetti, and M. Polini, Extractable work, the role of correlations, and asymptotic freedom in quantum batteries, Phys. Rev. Lett. {\bf 122}, 047702 (2019).

\bibitem{Deffner2021} S. Deffner, Energetic cost of Hamiltonian quantum gates, Europhys. Lett. {\bf 134}, 40002 (2021).

\bibitem{Bennett1982} C. H. Bennett, The thermodynamics of computation--a review, Int. J. Theor. Phys. {\bf 21}, 905 (1982).

\bibitem{Nielsen2010} M. A. Nielsen and I. L. Chuang, {\it Quantum Computation and Quantum Information
Information}, (Cambridge University Press, New York, 2010).

\bibitem{Heinosaari2012} T. A. Heinosaari and M. Ziman, {\it The Mathematical Language of Quantum Theory}, (Cambridge University Press, New York, 2012).

\bibitem{Allahverdyan2005} A. E. Allahverdyan and Th. M. Nieuwenhuizen, Fluctuations of work from quantum subensembles: The case against quantum work-fluctuation theorems, Phys. Rev. E {\bf 71}, 066102 (2005).

\bibitem{Gelin2008} M. F. Gelin and D. S. Kosov, Unified approach to the derivation of work theorems for equilibrium and steady-state, classical and quantum Hamiltonian systems, Phys. Rev. E {\bf 78}, 011116 (2008).

\bibitem{Lindblad1983} G. Lindblad, {\it Non-equilibrium Entropy and Irreversibility} (D. Reidel Publishing Company, Dordrecht, 1983).

\bibitem{Deffner2008} S. Deffner and E. Lutz, Nonequilibrium work distribution of a quantum harmonic oscillator, Phys. Rev. E {\bf 77}, 021128 (2008).

\bibitem{Deffner2010} S. Deffner, O. Abah, and E. Lutz, Quantum work statistics of linear and nonlinear parametric oscillators, Chem. Phys. E {\bf 375}, 200 (2010).

\bibitem{Chen2021} J. F. Chen, T. Qiu, and H. T. Quan, Quantum-Classical Correspondence Principle for Heat Distribution in Quantum Brownian Motion, Entropy {\bf 23}, 1602 (2021).

\bibitem{Zhu2016} L. Zhu, Z. Gong, B. Wu, and H. T. Quan, Quantum-classical correspondence principle for work distributions in a chaotic system, Phys. Rev. E {\bf 93}, 062108 (2016).

\bibitem{Garcia-Mata2017} I. Garc\'ia-Mata,  A. J. Roncaglia, and  D. A. Wisniacki, Chaos signatures in the short and long time behavior of the out-of-time ordered correlator, Phys. Rev. E {\bf 95}, 050102 (2017).

\bibitem{EPR1935} A. Einstein, B. Podolsky, and N. Rosen, Can quantum-mechanical description of physical reality be considered complete?, Phys. Rev. {\bf 47}, 777 (1935).

\bibitem{Bell1964} J. S. Bell, On the Einstein Podolsky Rosen paradox, Physics Physique Fizika {\bf 1}, 195 (1964).

\bibitem{Zurek2009} W. H. Zurek, Quantum darwinism, Nat. Phys. {\bf 5}, 181 (2009).

\bibitem{Bilobran2015} A. L. O. Bilobran and R. M. Angelo, A measure of physical reality, Europhys. Lett. {\bf 112}, 40005 (2015).

\bibitem{Dieguez2018} P. R. Dieguez and R. M. Angelo, Information-reality complementarity: The role of measurements and quantum reference frames, Phys. Rev. A {\bf 97}, 022107 (2018).

\bibitem{Dieguez2022} P. R. Dieguez, J. R. Guimar{\~a}es, J. P. S. Peterson, R. M. Angelo, and R. M. Serra, Experimental assessment of physical realism in a quantum-controlled device, Comm. Phys. {\bf 5}, 82 (2022).

\bibitem{Mari2011} A. Mari, K. Kieling, B. Melholt Nielsen, E. S. Polzik, and J. Eisert, Directly estimating nonclassicality, Phys. Rev. Lett. {\bf 106}, 010403 (2011).

\bibitem{Kwon2019} H. Kwon,
K. C. Tan, T. Volkoff, and H. Jeong, Nonclassicality as a quantifiable resource for quantum metrology, Phys. Rev. Lett. {\bf 122}, 040503 (2019).

\bibitem{Kenfack2004} A. Kenfack and K.
{\.Z}yczkowski, Negativity of the Wigner function as an indicator of non-classicality, J. Opt. B: Quantum Semiclass. Opt. {\bf 6}, 396 (2004).

\bibitem{Chabaud2021} U. Chabaud, P.-E. Emeriau, and F.
Grosshans, Witnessing Wigner negativity, Quantum {\bf 5}, 471 (2021).

\bibitem{Lemos2018} H. C. F. Lemos, A. C. L. Almeida, B. Amaral, and A. C. Oliveira, Roughness as classicality indicator of a quantum state, Phys. Lett. A {\bf 382}, 823 (2018).

\bibitem{Campisi2011a}  M. Campisi, P. Talkner, and P. H{\"a}nggi, Quantum Bochkov-Kuzovlev work fluctuation theorems, Phil. Trans. R. Soc. A {\bf 369}, 291 (2011).

\bibitem{Hartle2018} R. H{\"a}rtle, C. Schinabeck, M. Kulkarni, D. Gelbwaser-Klimovsky, M. Thoss, and U. Peskin, Cooling by heating in nonequilibrium nanosystems, Phys. Rev. B {\bf 98}, 081404(R) (2018).

\bibitem{Gelbwaser2018} D. Gelbwaser-Klimovsky, A. Bylinskii, D. Gangloff, R. Islam, A. Aspuru-Guzik, and V. Vule, Single-atom heat machines enabled by energy quantization, Phys. Rev. Lett. {\bf 120}, 170601 (2018).

\bibitem{Wigner1932} E. Wigner, On the quantum correction for thermodynamic equilibrium, Phys. Rev. {\bf 40}, 749 (1932).

\bibitem{Baumgratz2014} T. Baumgratz, M. Cramer, and M. B. Plenio, Quantifying coherence, Phys. Rev. Lett. {\bf 113}, 140401 (2014).

\bibitem{SMref} See Supplemental Material at [URL will be inserted by publisher], which includes Refs. [21, 67, 110, 112-114, 116, 120-126], for detailed proofs of results 1 and 2 and an in-depth analysis of the model described in Eqs. (4) and (5).

\bibitem{Cohen2020} C. Cohen-Tannoudji, B. Diu, and F. Lalo{\"e} {\it Quantum
Mechanics, Volume I: Basic Concepts, Tools, and Applications} Second Edition (Wiley-VCH Verlag GmbH \& Co, Weinheim, 2020).

\bibitem{Suslov2005} S. K. Suslov, {\it An analog of the Cauchy-Hadamard formula for expansions in q-polynomials.} in {\it Theory and Applications of Special Functions: A Volume Dedicated to Mizan Rahman}, 2nd ed. (Springer Science \& Business Media, New York, 2005).

\bibitem{Ahlfors1979} L. V. Ahlfors, {\it Complex Analysis: An Introduction to the Theory of Analytic Functions of One Complex Variable}, 3rd ed. (McGraw-Hill, New York, 1979).

\bibitem{Fortes2019} E. M. Fortes, I. Garc\'ia-Mata, R. A. Jalabert, and D. A. Wisniacki, Gauging classical and quantum integrability through out-of-time-ordered correlators, Phys. Rev. E {\bf 100}, 042201 (2019).

\bibitem{Schleich2001} W. P. Schleich {\it Quantum optics in phase space} (Wiley-VCH Verlag GmbH \& Co, Berlin, 2001).

\bibitem{Upadhyaya2023} T. Upadhyaya, W. F. Braasch, Jr., G. T. Landi, and N. Y. Halpern, What happens to entropy production when conserved quantities fail to commute with each other, \emph{arXiv:2305.15480}.

\bibitem{Alicki2023} R. Alicki, M. {\u S}indelka, and D. Gelbwaser-Klimovsky, Violation of detailed balance in quantum open systems, Phys. Rev. Lett. {\bf 131}, 040401 (2023).

\bibitem{Gelbwaser2021} D. Gelbwaser-Klimovsky, N. Graham, M. Kardar, and M. Kr{\"u}ger, Near field propulsion forces from nonreciprocal media, Phys. Rev. Lett. {\bf 126}, 170401 (2021).

\bibitem{Athreya2006} K. B. Athreya and  S. N. Lahiri, {\it Measure theory and probability theory}, (Springer, New York, 2006).

\bibitem{Transtrum2005} M. K. Transtrum and J. F. S. Van Huele, Commutation relations for functions of operators, J. Math. Phys. {\bf 46}, 6 (2005).

\bibitem{Fristedt2013} B. E. Fristedt and  L. F. Gray, {\it A modern approach to probability theory}, (Springer Science \& Business Media, New York, 2013).

\bibitem{Zhang2016} Y. R. Zhang,  L. H. Shao, Y. Li, and H. Fan, Quantifying coherence in infinite-dimensional systems, Phys. Rev. A {\bf 93}, 012334 (2016).

\bibitem{Eisert2002} J. Eisert, C. Simon, and M. B. Plenio, On the quantification of entanglement in infinite-dimensional quantum systems, J. Phys. A Math. Theor. {\bf 35}, 3911 (2002).

\bibitem{Streltsov2017} A. Streltsov,  G. Adesso, and M. B. Plenio, \emph{Colloquium}: Quantum coherence as a resource, Rev. Mod. Phys.{\bf 89}, 041003 (2017).

\bibitem{Lighthill1959} M. J. Lighthill {\it Introduction to Fourier analysis and generalised functions} (Cambridge University Press, London, 1959).

\end{thebibliography}
\end{document}


\title{Supplemental Material for\\ ``Quantum work: reconciling quantum mechanics and thermodynamics''}

\author{Thales A. B. Pinto Silva}
\email{pinto\_silva@campus.technion.ac.il}
\author{David Gelbwaser-Klimovsky}
\affiliation{Schulich Faculty of Chemistry and Helen Diller Quantum Center, Technion-Israel Institute of Technology, Haifa 3200003, Israel}

\maketitle

We provide detailed proofs of the results presented in the main text. Specifically, we derive results 1 and 2, followed by an in-depth analysis of the model described in Equations (4) and (5).

Throughout this supplement, we consider that a system
initially in a state $\rho$ ($t=0$) undergoes a quantum process of duration $\tau$. The processes' dynamics at time $t$ is described by the unitary evolution operator $U_t$, which is the solution of the Schr{\"o}dinger equation $i\hbar \partial_t U_t=H(t)U_t$. Here, $H(t)=\sum_j e_{j}(t) \ket{e_j (t)}\bra{e_j (t)}$ is a time-dependent Schr{\"o}dinger Hamiltonian whose eigenstates and eigenvalues are represented by $\{\ket{e_j(t)}\}$ and $\{e_j (t)\}$. In the Heisenberg picture $H(t)$ is thus defined as $H_{h}(t)=U_{t}^{\dagger}H(t)U_t$. The  quantum process is completely characterized by the quadruple $\{H(t),U_t,[0,\tau],\rho\}$.

\section{Derivation of result 1}

To derive our first result, we focus on the work statistics of the quantum processes described by $\{H(t),U_t,[0,\tau],\rho\}$. Specifically, we analyze the observable (OBS) and two-point measurement (TPM) statistics. As discussed in the main text, the TPM work probability is computed as
\begin{equation}
    P_{\text{\tiny TPM}}(w)=\Tr[M_{\text{\tiny TPM}}(w)\,\rho], \quad M_{\text{\tiny TPM}}(w)=\sum_{mn}\delta[w-(e_{m}(\tau)-e_{n}(0))]\mf{p}_{m|n}\ket{e_{n}(0)}\bra{e_{n}(0)}\label{TPMsm}
\end{equation}  
where $M_{\text{\tiny TPM}}(w)$ are the elements of the TPM POVM, satisfying $\int_{-\infty}^{\infty}\,dw\, M_{\text{\tiny TPM}}(w)=\mathbbm{1}$, and $\delta$ is the Dirac's delta. On the other hand, the OBS probability distribution is described by
\begin{equation}
    P_{\text{\tiny OBS}}(w)=\Tr[M_{\text{\tiny OBS}}(w)\rho], \quad  M_{\text{\tiny OBS}}(w)=\sum_{j}\delta[w-w_{j}(\tau,0)]\ket{w_{j}(\tau,0)}\bra{w_{j}(\tau,0)}
\end{equation}
where $M_{\text{\tiny OBS}}(w)$ are the elements of the OBS POVM ($\int_{-\infty}^{\infty}\,dw\, M_{\text{\tiny TPM}}(w)=\mathbbm{1}$) and $\{\ket{w_{j}(\tau,0)}\}$ and $\{w_{j}(\tau,0)\}$ are respectively the eigenvectors and eigenvalues of the work operator 
\begin{equation}
    W(\tau,0)=H_{h}(\tau)-H(0).\label{Woperator}
\end{equation}

To make the comparisons between $P_{\text{\tiny TPM}}(w)$ and $P_{\text{\tiny OBS}}(w)$ statistics, we considered in our proofs their associated characteristic functions, defined as \cite{Athreya2006,Fristedt2013,Talkner2007}
\begin{equation}
    \chi_{\text{\tiny TPM}}(u)=\int_{-\infty}^{\infty}\,dw\,\mathrm{e}^{iuw} P_{\text{\tiny TPM}}(w)=\left\langle\Phi_{H(0)}\left[\mathcal{T}_{>}\left(\mathrm{e}^{iuW(\tau,0)}\right)\right]\right\rangle\quad\text{and}\quad  \chi_{\text{\tiny OBS}}(u)=\int_{-\infty}^{\infty}\,dw\,\mathrm{e}^{iuw} P_{\text{\tiny OBS}}(w)=\braket{\mathrm{e}^{iuW(\tau,0)}}.\label{Guquantum}
\end{equation}
where, for any operator $A$, we denote 
\begin{equation}
    \Phi_{H(t)}(A)=\sum_{j}\ket{e_j(t)}\bra{e_j(t)}A\ket{e_j(t)}\bra{e_j(t)}
\end{equation} 
as the dephasing map of $A$ with respect to the energy basis at time $t$ and
\begin{equation}
    \braket{A}\coloneqq \Tr [A\rho] 
\end{equation}
 as its average with respect to the initial state. $\mathcal{T}_{>}$ is the time ordering operator, such that for any two arbitrary operators $A(t')$ and $B(t'')$ defined at times $t''>t'$, $\mathcal{T}_{>}[A(t')B(t'')]=B(t'')A(t')$ and $\mathcal{T}_{>}[B(t'')A(t')]=B(t'')A(t')$. 

 The use of characteristic functions is a well-established strategy in quantum thermodynamics literature \cite{Talkner2007,Campisi2011a,Campisi2011}. Notably, characteristic functions possess important properties worth highlighting. Firstly, for any work distribution $P(w)$,
 \begin{equation}
    P(w)=\frac{1}{2\pi}\int_{-\infty}^{\infty}du\,\mathrm{e}^{-iuw} \chi(u),\label{discf}
\end{equation}
i.e. the probability distribution is the Fourier transform of its characteristic function. In fact, it was proven in \cite{Talkner2007} that whenever the initial state is initially incoherent $\rho=\Phi_{H(0)}[\rho(t)]$, and $[H_h(\tau),H(0)]=0$, then $\chi_{\text{\tiny TPM}}(u)=\chi_{\text{\tiny OBS}}(u)$ and, from Eq. \eqref{discf}, $P_{\text{\tiny TPM}}(w)=P_{\text{\tiny OBS}}(w)$.

It follows directly from Eqs. \eqref{Guquantum} and \eqref{discf} and the fact that $P_{\text{\tiny TPM}}(w)$ and $P_{\text{\tiny OBS}}(w)$ are non-negative and normalized, that 
\begin{equation}
    |\chi_{\text{\tiny TPM}}(u)|\leq 1,\quad |\chi_{\text{\tiny OBS}}(u)|\leq 1,\quad \int_{-\infty}^{\infty}\,dw\,\left|\int_{-\infty}^{\infty}du\,\mathrm{e}^{-iuw} [\chi_{\text{\tiny TPM}}(u)-\chi_{\text{\tiny OBS}}(u)]\right|\leq 4\pi.\label{cfprop}
\end{equation}
It is important to highlight the fact that Eq. \eqref{cfprop} is not exclusive for the TPM or OBS statistics, but hold for general pairs of consistent characteristic functions.

Finally, we notice that we can rewrite the OBS characteristic function as 
\begin{equation}
    \chi_{\text{\tiny OBS}}(u)=\braket{\mathrm{e}^{iuW(\tau,0)}}=\sum_{n=0}^\infty\frac{(iu)^n}{n!}\braket{W^n(\tau,0)}\label{Charactobs2}
\end{equation}
where $\braket{W^{n}(\tau,0)}$ can be written, for any power $n$, as
\begin{equation}
    \braket{W^{n}(\tau,0)}=\sum_{i=1}^{2^{n}}a_{ni}\braket{H_{h}^{I_{1i}}(t)H^{J_{1i}}(0)\cdots H_{h}^{I_{s(i)i}}(t)H^{J_{s(i)i}}(0)}\label{Wnweird}
\end{equation}
where $a_{ni}=\pm 1$ and the sum over $i$ covers \emph{all} the $2^n$ possible combinations of product of powers $\mathbf{I}_{i}=\{I_{1i},I_{2i}\cdots,I_{s(i)i}\}\subset \mathbbm{Z}$ and $\mathbf{J}_{i}=\{J_{1i},J_{2i}\cdots,J_{s(i)i}\}\subset \mathbbm{Z}$, such that $\mathbf{I}_{i}+\mathbf{J}_{i}\coloneqq \sum_{j}(I_{ji}+J_{ji})=n$. Here, $s(i)\leq n$ denotes the maximum number of powers $I_{s(i)i}$ and $J_{s(i)i}$ for each $i$-th combination. The form of Eq. \eqref {Wnweird} of writing $\braket{W^{n}(\tau,0)}$ can be proved by induction. For instance, notice that Eq. \eqref {Wnweird} holds for $n=2$, since $\braket{W^{2}(\tau,0)}=\braket{H_{h}^2(t)}-\braket{H_{h}(t)H(0)}-\braket{H(0)H_{h}(t)}+\braket{H^2(0)}$. Then, by induction, we can show that it is valid for $n=3,4,\cdots$ and so on. Similarly, the TPM characteristic function can be written as
\begin{equation}
    \chi_{\text{\tiny TPM}}(u)=\left\langle\Phi_{H(0)}\left[\mathcal{T}_{>}\left(\mathrm{e}^{iuW(\tau,0)}\right)\right]\right\rangle=\sum_{n=0}^\infty\frac{(iu)^n}{n!}\left\langle\Phi_{H(0)}\left[\mathcal{T}_{>}\left(W^{n}\right)\right]\right\rangle\label{CharacTPM2}
\end{equation}\
where
\begin{equation}
    \left\langle\Phi_{H(0)}\left[\mathcal{T}_{>}\left(W^{n}\right)\right]\right\rangle=\sum_{i=1}^{2^{n}}a_{ni}\left\langle\Phi_{H(0)}\left[H_{h}^{\sum_{j=1}^{s(i)}I_{ji}}(t)H^{\sum_{j=1}^{s(i)}I_{ji}}(0)\right]\right\rangle\label{Wnweirdtpm}
\end{equation}
with the same coefficients $a_{ni}$ as in Eq. \eqref{Wnweird}. As a last requirement to establish Result 1, we assume that the average energy is always bounded. This condition has been previously explored in the context of coherences to avoid singularities related to measures of coherence and entanglement \cite{Baumgratz2014,Zhang2016,Eisert2002}. Incorporating these properties into our analysis, we derive Result 1 presented in the main text:

\begin{result}
    Consider a process $\{H(t),U_t,[0,\tau],\rho\}$ in which, for any initial incoherent state $\rho=\Phi_{H(0)}[\rho(t)]$, the average of energy is bounded $\braket{H(t)}=\Tr\left[H_{h}(t)\rho\right]<\infty$ and the 1-norm measure of coherences \cite{Baumgratz2014} is bounded as follows
    \begin{equation}
        C(\rho(t))=\|\rho(t)-\Phi_{H(t)}(\rho(t)) \|_1=\sum_{j\neq k}|\bra{e_j (t)}\rho(t)\ket{e_k (t)}|\leq \epsilon_1,\label{Crhota1}
    \end{equation}
     for all $t\in[0,\tau]$, where $\epsilon_1$ is a non-null real scalar. Then, it follows that
     \begin{equation}
         \int_{-\infty}^{\infty}dw\,|P_{\text{\tiny TPM}}(w)- P_{\text{\tiny OBS}}(w)|\leq C_{1}\epsilon_1,\label{Res1norm1}
     \end{equation}
     where $C_1$ is a bounded positive real scalar that is independent of $\epsilon_1$. Therefore,
     \begin{equation}
         \lim_{\epsilon\to 0} P_{\text{\tiny TPM}}(w)= P_{\text{\tiny OBS}}(w).\label{Res1lim}
     \end{equation}
\end{result}
\begin{proof}
Let us first consider the case in which $C(\rho(t))= 0$, i.e. $\epsilon_1= 0$, for every initially incoherent state. It then follows that $C(\rho(t))= 0$ for $\rho=\ket{e_l(0)}\bra{e_l(0)}$.  From Eq. \eqref{Res1norm1}, it follows that $|\bra{e_j(t)}U_t\ket{e_l(0)}\bra{e_l(0)}U_t^\dagger\ket{e_k(t)}|=0$ for all $l$ and $j\neq k$, and $t\in[0,\tau]$. Now, for any $j,k$, $\bra{e_j(t)}[H(t),U_\tau H (0)U_{t}^\dagger]\ket{e_k(t)}=\sum_{l}e_{l}(0) \bra{e_j(t)}U_t\ket{e_l(0)}\bra{e_l(0)}U_t^\dagger\ket{e_k(t)}(e_j(t) -e_k (t))$. Moreover, since the average of energy is bounded for any time $t\in[0,\tau]$, it follows that $\bra{e_j(t)}U_t\ket{e_l(0)}$ is non-null only for bounded values of $e_l(0)$ and $e_k(t)$. As a result, $[H(t),U_\tau H (0)U_{t}^\dagger]= 0$ and $[H_h(t),H(0)]= 0$. The proof for the case in which $\epsilon_1\to 0$ is completed employing the results of reference \cite{Talkner2007}, where it was shown that whenever $[H_{h}(t),H(0)]=0$, then $P_{\text{\tiny OBS}}(w)=P_{\text{\tiny TPM}}(w)$.

Now, consider the case in which $\epsilon_1>0$. Since Eq. \eqref{Crhota1} holds for any incoherent initial state $\rho$ at $t=0,\tau$, then it must be valid for any initial arbitrary eigenstate $\rho=\ket{e_{l} (0)}\bra{e_{l} (0)}$, for any $l$. Consequently, 
\begin{equation}
    \sum_{j\neq k}|\alpha_{lj}(t)||\alpha_{lk}(t)|\leq \epsilon_1,\,\, \forall(l,j\neq k)\,\,\text{and}\,\,t=0,\tau.\label{aux1}
\end{equation}
where $\alpha_{lj}(t)=\bra{e_l(0)}U_t^\dagger\ket{e_j (t)}$ and $\alpha_{lj}^{*}$ its conjugate. 

Taking into account Eq. \eqref{aux1}, let us compare the characteristic functions $\chi_{\text{\tiny OBS}}(u)$ and $\chi_{\text{\tiny TPM}}(u)$ considering an arbitrary initial incoherent state $\rho$ and Eqs. \eqref{Charactobs2} and \eqref{CharacTPM2}. We first notice, from Eq. \eqref{aux1}, that the components of the commutator $[ H^m (t),U_\tau H^n (0)U_{t}^\dagger]$ for any power $m$ and $n$ can be written as
\begin{equation}
    \bra{e_j (t)}[ H^m(t),UH^n(0)U^\dagger]\ket{e_k (t)}=\sum_{l}e_{l}^n(0) \alpha_{lj}(t)\alpha_{lk}^*(t)(e_j^m (t) -e_k^m (t))=\epsilon_1\sum_{l}e_{l}^n(0) \gamma_{jkl}(t,\epsilon_1)(e_j^m (t) -e_k^m (t)).\label{aux2}
\end{equation}
Here, we considered that $j=k\implies \bra{e_j (t)}[H^m(t),UH^n(0)U^\dagger]\ket{e_k (t)}=0$ and $j\neq k\implies |\alpha_{lj}(t)\alpha_{lk}^*(t)|\leq \epsilon_1$. From Eq. \eqref{aux1}, we defined 
\begin{equation}
    \alpha_{lj}(t)\alpha_{lk}^*(t)=\gamma_{jkl}(t,\epsilon_1)\epsilon_1,\label{gammadef}
\end{equation}
since the upper bound of $|\alpha_{lj}(t)\alpha_{lk}^*(t)|$ decays linearly with $\epsilon_1$. $\gamma_{jkl}(t,\epsilon_1)$ is a complex-valued function satisfying $\sum_{j\neq k}|\gamma_{jkl}(t,\epsilon_1)|\leq 1$ and  $\lim_{\epsilon_1\to 0}\gamma_{jkl}(t,\epsilon_1)\epsilon_1=0$ (see Eq. \eqref{aux1}). Since $H_{h}^{m}(t)=U_{t}^{\dagger}H^{m}(t)U_t$ and $UH^n(0) U^\dagger H^m(t)=H^m(t)UH^n(0) U^\dagger+[UH^n(0) U^\dagger, H^m(t)]$, then it follows that, for any initial incoherent state $\rho$, the powers in Eq. \eqref{Wnweird} can be written from Eq. \eqref{aux2} as
\begin{equation}
\ba{rl}
    \displaystyle\braket{H_{h}^{I_{1i}}(t)H^{J_{1i}}(0)\cdots H_{h}^{I_{s(i)i}}(t)H^{J_{s(i)i}}}&\displaystyle=\left\langle\left(\prod_{\mu=1}^{s(i)-2}H_{h}^{I_{\mu i}}(t)H^{J_{\mu i}}(0)\right)U_{t}^\dagger H^{I_{(s(i)-1)i}}(t) H^{I_{s(i)i}}(t)U_{t}H^{J_{(s(i)-1)i}}(0)U_{t}^\dagger U_{t}H^{J_{s(i)i}}(0)\right\rangle+\\
    &\displaystyle+\left\langle\left(\prod_{\mu=1}^{s(i)-2}H_{h}^{I_{\mu i}}(t)H^{J_{\mu i}}(0)\right)U_{t}^\dagger H^{I_{(s(i)-1)i}}(t)[U_{t}H^{J_{(s(i)-1)i}}(0)U_{t}^\dagger, H^{I_{s(i)i}}(t)]U_{t}H^{J_{s(i)i}}(0)\right\rangle\\
    &\displaystyle=\left\langle\left(\prod_{\mu=1}^{s(i)-2}H_{h}^{I_{\mu i}}(t)H^{J_{\mu i}}(0)\right)U_{t}^\dagger H^{I_{(s(i)-1)i}}(t) H^{I_{s(i)i}}(t)U_{t}H^{J_{(s(i)-1)i}}(0)U_{t}^\dagger U_{t}H^{J_{s(i)i}}(0)\right\rangle+C_{1i}(\epsilon_1)\epsilon_1\label{aux3}
    \ea
\end{equation}
where
\begin{equation}
    C_{1i}(\epsilon_1)=\sum_{l,j\neq k}e_{l}^{J_{(s(i)-1)i}}(0) \gamma_{jkl}(t,\epsilon_1) (e_k^{I_{s(i)i}} (t) -e_j^{I_{s(i)i}} (t))\left\langle\left(\prod_{\mu=1}^{s(i)-2}H_{h}^{I_{\mu i}}(t)H^{J_{\mu i}}(0)\right)U_{t}^\dagger H^{I_{(s(i)-1)i}}(t)\ket{e_{j}(t)}\bra{e_{k}(t)}U_{t}H^{J_{s(i)i}}(0)\right\rangle.\label{C1i}
\end{equation}

A similar procedure as the one used to obtain Eq. \eqref{aux3} can be considered to deduce
\begin{equation}
\ba{rl}
    \displaystyle\braket{H_{h}^{I_{1i}}(t)H^{J_{1i}}(0)\cdots H_{h}^{I_{s(i)i}}(t)H^{J_{s(i)i}}}&\displaystyle=\left\langle\left(\prod_{\mu=1}^{s(i)-3}H_{h}^{I_{\mu i}}(t)H^{J_{\mu i}}(0)\right)U_{t}^\dagger H^{I_{(s(i)-2)i}+I_{(s(i)-1)i}+I_{s(i)i}}(t) U_{t}H^{J_{(s(i)-2)i}+J_{(s(i)-1)i}+J_{s(i)i}}(0)\right\rangle+\\
    &\displaystyle+(C_{1i}(\epsilon_1)+C_{2i}(\epsilon_1))\epsilon_1,\label{aux4}
    \ea
\end{equation}
where
\begin{equation}
    C_{2i}(\epsilon_1)=\sum_{l,j\neq k}e_{l}^{J_{(s(i)-2)i}}(0) \gamma_{jkl}(t,\epsilon_1) (e_k^{I_{(s(i)-1)i}+I_{s(i)i}} (t) -e_j^{I_{(s(i)-1)i}+I_{s(i)i}}(t))\left\langle\left(\prod_{\mu=1}^{s(i)-3}H_{h}^{I_{\mu i}}(t)H^{J_{\mu i}}(0)\right)U_{t}^\dagger H^{I_{(s(i)-2)i}}(t)\ket{e_{j}(t)}\bra{e_{k}(t)}U_{t}H^{J_{(s(i)-1)i}+J_{s(i)i}}(0)\right\rangle.
\end{equation}
Repeating this procedure $s-1$ times, we obtain
\begin{equation}
    \braket{H_{h}^{I_{1i}}(t)H^{J_{1i}}(0)\cdots H_{h}^{I_{s(i)i}}(t)H^{J_{s(i)i}}}=\braket{H_{h}^{\sum_{j=1}^{s(i)}I_{ji}}(t)H^{\sum_{j=1}^{s(i)}I_{ji}}(0)}+C_{\mathbf{I}_i\mathbf{J}_i}(\epsilon_1)\epsilon_1,\label{induction}
\end{equation}
where $C_{\mathbf{I}_i\mathbf{J}_i}(\epsilon_1)=\sum_{m=1}^{s(i)-1}C_{mi}(\epsilon_1)$ and 
\begin{equation}
    \ba{rl}
    \displaystyle C_{mi}(\epsilon_1)&\displaystyle=\sum_{l,j\neq k}e_{l}^{J_{(s(i)-m)i}}(0) \gamma_{jkl}(t,\epsilon_1) (e_k^{\sum_{\nu=0}^{m-1}I_{(s(i)-\nu)i}} (t) -e_j^{\sum_{\nu=0}^{m-1}I_{(s(i)-\nu)i}}(t))\times\\
    &\displaystyle\times\left\langle\left(\prod_{\mu=1}^{s(i)-(m+1)}H_{h}^{I_{\mu i}}(t)H^{J_{\mu i}}(0)\right)U_{t}^\dagger H^{I_{(s(i)-m)i}}(t)\ket{e_{j}(t)}\bra{e_{k}(t)}U_{t}H^{\sum_{\nu=0}^{m-1}J_{(s(i)-\nu)i}}(0)\right\rangle.
    \ea\label{Cm}
\end{equation}
Considering the OBS and TPM characteristic functions in Eqs. \eqref{Charactobs2} and Eq. \eqref{CharacTPM2}, the cyclic property of the trace, and Eqs. \eqref{Wnweird}, \eqref{Wnweirdtpm}, and \eqref{induction}, we obtain
    \begin{equation}
            \chi_{\text{\tiny OBS}}(u)-\chi_{\text{\tiny TPM}}(u)=\sum_{n=0}^\infty\frac{(iu)^n}{n!}\sum_{i=1}^{2^{n}}a_{ni}\left[\braket{H_{h}^{I_{1i}}(t)H^{J_{1i}}(0)\cdots H_{h}^{I_{s(i)i}}(t)H^{J_{s(i)i}}}-\braket{H_{h}^{\sum_{j=1}^{s(i)}I_{ji}}(t)H^{\sum_{j=1}^{s(i)}I_{ji}}(0)}\right]=y(u,t,\epsilon_1)\epsilon_1,\label{Charactobslemma}
    \end{equation}
    with the function
    \begin{equation}
        y(u,t,\epsilon_1)=\sum_{n=0}^\infty\frac{(iu)^n}{n!}\sum_{i=1}^{2^{n}}a_{ni}C_{\mathbf{I}_i\mathbf{J}_i}(\epsilon_1).\label{yute10}
    \end{equation}
    As a result, 
    \begin{equation}
        \int_{-\infty}^{\infty}dw\,|P_{\text{\tiny TPM}}(w)- P_{\text{\tiny OBS}}(w)|=y'(t,\epsilon_1) \epsilon_1 .\label{relationpylinha}
     \end{equation}
     where 
    \begin{equation}
         y'(t,\epsilon_1)=\frac{1}{2\pi}\int_{-\infty}^{\infty}\,dw\,\left|\int_{-\infty}^{\infty}du\,\mathrm{e}^{-iuw} y(u,t,\epsilon_1)\right|.\label{ylinha}
     \end{equation}
     The proof of the result is finished by showing that $y'(t,\epsilon_1)$ is bounded for every $\epsilon_1$. This is proved in the subsection ``\emph{Boundness of $y'(t,\epsilon_1)$}'' below so that we can define
     \begin{equation}
         C_1=\max_{\epsilon_1}|y'(t,\epsilon_1)|,\label{C1def}
     \end{equation}
     where $C_1<\infty$ as a positive bounded real scalar. As a result, it follows that
     \begin{equation}
         \int_{-\infty}^{\infty}dw\,|P_{\text{\tiny TPM}}(w)- P_{\text{\tiny OBS}}(w)|=y'(t,\epsilon_1)\epsilon_1\leq |y'(t,\epsilon_1)|\epsilon_1\leq C_{1}\epsilon_1.\label{Res1norm10}
     \end{equation}
     Consequently,
    \begin{equation}
        \lim_{\epsilon_1\to 0}\int_{-\infty}^{\infty}dw\,|P_{\text{\tiny TPM}}(w)- P_{\text{\tiny OBS}}(w)|=0,
    \end{equation}
    which implies in Eq. \eqref{Res1lim}. 

 \end{proof}

It is important to note that the $l_1$ norm measure of coherences, $C[\rho(t)]$ in Eq.~\eqref{Crhota1}, is commonly employed in finite-dimensional cases. However, in the context of infinite-dimensional systems, certain challenges may arise with this measure, as discussed in \cite{Baumgratz2014,Zhang2016}. Remarkably, our results extend naturally to infinite-dimensional scenarios when considering the relative entropy of coherence \cite{Streltsov2017}
\begin{equation}
    C_{\text{\tiny REL}}[\rho(t)] = S[\Phi_{H(t)}(\rho(t))] - S[\rho(t)] \leq \epsilon_1,
\end{equation}
where $S(\rho) = -\Tr [\ln(\rho)\rho]$ represents the von Neumann entropy. This measure was shown to be consistent in both the finite- and infinite-dimensional cases \cite{Baumgratz2014,Zhang2016,Streltsov2017}. By employing an argument similar to that used to derive Eq.~\eqref{aux1}, we can establish
\begin{equation}
    -\sum_{j}|\alpha_{lj}(t)|^2\ln(|\alpha_{lj}(t)|^2) \leq \epsilon_1, \quad \forall l,
\end{equation}
which implies 
\begin{equation}
    |\alpha_{lj}(t)||\alpha_{lk}(t)|\leq \sqrt{\epsilon_1},\,\, \forall(l,j\neq k)\,\,\text{and}\,\,t\in[0,\tau].\label{aux12}
\end{equation} 
This can be proven by contradiction. Let us assume that Eq. \eqref{aux12} is not valid so that there is at least one $j$ and $k$ such that $j\neq k$ and $|\alpha_{lj}(t)||\alpha_{lk}(t)|> \sqrt{\epsilon_1}$. Since $\sum_{i}|\alpha_{li}(t)|^2=1$, then it follows that $|\alpha_{lj}(t)|^2|\alpha_{lk}(t)|^2\leq |\alpha_{lj}(t)|^2(1-|\alpha_{lj}(t)|^2)\leq 1/4$. Defining $\alpha_{\min}=\min\{|\alpha_{lj}(t)|,|\alpha_{lk}(t)|\}$ and $\alpha_{\max}=\max\{|\alpha_{lj}(t)|,|\alpha_{lk}(t)|\}$, then it follows that 
\begin{equation}
    \alpha_{\max}^2\alpha_{\min}^2=|\alpha_{lj}(t)|^2|\alpha_{lk}(t)|^2>\epsilon_1\geq -\sum_{j}|\alpha_{lj}(t)|^2\ln(|\alpha_{lj}(t)|^2) \geq -\alpha_{\min}^2[ \ln(\alpha_{\min}^2)+ \ln(\alpha_{\max}^2)]\geq \alpha_{\min}^2 \ln 4 
\end{equation}
which holds only if $\alpha_{\max}\geq \ln 4$, a contradction with $\sum_{i}|\alpha_{li}(t)|^2=1$. Consequently, Result 1 can be derived, substituting $\epsilon_1$ by $\sqrt{\epsilon_1}$, using the same arguments and equations as presented after Eq.~\eqref{aux1} in the previous proof.
\subsection{Boundness of $y'(t,\epsilon_1)$}

To prove that $y'(t,\epsilon_1)$ is bounded, we first compute its limit when $\epsilon_1\to 0$. For that, let us first consider the limit $\epsilon_1\to 0$ for $y(u,t,\epsilon_1)$ as defined in Eq. \eqref{yute10}:
\begin{equation}
        \lim_{\epsilon_1\to 0}y(u,t,\epsilon_1)=\lim_{\epsilon_1\to 0}\sum_{n=0}^\infty\frac{(iu)^n}{n!}\sum_{i=1}^{2^{n}}a_{ni}C_{\mathbf{I}_i\mathbf{J}_i}(\epsilon_1)=\sum_{n=0}^\infty\frac{(iu)^n}{n!}\sum_{i=1}^{2^{n}}a_{ni}\lim_{\epsilon_1\to 0}C_{\mathbf{I}_i\mathbf{J}_i}(\epsilon_1).\label{yute102}
    \end{equation}
where the indexes $\mathbf{I}_i=\{I_{1i},I_{2i},\cdots,I_{s(i)i}\}$ and $\mathbf{J}_i=\{J_{1i},J_{2i},\cdots,I_{s(i)i}\}$ describing $C_{\mathbf{I}_i\mathbf{J}_i}(\epsilon_1)$ satisfy $\sum_{\mu=1}^{s(i)}(I_{\mu i}+J_{\mu i})=n$, $C_{\mathbf{I}_i\mathbf{J}_i}(\epsilon_1)=\sum_{m=1}^{s(i)-1}C_{mi}(\epsilon_1)$, and $C_{mi}(\epsilon_1)$ is defined in Eq. \eqref{Cm}. Notice that, because $\sum_{n=0}^N\frac{(iu)^n}{n!}\sum_{i=1}^{2^{n}}a_{ni}C_{\mathbf{I}_i\mathbf{J}_i}(\epsilon_1)$ converges to $y(u,t,\epsilon_1)$ in the limit $N\to \infty$ for any $\epsilon_1$ (therefore pointwise in $\epsilon_1$), then we could exchange the limit $\epsilon_1\to 0$ and the sum $\sum_{n=0}^\infty$ from the second to the third equality \footnote{Consider any sequence of functions $f_{N}(\epsilon_1)=\Sigma_{n=0}^{N}a_{n}(\epsilon_1)$ that converges to $f(\epsilon_1)$ for any $\epsilon_1$. Then for any $\varepsilon>0$ there exist a natural finite number $N'$, such that for every $N>N'$, $|f_{N}(\epsilon_1)-f(\epsilon_1)|\leq \varepsilon$ for every $\epsilon_1>0$. Therefore, $\sup_{\epsilon_1}|f_{N}(\epsilon_1)-f(\epsilon_1)|\leq \varepsilon$. As a result, $|\lim_{\epsilon_1\to 0}f_{N}(\epsilon_1)-\lim_{\epsilon_1\to 0}f(\epsilon_1)|\leq \varepsilon$. Since $\varepsilon$ is arbitrary, then for every $\varepsilon>0$, for $N>N'$, $|\lim_{\epsilon_1\to 0}f_{N}(\epsilon_1)-\lim_{\epsilon_1\to 0}f(\epsilon_1)|\leq \varepsilon$. This means that $\lim_{N\to\infty}\lim_{\epsilon_1\to 0}f_{N}(\epsilon_1)=\lim_{\epsilon_1\to 0}f(\epsilon_1)$. This result is considered in many parts of the deductions in the section ``boundness of $y'(t,\epsilon_1)$.''}. Since $\lim_{\epsilon_1\to 0}C_{\mathbf{I}_i\mathbf{J}_i}(\epsilon_1)=\sum_{m=1}^{s(i)-1} \lim_{\epsilon_1\to 0} C_{mi}(\epsilon_1)$, our focus is now to compute $\lim_{\epsilon_1\to 0} C_{mi}(\epsilon_1)$, so that, by considering Eq. \eqref{Cm}, we aim to explicitly compute
\begin{equation}
    \ba{rl}
    \displaystyle \lim_{\epsilon_1\to 0}C_{mi}(\epsilon_1)&\displaystyle=\lim_{\epsilon_1\to 0}\sum_{l,j\neq k}e_{l}^{J_{(s(i)-m)i}}(0) \gamma_{jkl}(t,\epsilon_1) (e_k^{\sum_{\nu=0}^{m-1}I_{(s(i)-\nu)i}} (t) -e_j^{\sum_{\nu=0}^{m-1}I_{(s(i)-\nu)i}}(t))\times\\
    &\displaystyle\times\left\langle\left(\prod_{\mu=1}^{s(i)-(m+1)}H_{h}^{I_{\mu i}}(t)H^{J_{\mu i}}(0)\right)U_{t}^\dagger H^{I_{(s(i)-m)i}}(t)\ket{e_{j}(t)}\bra{e_{k}(t)}U_{t}H^{\sum_{\nu=0}^{m-1}J_{(s(i)-\nu)i}}(0)\right\rangle.\\
    &\displaystyle=\lim_{\epsilon_1\to 0}\lim_{\{N_l,N_j,N_k\}\to \{\infty,\infty,\infty\}}\sum_{l=1}^{N_l}\sum_{j=1}^{N_j}\sum_{k=1,k\neq j}^{N_k}e_{l}^{J_{(s(i)-m)i}}(0) \gamma_{jkl}(t,\epsilon_1) (e_k^{\sum_{\nu=0}^{m-1}I_{(s(i)-\nu)i}} (t) -e_j^{\sum_{\nu=0}^{m-1}I_{(s(i)-\nu)i}}(t))\times\\
    &\displaystyle\times\left\langle\left(\prod_{\mu=1}^{s(i)-(m+1)}H_{h}^{I_{\mu i}}(t)H^{J_{\mu i}}(0)\right)U_{t}^\dagger H^{I_{(s(i)-m)i}}(t)\ket{e_{j}(t)}\bra{e_{k}(t)}U_{t}H^{\sum_{\nu=0}^{m-1}J_{(s(i)-\nu)i}}(0)\right\rangle.\\
    &\displaystyle=\lim_{\{N_l,N_j,N_k\}\to \{\infty,\infty,\infty\}}\sum_{l=1}^{N_l}\sum_{j=1}^{N_j}\sum_{k=1,k\neq j}^{N_k}\lim_{\epsilon_1\to 0}e_{l}^{J_{(s(i)-m)i}}(0) \gamma_{jkl}(t,\epsilon_1) (e_k^{\sum_{\nu=0}^{m-1}I_{(s(i)-\nu)i}} (t) -e_j^{\sum_{\nu=0}^{m-1}I_{(s(i)-\nu)i}}(t))\times\\
    &\displaystyle\times\left\langle\left(\prod_{\mu=1}^{s(i)-(m+1)}H_{h}^{I_{\mu i}}(t)H^{J_{\mu i}}(0)\right)U_{t}^\dagger H^{I_{(s(i)-m)i}}(t)\ket{e_{j}(t)}\bra{e_{k}(t)}U_{t}H^{\sum_{\nu=0}^{m-1}J_{(s(i)-\nu)i}}(0)\right\rangle.
    \ea\label{Cm2}
\end{equation}
Since the sum $\sum_{l=1}^{N_l}\sum_{j=1}^{N_j}\sum_{k=1,k\neq j}^{N_k}$ converges to $C_{mi}(\epsilon_1)$ for every $\epsilon_1$ in the limit $\lim_{\{N_l,N_j,N_k\}\to \{\infty,\infty,\infty\}}$ (thus pointwise converging in terms of $\epsilon_1$), then we could exchange the limit $\epsilon_1$ with the limit $\lim_{\{N_l,N_j,N_k\}\to \{\infty,\infty,\infty\}}$ and the summations from the second to the third equality. Moreover, from now on, we order all the eigenvalues of energy in ascending order $e_{i}(0)\leq e_{i+1}(0)$ and $e_{i}(t)\leq e_{i+1}(t)$, where for each fixed finite value $i$, $e_{i}(t)$ is also finite, but can be unbounded when $i\to \infty$.

Just as deduced in the beginning of the proof of result 1, we notice that as $\epsilon_1\to 0$, $|\alpha_{lj}^{*}(t)\alpha_{lk}(t)|=|\bra{e_j(t)}U_t\ket{e_l(0)}\bra{e_l(0)}U_t^\dagger\ket{e_k(t)}|\to 0$ for all $l$ and $j\neq k$ and $t\in[0,\tau]$. Also, it follows that $[U_{t}^\dagger H(t)U_\tau, H (0)]=[H(t),U_\tau H (0)U_{t}^\dagger]\to 0$. Taking these relations into account in Eq. \eqref{Cm2}, we obtain 
\begin{equation}
    \ba{rl}
    \displaystyle \lim_{\epsilon_1\to 0}C_{mi}(\epsilon_1)&\displaystyle=\lim_{\{N_l,N_j,N_k\}\to \{\infty,\infty,\infty\}}\sum_{l=1}^{N_l}\sum_{j=1}^{N_j}\sum_{k=1,k\neq j}^{N_k}\lim_{\epsilon_1\to 0}e_{l}^{J_{(s(i)-m)i}}(0) \gamma_{jkl}(t,\epsilon_1) (e_k^{\sum_{\nu=0}^{m-1}I_{(s(i)-\nu)i}} (t) -e_j^{\sum_{\nu=0}^{m-1}I_{(s(i)-\nu)i}}(t))\times\\
    &\displaystyle\times\left\langle\left(H_{h}^{\sum_{\mu=1}^{s(i)-(m+1)}I_{\mu i}}(t)H^{\sum_{\mu=1}^{s(i)-(m+1)}J_{\mu i}}(0)\right)U_{t}^\dagger H^{I_{(s(i)-m)i}}(t)\ket{e_{j}(t)}\bra{e_{k}(t)}U_{t}H^{\sum_{\nu=0}^{m-1}J_{(s(i)-\nu)i}}(0)\right\rangle.
    \ea
\end{equation}
Considering the expansion of the initial incoherent state as $\rho=\sum_{l'}p_l\ket{e_{l'}(0)}\bra{e_{l'}(0)}$, we deduce
\begin{equation}
    \ba{rl}
    \displaystyle \lim_{\epsilon_1\to 0}C_{mi}(\epsilon_1)&\displaystyle=\lim_{N_{l,l',j,k}\to \infty}\sum_{l,l',j\neq k}\sum_{l,l',j\neq k}\lim_{\epsilon_1\to 0}p_{l'}e_{l}^{J_{(s(i)-m)i}}(0) \gamma_{jkl}(t,\epsilon_1) (e_k^{\sum_{\nu=0}^{m-1}I_{(s(i)-\nu)i}} (t) -e_j^{\sum_{\nu=0}^{m-1}I_{(s(i)-\nu)i}}(t))C_{mi}'(\epsilon_1),
    \ea\label{C1ibounded2}
\end{equation}
where we considered the notation $\lim_{N_{l,l',j,k}\to \infty}\sum_{l,l',j\neq k}\equiv \lim_{\{N_l,N_j,N_k\}\to \{\infty,\infty,\infty\}}\sum_{l=1}^{N_l}\sum_{l=1}^{N_{l'}}\sum_{j=1}^{N_j}\sum_{k=1,k\neq j}^{N_k}$ and the following definition
\begin{equation}
    C_{mi}'(\epsilon_1)=e_{j}^{I_{(s(i)-m)i}}(t)e_{l'}^{\sum_{\nu=0}^{m-1}J_{(s(i)-\nu)i}}(0)\bra{e_{l'}(0)}\left(H_{h}^{\sum_{\mu=1}^{s(i)-(m+1)}I_{\mu i}}(t)H^{\sum_{\mu=1}^{s(i)-(m+1)}J_{\mu i}}(0)\right)U_{t}^\dagger \ket{e_{j}(t)}\bra{e_{k}(t)}U_{t}\ket{e_{l'}(0)}.
\end{equation}
Considering the definition $\alpha_{lk}(t)=\bra{e_l(0)}U_t^\dagger\ket{e_k(t)}$ and the expansions $H_{h}^{\sum_{\mu=1}^{s(i)-2}I_{\mu i}}(t)=\sum_{p}e_{p}^{\sum_{\mu=1}^{s(i)-2}I_{\mu i}}(t)U_{t}^{\dagger}\ket{e_{p}(t)}\bra{e_{p}(t)}U_t$ and  $H^{\sum_{\mu=1}^{s(i)-2}J_{\mu i}}(0)=\sum_{q}e_q^{\sum_{\mu=1}^{s(i)-2}J_{\mu i}}(0)\ket{e_{q}(0)}\bra{e_{q}(0)}$, we can rewrite $C_{mi}'(\epsilon_1)$ as
\begin{equation}
    C_{mi}'(\epsilon_1)=e_{j}^{I_{(s(i)-m)i}}(t)e_{l'}^{\sum_{\nu=0}^{m-1}J_{(s(i)-\nu)i}}(0)\lim_{N_{q,p}\to \infty}\sum_{q,p}\left(e_p^{\sum_{\mu=1}^{s(i)-(m+1)}I_{\mu i}}(t)e_q^{\sum_{\mu=1}^{s(i)-(m+1)}J_{\mu i}}(0)\right)\alpha_{l'p}(t)\alpha_{qp}^{*}(t)\alpha_{qj}(t)\alpha_{l'k}^{*}(t), \label{C1ilinha2}
\end{equation}
Inserting Eq.\eqref{C1ilinha2} in Eq. \eqref{C1ibounded2}, we obtain
\begin{equation}
    \ba{rl}\displaystyle
     \lim_{\epsilon_1\to 0}C_{mi}(\epsilon_1)=\lim_{N_{l,l',j,k,q,p}\to \infty}\sum_{l,l',j\neq k,q,p} \lim_{\epsilon_1\to 0} E_{ll' jkqp}^{\mathbf{I}_i\mathbf{J}_i} C_{ll'jkqp}''(\epsilon_1)\label{Cmshort}
    \ea
\end{equation}
where
\begin{equation}
    E_{ll' jkqp}^{\mathbf{I}_i\mathbf{J}_i}=e_{l}^{J_{(s(i)-m)i}}(0)\,(e_k^{\sum_{\nu=0}^{m-1}I_{(s(i)-\nu)i}} (t) -e_j^{\sum_{\nu=0}^{m-1}I_{(s(i)-\nu)i}}(t))\,e_{j}^{I_{(s(i)-m)i}}(t)\,e_{l'}^{\sum_{\nu=0}^{m-1}J_{(s(i)-\nu)i}}(0)\,e_p^{\sum_{\mu=1}^{s(i)-(m+1)}I_{\mu i}}(t)\,e_q^{\sum_{\mu=1}^{s(i)-(m+1)}J_{\mu i}}(0))\label{Eindexes}
\end{equation}
and
\begin{equation}
    C_{ll'jkqp}''(\epsilon_1)=p_{l'} \gamma_{jkl}(t,\epsilon_1)\alpha_{l'k}^{*}(t)\alpha_{l'p}(t)\alpha_{qp}^{*}(t)\alpha_{qj}(t)
\end{equation}
In the limit $\epsilon_1\to 0$, $|\alpha_{l'p}(t)\alpha_{l'k}^{*}(t)|\to 0 $ for all $p\neq k $ and $|\alpha_{qp}^{*}\alpha_{qj}(t)|\to 0$ for all $p\neq j$ (see Eq. \eqref{aux1}). Therefore, when $\epsilon_1\to 0$ the product $\alpha_{qp}^{*}\alpha_{qj}(t)\alpha_{l'p}(t)\alpha_{l'k}^{*}(t)$ can only be non-null when $k=p$ \emph{and} $p=j$. However, since in the sum over $k$ and $j$ we are considering only cases in which $k\neq j$, then it cannot be the case where $k=p$ \emph{and} $p=j$. Therefore, $\lim_{\epsilon_1\to 0}\alpha_{qp}^{*}\alpha_{qj}(t)\alpha_{l'p}(t)\alpha_{l'k}^{*}(t)=0$ for $k\neq j$. Moreover, since $p_{l'}\leq 1$ and $|\gamma_{jkl}(t,\epsilon_1)|\leq 1$, then it follows that
\begin{equation}
    \lim_{\epsilon_1\to 0}C_{ll'jkqp}''(\epsilon_1)=0.\label{cruxbound}
\end{equation}
Moreover, since for every fixed natural $i\in\{l,l',j,k,q,p\}$, $e_i\leq e_{N_i}$ is finite, then $E_{ll' jkqp}^{\mathbf{I}_i\mathbf{J}_i}$ is finite as well and
\begin{equation}
    \lim_{\epsilon_1\to 0}E_{ll' jkqp}^{\mathbf{I}_i\mathbf{J}_i} C_{ll'jkqp}''(\epsilon_1)=0\quad \implies \quad \lim_{\epsilon_1\to 0}C_{mi}(\epsilon_1)=\lim_{N_{l,l',j,k,q,p}\to \infty}\sum_{l,l',j\neq k,q,p} \lim_{\epsilon_1\to 0} E_{ll' jkqp}^{\mathbf{I}_i\mathbf{J}_i} C_{ll'jkqp}''(\epsilon_1)=0.\label{cruxbound2}
\end{equation}
To deduce the second part of Eq. \eqref{cruxbound2}, we considered the following reasoning. Using the notation $N_l,N_{l'},N_j,N_k,N_q,N_p\to N_{l,l',j,k,q,p}$, we can define the sum 
\begin{equation}
    C_{mi}^{l,l',j,k,q,p}(\epsilon_1)=\sum_{l,l',j\neq k,q,p}^{N_{l,l',j,k,q,p}} \lim_{\epsilon_1\to 0} E_{ll' jkqp}^{\mathbf{I}_i\mathbf{J}_i} C_{ll'jkqp}''(\epsilon_1)
\end{equation} 
which converges to $C_{mi}(\epsilon_1)$ for every $\epsilon_1$ whenever $N_{l,l',j,k,q,p}\to \infty$. Therefore, for every $\varepsilon>0$, there exist finite natural numbers $N_{l,l',j,k,q,p}'$ such that $\sup_{\epsilon_1}|C_{mi}^{l,l',j\neq k,q,p}(\epsilon_1)-C_{mi}(\epsilon_1)|<\varepsilon$, for any $N_{l,l',j\neq k,q,p}>N_{l,l',j\neq k,q,p}'$. Since the limits can be exchanged whenever the sequence converges for every $\epsilon_1$ (see footnote [16]), then for any naturals $N_{l,l',j\neq k,q,p}>N_{l,l',j\neq k,q,p}'$, it follows that $|\lim_{\epsilon_1\to 0}C_{mi}(\epsilon_1)|=|\lim_{\epsilon_1\to 0}C_{mi}^{l,l',j\neq k,q,p}(\epsilon_1)-\lim_{\epsilon_1\to 0}C_{mi}(\epsilon_1)|<\varepsilon$. By definition, this means that 
\begin{equation}
    \ba{rl}
    0&\displaystyle =\lim_{N_{l,l',j,k,q,p}\to \infty}\left|\sum_{l,l',j\neq k,q,p}^{N_{l,l',j,k,q,p}} \lim_{\epsilon_1\to 0} E_{ll' jkqp}^{\mathbf{I}_i\mathbf{J}_i} C_{ll'jkqp}''(\epsilon_1)-\lim_{\epsilon_1\to 0}C_{mi}(\epsilon_1)\right|\\
    &\displaystyle =\,\left|\lim_{N_{l,l',j,k,q,p}\to \infty}\sum_{l,l',j\neq k,q,p}^{N_{l,l',j,k,q,p}} \lim_{\epsilon_1\to 0} E_{ll' jkqp}^{\mathbf{I}_i\mathbf{J}_i} C_{ll'jkqp}''(\epsilon_1)-\lim_{\epsilon_1\to 0}C_{mi}(\epsilon_1)\right|=|\lim_{\epsilon_1\to 0}C_{mi}(\epsilon_1)|.
    \ea\label{responseref2}
\end{equation}
Therefore, given that $\lim_{\epsilon_1\to 0}C_{\mathbf{I}_i\mathbf{J}_i}(\epsilon_1)=\sum_{m=1}^{s(i)-1} \lim_{\epsilon_1\to 0} C_{mi}(\epsilon_1)=0$, and inserting this result in the sum \eqref{yute102}, we deduce, analogously as was done for $\lim_{\epsilon_1\to 0}C_{mi}(\epsilon_1)$ in Eq. \eqref{responseref2}, the following result
\begin{equation}
    \lim_{\epsilon_1\to 0}y(u,t,\epsilon_1)=\lim_{N_n\to \infty}\sum_{n=0}^{N_n}\frac{(iu)^n}{n!}\sum_{i=1}^{2^{n}}a_{ni}\lim_{\epsilon_1\to 0}C_{\mathbf{I}_i\mathbf{J}_i}(\epsilon_1)=0,\label{yute1022}
\end{equation}
where we considered the fact that $u$ is bounded, since from the Riemman-Lebesgue Lemma (Lemma 7 in section 15.6 in \cite{Fristedt2013}), $|u|\to \infty\,\,\implies\,\,y(u,t,\epsilon_1)=0$ \footnote{ From the Riemman-Lebesgue Lemma (Lemma 7 in section 15.6 in \cite{Fristedt2013}), it follows that for $|u|\to \infty$, the characteristic functions satisfy $\lim_{|u|\to \infty}\chi_{\text{\tiny TPM}}(u)=0$ and $\lim_{|u|\to \infty}\chi_{\text{\tiny OBS}}(u)=0$ for every $\epsilon_1$. From Eq. \eqref{Charactobslemma}, this implies $0=\lim_{|u|\to \infty}[\chi_{\text{\tiny OBS}}(u)-\chi_{\text{\tiny TPM}}(u)]=\lim_{|u|\to \infty}y(u,t,\epsilon_1)\epsilon_1$ for every $\epsilon_1>0$. Therefore, $\lim_{\epsilon_1\to 0}\lim_{u\to \infty}y(u,t\epsilon_1)=0$.}.

Next, let us rewrite the expansion described in Eq. \eqref{relationpylinha}
\begin{equation}
    \int_{-\infty}^{\infty}dw\,|P_{\text{\tiny TPM}}(w)- P_{\text{\tiny OBS}}(w)|=y'(t,\epsilon_1) \epsilon_1 .
\end{equation}
where 
\begin{equation}
         y'(t,\epsilon_1)=\frac{1}{2\pi}\int_{-\infty}^{\infty}\,dw\,\left|\int_{-\infty}^{\infty}du\,\mathrm{e}^{-iuw} y(u,t,\epsilon_1)\right|.\label{ylinhaboundedscheffes}
     \end{equation}
Because 
\begin{equation}
        2 = \int_{-\infty}^{\infty}dw\,(|P_{\text{\tiny TPM}}(w)|+|P_{\text{\tiny OBS}}(w)|)\geq \int_{-\infty}^{\infty}dw\,|P_{\text{\tiny TPM}}(w)- P_{\text{\tiny OBS}}(w)|=y'(t,\epsilon_1) \epsilon_1,\label{boundylinhaepsilonzero}
     \end{equation}
for every $\epsilon_1$, then for every $\epsilon_1>0$, $y'(t,\epsilon_1)$ should be bounded \footnote{Notice that the fact that $y'(t,\epsilon_1)$ should be bounded does not immediately implies that $\lim_{\epsilon_1}y'(t,\epsilon_1)$ is bounded. An instance of this case occurs when $y'(t,\epsilon_1)=\epsilon_1^{-k}$ with $0< k<1$.}. Therefore, we can consider Scheffe's theorem (Theorem 2.5.4 in \cite{Athreya2006}), to show that $\lim_{\epsilon_1\to 0}y(u,t,\epsilon_1)=0$ implies, from Eq. \eqref{ylinhaboundedscheffes}, that
\begin{equation}
         \lim_{\epsilon_1\to 0}y'(t,\epsilon_1)=\lim_{\epsilon_1\to 0}\frac{1}{2\pi}\int_{-\infty}^{\infty}\,dw\,\left|\int_{-\infty}^{\infty}du\,\mathrm{e}^{-iuw} y(u,t,\epsilon_1)\right|=0.\label{ylinhascheffespos}
     \end{equation}
Therefore, in the limit where $\epsilon_1\to 0$, $y'(t,\epsilon_1)$ is null and therefore bounded. Moreover, because of Eq. \eqref{boundylinhaepsilonzero}, it follows that for every $\epsilon_1>0$, with $\epsilon_1$ not tending to $0$, $y'(t,\epsilon_1)\leq 2/\epsilon_1$, which is also bounded. As a result, for any $\epsilon_1$, converging or not to $0$, $y'(t,\epsilon_1')$ is bounded. Consequently, there exist a bounded real positive constant $C_1$ such that
\begin{equation}
    C_1=\max_{\epsilon_1'}y'(t,\epsilon_1').
\end{equation}
\section{Derivation of result 2}

To establish Result 2, we investigate processes described by $\{H(t),U_t,[0,\tau],\rho\}$, where the Hamiltonian operator is associated with a one-dimensional quantum system dependent on a Schr{\"o}dinger coordinate observable $X$ and its conjugate momentum $P$ ($[X,P]=i\hbar$). Consequently, both the unitary evolution operator $U_t$ and the Heisenberg version of the Hamiltonian $H_h(t)$ can be expressed as continuous functions of $X$, $P$, and $t$. We thus identify
\begin{equation}
    H_h(t) = G(X, P, t),
\end{equation}
where $G$ is a \emph{analytic} continuous function that can be represented as a series of powers of $X$, $P$, and $t$.

In parallel, we consider the classical counterpart of this scenario. We introduce the classical function $\varrho(\Gamma_0)$, representing the probability distribution of locating a classical system at the phase point $\Gamma_0=(x_0,p_0)$ at time $0$, and a classical Hamiltonian function $H_{\text{\tiny CL}}(\Gamma_0,t)$, corresponding to the energy at time $t$, associated with the phase point at $\Gamma_0$ at time $0$. Throughout this section, we employ the notation
\begin{equation}
    \braket{A} \coloneqq \Tr [A\rho] \quad \text{and} \quad \braket{a}_{\text{\tiny CL}} \coloneqq \int_{\Gamma_0}d\Gamma_0\, a(\Gamma_0) \varrho(\Gamma_0) \equiv \int_{-\infty}^{\infty}\int_{-\infty}^{\infty}dx_0\,dp_0\, a(x_0,p_0) \varrho(x_0,p_0),
\end{equation}
for the expectation values referring to the same physical observable in the quantum ($A$) and the classical ($a$) regimes, concerning the initial state and distribution, respectively. Furthermore, we adopt an identical functional form for both classical and quantum scenarios. Consequently, if $F(A)$ represents a function of the operator $A$ in the quantum scenario, then $F(a)$ takes on the same form, with $A$ being replaced by $a$ in $F(A)$. Within this framework, we review the necessary conditions discussed in the main text for the quantum process $\{X,P,\rho,H_h(t)\}$ to approximate the classical description in the classical limit:
\vspace{0.3cm}

(A) \emph{There exists at least one classical distribution $\varrho(\Gamma_0)$ and a Hamiltonian function $H_{\text{\tiny CL}}(\Gamma_0,t)$ characterizing the energy of a classical system such that}
    \begin{equation}
        \left|\Tr\left[P^{m}X^{n}\rho \right]-\int d\Gamma_0\, p_0^{m}x_{0}^n \,\varrho(\Gamma_0)\right|\leq \epsilon_{A} \Tr\left[P^{m}X^{n}\rho \right]
    \end{equation}
\emph{with $\epsilon_B\ll 1$ for any integers $m$ and $n$, and}
\begin{equation}
    H_{\text{\tiny CL}}(\Gamma_0,t')=G(x_0,p_0,t'),\label{CondBsecondpart}
    \end{equation}
\emph{where $G$ possesses the same functional form as the function defining the Heisenberg Hamiltonian $H_h(t)$, with the substitution of operators $X$ and $P$ by $x_0$ and $p_0$, respectively. The parameter $t'$ can take on values of either $0$ or $\tau$.} 

\vspace{0.3cm}    

(B) \emph{For any smooth functions $g_{l}\equiv g_l(X,P)$ and $g_r\equiv g_r(X,P)$ which can be written as a sum of powers of $X$ and $P$, it follows that}
\begin{equation}
    |\Tr[g_l\,[X,P]\,g_r\,\rho]|\leq \epsilon_B|\Tr[g_l\, XP\, g_r\,\rho]|,\quad \text{and}\quad |\Tr[g_l\,[X,P]\,g_r\,\rho]|\leq \epsilon_{B}|\Tr[g_l\, PX\, g_r\,\rho]|,
\end{equation}
\emph{where $\epsilon_B\ll 1$}.

Our approach to establish Result 2 as presented in the main text is as follows: we demonstrate that when conditions (A) and (B) are satisfied, the difference between the characteristic functions of the OBS protocol and the classical work statistics are both bounded by $\epsilon_{A}$ and $\epsilon_{B}$. Subsequently, we illustrate how this boundedness affects into probability distributions of work.

As stated in the Eq. \eqref{Guquantum} in the previous section, the OBS characteristic function is 
\begin{equation}
    \chi_{\text{\tiny OBS}}(u)=\int_{-\infty}^{\infty}\,dw\,\mathrm{e}^{iuw} P_{\text{\tiny OBS}}(w)=\braket{\mathrm{e}^{iuW(\tau,0)}}.\label{Guquantum2}
\end{equation}
Taking into account that the hamiltonian operators are described in terms of sufficiently smooth functions such that $H_{h}(t)=G(X,P,t)$, then $W(\tau,0)=G(X,P,\tau)-G(X,P,0)$, and its exponential can be written as a power series in the general form
\begin{equation}
    \mathrm{e}^{iuW(\tau,0)}=\sum_{\mathbf{I}\mathbf{J}}a_{\mathbf{I}\mathbf{J}}(u)X^{I_1}P^{J_{1}}\cdots X^{I_n}P^{J_{n}},\label{expiuw}
\end{equation}
where $a_{\mathbf{I}\mathbf{J}}(u)$ are $u$-dependent complex-valued functions, indexed by $\mathbf{I}\mathbf{J}\equiv \{I_1,I_2,\cdots I_n,J_1,J_2,\cdots J_n\}\subset \mathbbm{Z}$. The summation $\sum_{\mathbf{I}\mathbf{J}}$ is done over all values of the integer powers $\mathbf{I}\mathbf{J}$. Inserting Eq. \eqref{expiuw} in Eq. \eqref{Guquantum2}, it follows that
\begin{equation}
    \chi_{\text{\tiny OBS}}(u)=\sum_{\mathbf{I}\mathbf{J}}a_{\mathbf{I}\mathbf{J}}(u)\braket{X^{I_1}P^{J_{1}}\cdots X^{I_n}P^{J_{n}}}.\label{charactQ}
\end{equation}
The work distribution for the classical scenario initiated in a classical distribution $\varrho(\Gamma_0)$ was introduced in the main text as
\begin{equation}
P_{\text{\tiny CL}}(w)=\int_{\text{\tiny $\Gamma_0$}}\,d\Gamma_0\, \varrho(\Gamma_0) \,\delta[w-W_{\text{\tiny CL}}(\Gamma_0,\tau,0)].\label{pcl}
\end{equation}
and the associated classical characteristic function is
\begin{equation}
    \chi_{\text{\tiny CL}}(u)=\int_{-\infty}^{\infty}dw \mathrm{e}^{iuw}P_{\text{\tiny CL}}(w)=\int_{\Gamma_0}\,d\Gamma_0\, \varrho(\Gamma_0)\,\mathrm{e}^{iuW_{\text{\tiny CL}}(\Gamma_0,\tau,0)}\equiv\braket{\mathrm{e}^{iuW_{\text{\tiny CL}}(\tau,0)}}_{\text{\tiny CL}}.\label{Gucl}
\end{equation}
Under the assumption that condition (A) is satisfied, we have $H_{\text{\tiny CL}}(\Gamma_0,t) = G(x_0,p_0,t)$, where the function $G$ maintains the same functional form as that defining $H_{h}(t)$. Consequently, the classical work assumes a form akin to the operator $W(\tau,0)$, substituting the operators $X$ and $P$ by the variables $x_0$ and $p_0$, respectively. Accordingly, $W_{\text{\tiny CL}}(\Gamma_0,\tau,0) = G(x_0,p_0,\tau) - G(x_0,p_0,0)$. Furthermore, the classical characteristic function can be expressed as follows:
\begin{equation}
    \chi_{\text{\tiny CL}}(u)=\sum_{\mathbf{I}\mathbf{J}}a_{\mathbf{I}\mathbf{J}}(u)\braket{p_0^{\sum_{j=1}^{n} J_{j}} x_0^{\sum_{i=1}^{n}I_i}}_{\text{\tiny CL}}\label{expiuwcl}
\end{equation}
where $a_{\mathbf{I}\mathbf{J}}(u)$ refer to the same coefficient functions and powers $\mathbf{I}\mathbf{J}$ as in the quantum case in Eq. \eqref{charactQ}. Taking into account Eqs. \eqref{expiuw} and \eqref{expiuwcl}, we deduce the following result:

\begin{proposition}
If conditions (A) and (B) hold, then 
   \begin{equation}
        \chi_{\text{\tiny OBS}}(u)-\chi_{\text{\tiny CL}}(u)=\epsilon_{\max}g(u,\epsilon_{A},\epsilon_{B})
    \end{equation}
    where $\epsilon_{\max}=\max\{\epsilon_{A},\epsilon_{B}\}$,
    \begin{equation}
        g(u,\epsilon_{A},\epsilon_{B})=\sum_{\mathbf{I}\mathbf{J}}a_{\mathbf{I}\mathbf{J}}(u)\left[\hat{\xi}'_{\mathbf{I}\mathbf{J}}(1+\epsilon_{B})^{\sum_{j=1}^{n}I_{j}(\sum_{m=j}^{n}J_{m})}+\hat{\xi}_{\mathbf{I}\mathbf{J}}\left(\sum_{k=0}^{\sum_{j=1}^{n}I_{j}(\sum_{m=j}^{n}J_{m})-1}(1+\epsilon_{B})^{k}\right) \right]\braket{X^{I_{1}}P^{J_{1}}\cdots X^{I_{n}}P^{J_{n}}},\label{gutau}
    \end{equation}
    and $\hat{\xi}_{\mathbf{I}\mathbf{J}}$ and $\hat{\xi}_{\mathbf{I}\mathbf{J}}'$ are complex scalars satisfying $|\hat{\xi}_{\mathbf{I}\mathbf{J}}'|,|\hat{\xi}_{\mathbf{I}\mathbf{J}}|\leq 1$.
    \label{lemag1}
\end{proposition}
\begin{proof}
From condition (B), we obtain
\begin{equation}
\ba{rl}
    |\braket{X^{I_{1}}P^{J_{1}}\cdots X^{I_{n}}P^{J_{n}}}-\braket{P^{\text{\tiny $\sum_{i=1}^{n}J_i'$}}X^{\text{\tiny $\sum_{i=1}^{n}I_i'$}}}|&=|\braket{X^{I_{1}}P^{J_{1}}\cdots X^{I_{n}-1}[X,P]P^{J_{n}-1}}+\braket{X^{I_{1}}P^{J_{1}}\cdots X^{I_{n}-1}PXP^{J_{n}-1}}-\braket{P^{\text{\tiny $\sum_{i=1}^{n}J_i'$}}X^{\text{\tiny $\sum_{i=1}^{n}I_i'$}}}|\\
    &\leq |\braket{X^{I_{1}}P^{J_{1}}\cdots X^{I_{n}-1}[X,P]P^{J_{n}-1}}|+|\braket{X^{I_{1}}P^{J_{1}}\cdots X^{I_{n}-1}PXP^{J_{n}-1}}-\braket{P^{\text{\tiny $\sum_{i=1}^{n}J_i'$}}X^{\text{\tiny $\sum_{i=1}^{n}I_i'$}}}|\\
    &\leq \epsilon_{B} \braket{X^{I_{1}}P^{J_{1}}\cdots X^{I_{k}}P^{J_{k}}\cdots X^{I_{n}}P^{J_{n}}}  +|\braket{X^{I_{1}}P^{J_{1}}\cdots X^{I_{n}-1}PXP^{J_{n}-1}}-\braket{P^{\text{\tiny $\sum_{i=1}^{n}J_i'$}}X^{\text{\tiny $\sum_{i=1}^{n}I_i'$}}}|.
    \ea 
\end{equation}
Similarly, we derive
\begin{equation}
\ba{rl}
    |\braket{X^{I_{1}}P^{J_{1}}\cdots X^{I_{n}-1}PXP^{J_{n}-1}}-\braket{P^{\text{\tiny $\sum_{i=1}^{n}J_i'$}}X^{\text{\tiny $\sum_{i=1}^{n}I_i'$}}}|&\leq\epsilon_{B} \left(1+\epsilon_{B}\right)|\braket{X^{I_{1}}P^{J_{1}}\cdots X^{I_{k}}P^{J_{k}}\cdots X^{I_{n}}P^{J_{n}}}| \\
    &+|\braket{X^{I_{1}}P^{J_{1}}\cdots X^{I_{n}-1}P^2XP^{J_{n}-2}}-\braket{P^{\text{\tiny $\sum_{i=1}^{n}J_i'$}}X^{\text{\tiny $\sum_{i=1}^{n}I_i'$}}}|,
    \ea
\end{equation}
where we considered condition (B) to obtain
\begin{equation}
    |\braket{X^{I_{1}}P^{J_{1}}\cdots X^{I_{n}-1}P[X,P]P^{J_{n}-2}}|\leq \epsilon_{B} |\braket{X^{I_{1}}P^{J_{1}}\cdots X^{I_{n}-1}PXP^{J_{n}-1}}|\leq \epsilon_{B} \left(1+\epsilon_{B}\right)|\braket{X^{I_{1}}P^{J_{1}}\cdots X^{I_{k}}P^{J_{k}}\cdots X^{I_{n}}P^{J_{n}}}|.
\end{equation}
In fact, we can derive, by induction, that
\begin{equation}
   |\braket{X^{I_{1}}P^{J_{1}}\cdots X^{I_{n}-1}P^l X P^{J_{n}-l}}| \leq (1+\epsilon_{B})^l|\braket{X^{I_{1}}P^{J_{1}}\cdots X^{I_{n}}P^{J_{n}}}|\label{induc0}
\end{equation}
for $1\leq l\leq J_n$. This can be proved by noticing first that Eq. \eqref{induc0} holds for $l=1$:
\begin{equation}
\ba{rl}
    |\braket{X^{I_{1}}P^{J_{1}}\cdots X^{I_{n}-1}P X P^{J_{n}-1}}|&=|-\braket{X^{I_{1}}P^{J_{1}}\cdots X^{I_{n}-1}[X,P] P^{J_{n}-1}}+\braket{X^{I_{1}}P^{J_{1}}\cdots X^{I_{n}}P^{J_{n}}}|\\
    &\leq (\epsilon_{B}+1)|\braket{X^{I_{1}}P^{J_{1}}\cdots X^{I_{n}}P^{J_{n}}}|.
    \ea
\end{equation}
Also, if Eq. \eqref{induc0} holds for $l$, then it also holds for $l+1$: 
\begin{equation}
\ba{rl}
|\braket{X^{I_{1}}P^{J_{1}}\cdots X^{I_{n}-1}P^{l+1} X P^{J_{n}-l-1}}|&\leq |\braket{X^{I_{1}}P^{J_{1}}\cdots X^{I_{n}-1}P^{l} [X,P] P^{J_{n}-l-1}}|+|\braket{X^{I_{1}}P^{J_{1}}\cdots X^{I_{n}-1}P^{l} X P^{J_{n}-l}}|\\
&\leq (\epsilon_{B}+1)|\braket{X^{I_{1}}P^{J_{1}}\cdots X^{I_{n}-1}P^{l} X P^{J_{n}-l}}|\leq (1+\epsilon_{B})^{l+1}|\braket{X^{I_{1}}P^{J_{1}}\cdots X^{I_{n}}P^{J_{n}}}|.\label{induc1}
\ea
\end{equation}
Therefore, by induction from $l=1$ to $2$ and so on, then Eq. \eqref{induc0} should be valid for any  $1\leq l\leq J_n$. Another expression which can also be obtained by induction is the following:
\begin{equation}
    \braket{X^{I_{1}}P^{J_{1}}\cdots X^{I_{n}-1}P^{l}[X,P]P^{J_{n}-l-1}}\leq \epsilon_{B}(\epsilon_{B}+1)^l|\braket{X^{I_{1}}P^{J_{1}}\cdots X^{I_{n}}P^{J_{n}}}|.\label{induc2}
\end{equation}
Indeed, from Eqs. \eqref{induc0} and \eqref{induc2}, it follows that
\begin{equation}
    \braket{X^{I_{1}}P^{J_{1}}\cdots X^{I_{n}-1}P^{l+1}[X,P]P^{J_{n}-l-2}}\leq \epsilon_{B}(\epsilon_{B}+1)^{l+1}|\braket{X^{I_{1}}P^{J_{1}}\cdots X^{I_{n}}P^{J_{n}}}|
\end{equation}
Using the induction rule \eqref{induc2}, we deduce:
\begin{equation}
\ba{rl}
    |\braket{X^{I_{1}}P^{J_{1}}\cdots X^{I_{n}-1}P^l XP^{J_{n}-l}}-\braket{P^{\text{\tiny $\sum_{i=1}^{n}J_i'$}}X^{\text{\tiny $\sum_{i=1}^{n}I_i'$}}}|&\leq |\braket{X^{I_{1}}P^{J_{1}}\cdots X^{I_{n}-1}P^{l+1} [X,P]P^{J_{n}-(l+1)}}|\\
    &+|\braket{X^{I_{1}}P^{J_{1}}\cdots X^{I_{n}-1}P^{l+1} XP^{J_{n}-(l-1)}}-\braket{P^{\text{\tiny $\sum_{i=1}^{n}J_i'$}}X^{\text{\tiny $\sum_{i=1}^{n}I_i'$}}}|\\
    &\leq \epsilon_{B}(\epsilon_{B}+1)^l|\braket{X^{I_{1}}P^{J_{1}}\cdots X^{I_{n}}P^{J_{n}}}|\\
    &+|\braket{X^{I_{1}}P^{J_{1}}\cdots X^{I_{n}-1}P^{l+1} XP^{J_{n}-(l+1)}}-\braket{P^{\text{\tiny $\sum_{i=1}^{n}J_i'$}}X^{\text{\tiny $\sum_{i=1}^{n}I_i'$}}}|\\
    &\leq \epsilon_{B}[(\epsilon_{B}+1)^l+(\epsilon_{B}+1)^{l+1}]|\braket{X^{I_{1}}P^{J_{1}}\cdots X^{I_{n}}P^{J_{n}}}|\\
    &+|\braket{X^{I_{1}}P^{J_{1}}\cdots X^{I_{n}-1}P^{l+2} XP^{J_{n}-(l+2)}}-\braket{P^{\text{\tiny $\sum_{i=1}^{n}J_i'$}}X^{\text{\tiny $\sum_{i=1}^{n}I_i'$}}}|\label{res2aux0}
    \ea
\end{equation}
Repeating this procedure $J_n$ times, we obtain
    \begin{equation}
\ba{l}
    |\braket{X^{I_{1}}P^{J_{1}}\cdots X^{I_{n}}P^{J_{n}}}-\braket{P^{\text{\tiny $\sum_{i=1}^{n}J_i'$}}X^{\text{\tiny $\sum_{i=1}^{n}I_i'$}}}|\leq \epsilon_{B}\left(\sum_{k=0}^{J_{n}-1}(1+\epsilon_{B})^{k}\right) |\braket{X^{I_{1}}P^{J_{1}}\cdots X^{I_{n}}P^{J_{n}}}|
    +|\braket{X^{I_{1}}P^{J_{1}}\cdots X^{I_{n}-1}P^{J_{n}}X}-\braket{P^{\text{\tiny $\sum_{i=1}^{n}J_i'$}}X^{\text{\tiny $\sum_{i=1}^{n}I_i'$}}}|.
    \ea
\end{equation}
From the previous induction rules, it follows that
\begin{equation}
\ba{l}
   |\braket{X^{I_{1}}P^{J_{1}}\cdots X^{I_{n}-1}P^{J_n} X }| \leq (1+\epsilon_{B})^{J_n}|\braket{X^{I_{1}}P^{J_{1}}\cdots X^{I_{n}}P^{J_{n}}}|,\\
    \braket{X^{I_{1}}P^{J_{1}}\cdots X^{I_{n}-1}P^{J_n -1}[X,P]}\leq \epsilon_{B}(\epsilon_{B}+1)^{J_n-1}|\braket{X^{I_{1}}P^{J_{1}}\cdots X^{I_{n}}P^{J_{n}}}|.
\ea    
\end{equation}
Then, the following new induction rules can be derived:
\begin{equation}
\ba{l}
  \displaystyle|\braket{X^{I_{1}}P^{J_{1}}\cdots X^{I_{n}-2}P^{l}XP^{J_n -l} X }| \leq (1+\epsilon_{B})^{J_n+l}|\braket{X^{I_{1}}P^{J_{1}}\cdots X^{I_{n}}P^{J_{n}}}|,\\
\displaystyle\braket{X^{I_{1}}P^{J_{1}}\cdots X^{I_{n}-2}P^{l}[X,P]P^{J_n -l-1}X} \leq \epsilon_{B} (1+\epsilon_{B})^{J_n+l}|\braket{X^{I_{1}}P^{J_{1}}\cdots X^{I_{n}}P^{J_{n}}}|.\label{R2L1auxL2}
\ea
\end{equation}
We can use further these induction rules for $I_{n}$ power of $X$, obtaining
\begin{equation}
\ba{rl}
\displaystyle|\braket{X^{I_{1}}P^{J_{1}}\cdots X^{I_{n}}P^{J_{n}}}-\braket{P^{\text{\tiny $\sum_{i=1}^{n}J_i'$}}X^{\text{\tiny $\sum_{i=1}^{n}I_i'$}}}|&\leq \epsilon_{B} \left(\sum_{k=0}^{I_{n}\times J_{n}-1}(1+\epsilon_{B})^{k}\right) |\braket{X^{I_{1}}P^{J_{1}}\cdots X^{I_{n}}P^{J_{n}}}|\\
&\displaystyle+|\braket{X^{I_{1}}P^{J_{1}}\cdots X^{I_{n-1}}P^{J_{n-1}+J_{n}}X^{I_{n}}}-\braket{P^{\text{\tiny $\sum_{i=1}^{n}J_i'$}}X^{\text{\tiny $\sum_{i=1}^{n}I_i'$}}}|.
\ea
\end{equation}
Then
\begin{equation}
\ba{l}
  |\braket{X^{I_{1}}P^{J_{1}}\cdots X^{I_{n-1}-1}P^{l}XP^{J_n+J_{n-1} -l} X^{I_n} }| \leq (1+\epsilon_{B})^{I_n\times J_n+l}|\braket{X^{I_{1}}P^{J_{1}}\cdots X^{I_{n}}P^{J_{n}}}|,\\
  \braket{X^{I_{1}}P^{J_{1}}\cdots X^{I_{n-1}-1}P^{l}[X,P]P^{J_n -l-1}X^{I_{n}}} \leq \epsilon_{B} (1+\epsilon_{B})^{I_n\times J_n+l}|\braket{X^{I_{1}}P^{J_{1}}\cdots X^{I_{n}}P^{J_{n}}}|.
  \ea
  \label{R2lema1auxlema2}
\end{equation}
From a similar procedure, it follows that
\begin{equation}
\ba{rl}
    |\braket{X^{I_{1}}P^{J_{1}}\cdots X^{I_{n}}P^{J_{n}}}-\braket{P^{\text{\tiny $\sum_{i=1}^{n}J_i'$}}X^{\text{\tiny $\sum_{i=1}^{n}I_i'$}}}|&\leq \epsilon_{B}\left(\sum_{k=0}^{I_{n}J_{n}+I_{n-1}(J_{n} + J_{n-1})-1}(1+\epsilon_{B})^{k}\right) \braket{X^{I_{1}}P^{J_{1}}\cdots X^{I_{k}}P^{J_{k}}\cdots X^{I_{n}}P^{J_{n}}} \\
    &+|\braket{X^{I_{1}}P^{J_{1}}\cdots X^{I_{n-2}}P^{J_{n-2}+J_{n-1}+J_{n}}X^{I_{n-1}+I_{n}}}-\braket{P^{\text{\tiny $\sum_{i=1}^{n}J_i'$}}X^{\text{\tiny $\sum_{i=1}^{n}I_i'$}}}|    
    \ea
\end{equation}
Doing this iteratively we get
\begin{equation}
    |\braket{X^{I_{1}}P^{J_{1}}\cdots  X^{I_{n}}P^{J_{n}}}-\braket{P^{\text{\tiny $\sum_{i=1}^{n}J_i'$}}X^{\text{\tiny $\sum_{i=1}^{n}I_i'$}}}|\leq C_{\mathbf{I}\mathbf{J}}(\epsilon_{B})\epsilon_{B}\,|\braket{X^{I_{1}}P^{J_{1}}\cdots X^{I_{n}}P^{J_{n}}}|
\end{equation}
where 
\begin{equation}
    C_{\mathbf{I}\mathbf{J}}(\epsilon_{B})=\left(\sum_{k=0}^{(\sum_{j=1}^{n}I_{j}\sum_{m=j}^{n}J_{m})-1}(1+\epsilon_{B})^{k}\right)= \left(\sum_{j=1}^{n}\sum_{m=j}^{n}I_{j}J_{m}\right)+\mathcal{O}(\epsilon_{B}).\label{CIJdef}
\end{equation}
Therefore, it follows that
    \begin{equation}
        \braket{X^{I_{1}}P^{J_{1}}\cdots X^{I_{n}}P^{J_{n}}}=\braket{P^{\text{\tiny $\sum_{i=1}^{n}J_i'$}}X^{\text{\tiny $\sum_{i=1}^{n}I_i'$}}}+\xi_{\mathbf{I}\mathbf{J}}C_{\mathbf{I}\mathbf{J}}(\epsilon_{B})\epsilon_{B}\braket{X^{I_{1}}P^{J_{1}}\cdots  X^{I_{n}}P^{J_{n}}}\label{chidef}
    \end{equation}
    where $\xi_{\mathbf{I}\mathbf{J}}$ are complex scalars satisfying $|\xi_{\mathbf{I}\mathbf{J}}|\leq 1$. Consequently, the OBS characteristic function can be written as
    \begin{equation}
    \chi_{\text{\tiny OBS}}(u)=\braket{\mathrm{e}^{iuW(\tau,0)}}=\sum_{\mathbf{I}\mathbf{J}}a_{\mathbf{I}\mathbf{J}}(u)\braket{X^{I_1}P^{J_{1}}\cdots X^{I_n}P^{J_{n}}}=\sum_{\mathbf{I}\mathbf{J}}a_{\mathbf{I}\mathbf{J}}(u)\left[\braket{P^{\text{\tiny $\sum_{i=1}^{n}J_i'$}}X^{\text{\tiny $\sum_{i=1}^{n}I_i'$}}}+\xi_{\mathbf{I}\mathbf{J}}C_{\mathbf{I}\mathbf{J}}(\epsilon_{B})\epsilon_{B}\braket{X^{I_{1}}P^{J_{1}}\cdots  X^{I_{n}}P^{J_{n}}}\right].
    \end{equation}
    Moreover, considering condition (A), we have
    \begin{equation}
        |\braket{P^{\prod_{j=1}^{n} J_{j}} X^{\sum_{i=1}^{n}I_i}}-\braket{p^{\prod_{j=1}^{n} J_{j}} x^{\sum_{i=1}^{n}I_i}}_{\text{\tiny CL}}|\leq \epsilon_{A} |\braket{P^{\prod_{j=1}^{n} J_{j}} X^{\sum_{i=1}^{n}I_i}}|.
    \end{equation}
    Using a similar procedure to deduce Eq. \eqref{R2L1auxL2}, we derive via induction that
    \begin{equation}
        |\braket{P^{\prod_{j=1}^{n} J_{j}} X^{\sum_{i=1}^{n}I_i}}|\leq (1+\epsilon_{B})^{\sum_{j=1}^{n}I_{j}(\sum_{m=j}^{n}J_{m})}|\braket{X^{I_{1}}P^{J_{1}}\cdots X^{I_{n}}P^{J_{n}}}|.\label{R2L2aux2}
    \end{equation}
    Therefore, we can define complex scalars $\xi_{\mathbf{I}\mathbf{J}}'$ such that $|\xi_{\mathbf{I}\mathbf{J}}'|\leq 1$ and 
    \begin{equation}
        \braket{P^{\prod_{j=1}^{n} J_{j}} X^{\sum_{i=1}^{n}I_i}}=\braket{p^{\prod_{j=1}^{n} J_{j}} x^{\sum_{i=1}^{n}I_i}}_{\text{\tiny CL}}+\xi_{\mathbf{I}\mathbf{J}}'(1+\epsilon_{B})^{\sum_{j=1}^{n}I_{j}(\sum_{m=j}^{n}J_{m})}\braket{X^{I_{1}}P^{J_{1}}\cdots X^{I_{n}}P^{J_{n}}}\epsilon_{A},\label{chilinhadef}
    \end{equation}
    so that:
    \begin{equation}
        \chi_{\text{\tiny OBS}}(u)=\sum_{\mathbf{I}\mathbf{J}}a_{\mathbf{I}\mathbf{J}}(u)\left[\braket{p^{\prod_{j=1}^{n} J_{j}} x^{\sum_{i=1}^{n}I_i}}_{\text{\tiny CL}}+\left(\epsilon_{A}\xi_{\mathbf{I}\mathbf{J}}'(1+\epsilon_{B})^{\sum_{j=1}^{n}I_{j}(\sum_{m=j}^{n}J_{m})}+\xi_{\mathbf{I}\mathbf{J}}C_{\mathbf{I}\mathbf{J}}(\epsilon_{B})\epsilon_{B}\right)\braket{X^{I_{1}}P^{J_{1}}\cdots X^{I_{n}}P^{J_{n}}}\right].
    \end{equation}
    Taking into account the definition of the classical characteristic function in Eq. \eqref{expiuwcl}, it follows that
    \begin{equation}
        \chi_{\text{\tiny OBS}}(u)-\chi_{\text{\tiny CL}}(u)=\sum_{\mathbf{I}\mathbf{J}}a_{\mathbf{I}\mathbf{J}}(u)\left[\epsilon_{A}\xi_{\mathbf{I}\mathbf{J}}'(1+\epsilon_{B})^{\sum_{j=1}^{n}I_{j}(\sum_{m=j}^{n}J_{m})}+\xi_{\mathbf{I}\mathbf{J}}C_{\mathbf{I}\mathbf{J}}(\epsilon_{B})\epsilon_{B}\right]\braket{X^{I_{1}}P^{J_{1}}\cdots X^{I_{n}}P^{J_{n}}}.
    \end{equation}
    Defining $\epsilon_{\max}=\max\{\epsilon_{A},\epsilon_{B}\}$  and considering the explicit form of $C_{\mathbf{I}\mathbf{J}}(\epsilon_{B})$ in Eq. \eqref{CIJdef}, the result of the Proposition is obtained:
    \begin{equation}
        \chi_{\text{\tiny OBS}}(u)-\chi_{\text{\tiny CL}}(u)=\epsilon_{\max}g(u,\epsilon_{A},\epsilon_{B})\label{xdifference}
    \end{equation}
    where 
    \begin{equation}
        g(u,\epsilon_{A},\epsilon_{B})=\sum_{\mathbf{I}\mathbf{J}}a_{\mathbf{I}\mathbf{J}}(u)\left[\hat{\xi}'_{\mathbf{I}\mathbf{J}}(1+\epsilon_{B})^{\sum_{j=1}^{n}I_{j}(\sum_{m=j}^{n}J_{m})}+\hat{\xi}_{\mathbf{I}\mathbf{J}}\left(\sum_{k=0}^{\sum_{j=1}^{n}I_{j}(\sum_{m=j}^{n}J_{m})-1}(1+\epsilon_{B})^{k}\right) \right]\braket{X^{I_{1}}P^{J_{1}}\cdots X^{I_{n}}P^{J_{n}}},\label{gu2}
    \end{equation}
    and 
    \begin{equation}
        \hat{\xi}_{\mathbf{I}\mathbf{J}}'=\frac{\epsilon_A}{\epsilon_{\max}}\xi_{\mathbf{I}\mathbf{J}}',\quad \text{and}\quad \hat{\xi}_{\mathbf{I}\mathbf{J}}=\frac{\epsilon_{B}}{\epsilon_{\max}}\xi_{\mathbf{I}\mathbf{J}}, \label{chilinhahatdef}
    \end{equation}
    where $|\hat{\xi}_{\mathbf{I}\mathbf{J}}'|,|\hat{\xi}_{\mathbf{I}\mathbf{J}}|\leq 1$. 
\end{proof}  

To establish that the above result implies the convergence of the OBS characteristic function to classical statistics as $\epsilon_{A}$ and $\epsilon_{B}$ tend to zero, we must demonstrate the boundedness of the function $g(u,\epsilon_{A},\epsilon_{B})$ defined in the preceding result. To accomplish this, we invoke the following outcome, inspired by the Cauchy-Hadamard formula (see \cite{Suslov2005} and Theorem 2 on page 38 in \cite{Ahlfors1979}): 
\begin{proposition}
     Consider the series of the form
    \begin{equation}
        \sum_{\mathbf{I}\mathbf{J}}b_{\mathbf{I}\mathbf{J}}s_{\mathbf{I}\mathbf{J}},\label{bseries}
    \end{equation}
    where $b_{\mathbf{I}\mathbf{J}}$ and $s_{\mathbf{I}\mathbf{J}}$ are complex scalars indexed by $\mathbf{I}\mathbf{J}\equiv \{I_1,I_2,\cdots I_n,J_1,J_2,\cdots J_n\}\subset \mathbbm{Z}$, with $|b_{\mathbf{I}\mathbf{J}}|$ bounded. This series will be absolutely convergent (convergent and bounded) iff there are real bounded positive values $\{R_{\mathbf{I}\mathbf{J}}\}$ such that
    \begin{equation}
    R_{\mathbf{I}\mathbf{J}}^{\mathbf{I}+\mathbf{J}}>|s_{\mathbf{I}\mathbf{J}} |\qquad\text{and}\qquad\limsup_{\mathbf{I}+\mathbf{J}\to \infty} \left(|b_{\mathbf{I}\mathbf{J}}|^{1/(\mathbf{I}+\mathbf{J})}R_{\mathbf{I}\mathbf{J}}\right)\to 1,\label{limpowers}
    \end{equation}
    where $\mathbf{I}+\mathbf{J}=\sum_{j=1}^{n} I_j+J_{j}$.
    \label{lemaconvergent}
\end{proposition}
\begin{proof}
    First, suppose that $\{R_{\mathbf{I}\mathbf{J}}\}$ exist and satisfy conditions in Eq. \eqref{limpowers}. Then it follows that for any real $\zeta$, there is only a finite number of natural numbers $\{n_1,n_2\cdots n_{N}\}$ such that $\mathbf{I}+\mathbf{J}=n_i$, $i=1,2,\cdots N$, and $|b_{\mathbf{I}\mathbf{J}}|^{1/(\mathbf{I}+\mathbf{J})}>1/R_{\mathbf{I}\mathbf{J}}+\zeta$. To prove this statement, assume its negation, i.e. that there is a $\zeta$ such that for infinitely many natural numbers $\{n_1,n_2\cdots\}$, $n_i=\mathbf{I}+\mathbf{J}$ and $|b_{\mathbf{I}\mathbf{J}}|^{1/(\mathbf{I}+\mathbf{J})}> 1/R_{\mathbf{I}\mathbf{J}}+\zeta$. This implies $|b_{\mathbf{I}\mathbf{J}}|^{1/(\mathbf{I}+\mathbf{J})}R_{\mathbf{I}\mathbf{J}}-1>\zeta'$, where $ \zeta'=\zeta \min\{R_{\mathbf{I}\mathbf{J}}\}$ (notice that $R_{\mathbf{I}\mathbf{J}}>0$). Then it follows that there is no natural number $N'$ such that, for any natural $n>N'$, $n=\mathbf{I}+\mathbf{J}$ 
    \begin{equation}
        ||b_{\mathbf{I}\mathbf{J}}|^{1/(\mathbf{I}+\mathbf{J})}R_{\mathbf{I}\mathbf{J}}-1|\leq \zeta'.
    \end{equation}
    This is in direct contradiction with the limit in \eqref{limpowers}. Therefore, for every $\zeta$, there is only a finite number of natural numbers $\{n_1,n_2\cdots n_{N}\}$ such that $\mathbf{I}+\mathbf{J}=n_i$, $i=1,2,\cdots N$, and $|b_{\mathbf{I}\mathbf{J}}|^{1/(\mathbf{I}+\mathbf{J})}>1/R_{\mathbf{I}\mathbf{J}}+\zeta$. Consequently, $|b_{\mathbf{I}\mathbf{J}}|\leq (1/R_{\mathbf{I}\mathbf{J}}+\zeta)^{\mathbf{I}+\mathbf{J}}$ for all but a finite number of $|b_{\mathbf{I}\mathbf{J}}|$. As a result, one can rewrite the series \eqref{bseries} as
    \begin{equation}
        \sum_{>}b_{\mathbf{I}\mathbf{J}}s_{\mathbf{I}\mathbf{J}}+\sum_{\leq}b_{\mathbf{I}\mathbf{J}}s_{\mathbf{I}\mathbf{J}}\label{maisleq}
    \end{equation}
     where $\sum_{>}$ and $\sum_{\leq}$ represents the sum over all $\mathbf{I}\mathbf{J}$ indexes in which $|b_{\mathbf{I}\mathbf{J}}|^{1/(\mathbf{I}+\mathbf{J})}>1/R_{\mathbf{I}\mathbf{J}}+\zeta$ and $|b_{\mathbf{I}\mathbf{J}}|^{1/(\mathbf{I}+\mathbf{J})}\leq 1/R_{\mathbf{I}\mathbf{J}}+\zeta$, respectively. Because the sum $\sum_{>}$ is finite and $|b_{\mathbf{I}\mathbf{J}}|$ is bounded, then there exists a real non-negative value $C_{>}$ such that
    \begin{equation}
        \left|\sum_{>}b_{\mathbf{I}\mathbf{J}}s_{\mathbf{I}\mathbf{J}}\right|\leq C_{>}.
    \end{equation}
    On the other hand, if 
    \begin{equation}
        |s_{\mathbf{I}\mathbf{J}} |< \left(1/R_{\mathbf{I}\mathbf{J}}+\zeta\right)^{-(\mathbf{I}+\mathbf{J})},
    \end{equation}
    then it follows that
    \begin{equation}
        \left|\sum_{\leq}b_{\mathbf{I}\mathbf{J}}s_{\mathbf{I}\mathbf{J}}\right|\leq \sum_{\leq}(1/R_{\mathbf{I}\mathbf{J}}+\zeta)^{\mathbf{I}+\mathbf{J}}|s_{\mathbf{I}\mathbf{J}}|\leq \sum_{\leq} S^{\mathbf{I}+\mathbf{J}}\leq C_{\leq} \label{Cleq}
    \end{equation}
    where $C_{\leq}$ and $S<1$ are bounded real non-negative scalars. Notice that the infinity series of the type $\sum_{\leq} S^{\mathbf{I}+\mathbf{J}}$ is always convergent for $S< 1$, justifying the last inequality in Eq. \eqref{Cleq}. Because this result holds for any $\zeta$, then it follows that the series is convergent when $|s_{\mathbf{I}\mathbf{J}}|<R_{\mathbf{I}\mathbf{J}}^{\mathbf{I}+\mathbf{J}}$.

    For the converse, we prove by contradiction. Let us thus suppose that the series in Eq. \eqref{bseries} converge but for all set $\{R_{\mathbf{I}\mathbf{J}}\}$ such that $|s_{\mathbf{I}\mathbf{J}}|\leq R_{\mathbf{I}\mathbf{J}}^{\mathbf{I}+\mathbf{J}}$, the limit in \eqref{limpowers} does not hold. Then there is a real scalar $\zeta>0$, such that $|b_{\mathbf{I}\mathbf{J}}|^{1/(\mathbf{I}+\mathbf{J})}>(1/R_{\mathbf{I}\mathbf{J}}+\zeta)$ for all but a finite number of $|b_{\mathbf{I}\mathbf{J}}|$. Then, it follows that
    \begin{equation}
        \left|b_{\mathbf{I}\mathbf{J}}s_{\mathbf{I}\mathbf{J}}\right|>\left|(1/R_{\mathbf{I}\mathbf{J}}+\zeta)(s_{\mathbf{I}\mathbf{J}})^{1/(\mathbf{I}+\mathbf{J})}\right|^{\mathbf{I}+\mathbf{J}}.\label{contra1}
    \end{equation}
    Since this relation holds for any set  $\{R_{\mathbf{I}\mathbf{J}}\}$, then we can choose $R_{\mathbf{I}\mathbf{J}}^{\mathbf{I}+\mathbf{J}}=|s_{\mathbf{I}\mathbf{J}}|(1+\hat{\zeta}_{\mathbf{I}\mathbf{J}})^{\mathbf{I}+\mathbf{J}}>|s_{\mathbf{I}\mathbf{J}}|$, where 
    \begin{equation}
        \begin{array}{ll}
            0<\hat{\zeta}_{\mathbf{I}\mathbf{J}}, & \text{if}\,\,\zeta \geq 1   \\
             0<\hat{\zeta}_{\mathbf{I}\mathbf{J}}\leq \frac{\zeta|s_{\mathbf{I}\mathbf{J}}|^{1/(\mathbf{I}+\mathbf{J})}}{1-\zeta|s_{\mathbf{I}\mathbf{J}}|^{1/(\mathbf{I}+\mathbf{J})}}&\text{if}\,\, \zeta <  1,\label{hatzeta}
        \end{array}
    \end{equation}
    so that from Eq. \eqref{contra1}, it follows that for any $\mathbf{I}\mathbf{J}$,
    \begin{equation}
        \left|b_{\mathbf{I}\mathbf{J}}s_{\mathbf{I}\mathbf{J}}\right|>1,
    \end{equation}
    and the series therefore cannot converge, a contradiction. The converse is thus proved.
\end{proof}

We expect the series that define $\chi_{\text{\tiny OBS}}(u)$ and $\chi_{\text{\tiny OBS}}(u)$ to absolutely converge, leading to proper characteristic functions, satisfying:
\begin{equation}
    |\chi_{\text{\tiny OBS}}(u)|=\left|\sum_{\mathbf{I}\mathbf{J}}a_{\mathbf{I}\mathbf{J}}(u)\braket{X^{I_1}P^{J_{1}}\cdots X^{I_n}P^{J_{n}}}\right|\leq 1 \quad \text{and}\quad |\chi_{\text{\tiny CL}}(u)|=\left|\sum_{\mathbf{I}\mathbf{J}}a_{\mathbf{I}\mathbf{J}}(u)\braket{p_0^{\sum_{j=1}^{n} J_{j}} x_0^{\sum_{i=1}^{n}I_i}}_{\text{\tiny CL}}\right|\leq 1.
\end{equation}
and $\lim_{u\to \infty}\chi_{\text{\tiny OBS}}(u)=\lim_{u\to \infty}\chi_{\text{\tiny CL}}(u)=0$. It thus follows that $a_{\mathbf{I}\mathbf{J}}(u)$ are bounded functions, since they are sum of polynomials of $u<\infty$. As consequence of the Proposition \ref{lemaconvergent}, this implies, after the substitutions $b_{\mathbf{I}\mathbf{J}}\to a_{\mathbf{I}\mathbf{J}}(u)$ and $s_{\mathbf{I}\mathbf{J}}\to \braket{X^{I_1}P^{J_{1}}\cdots X^{I_n}P^{J_{n}}},\braket{p_0^{\sum_{j=1}^{n} J_{j}} x_0^{\sum_{i=1}^{n}I_i}}_{\text{\tiny CL}}$, that there exist real bounded non-negative scalars $\{R_{\mathbf{I}\mathbf{J}}\}$ and  $\{R_{\mathbf{I}\mathbf{J}}'\}$ such that 
\begin{equation}
    R_{\mathbf{I}\mathbf{J}}^{\mathbf{I}+\mathbf{J}}>|\braket{X^{I_1}P^{J_{1}}\cdots X^{I_n}P^{J_{n}}}|\qquad\text{and}\qquad\limsup_{\mathbf{I}+\mathbf{J}\to \infty} \left(|a_{\mathbf{I}\mathbf{J}}(u)|^{1/(\mathbf{I}+\mathbf{J})}R_{\mathbf{I}\mathbf{J}}\right)\to 1\label{limpowers2}
    \end{equation}
and 
\begin{equation}
    R_{\mathbf{I}\mathbf{J}}^{'\mathbf{I}+\mathbf{J}}>|\braket{p_0^{\sum_{j=1}^{n} J_{j}} x_0^{\sum_{i=1}^{n}I_i}}_{\text{\tiny CL}}|\qquad\text{and}\qquad\limsup_{\mathbf{I}+\mathbf{J}\to \infty} \left(|a_{\mathbf{I}\mathbf{J}}(u)|^{1/(\mathbf{I}+\mathbf{J})}R_{\mathbf{I}\mathbf{J}}'\right)\to 1.\label{limpowers3}
    \end{equation}
Taking into account these facts, we thus deduce the following important result.
\begin{proposition}
    The function $g(u,\epsilon_{A},\epsilon_{B})$ defined in Eq. \eqref{gu2} absolutely converges for sufficiently small $\epsilon_{\max}=\max\{\epsilon_{A},\epsilon_{B}\}\ll 1$. Therefore, if conditions (A) and (B) hold, then there is a bounded real non-negative scalar $C'$, such that
    \begin{equation}
        |\chi_{\text{\tiny OBS}}(u)-\chi_{\text{\tiny CL}}(u)|\leq C'\epsilon_{\max}.\label{boundcharac}
    \end{equation}
    \label{lemagutau}
\end{proposition}
\begin{proof}
    Since $|\hat{\xi}'_{\mathbf{I}\mathbf{J}}|,|\hat{\xi}_{\mathbf{I}\mathbf{J}}|\leq 1$ in the definition of $g(u,\epsilon_{A},\epsilon_{B})$ in Eq. \eqref{gu2}, then it follows that
    \begin{equation}
        \ba{rl}
        |g(u,\epsilon_{A},\epsilon_{B})|&=\left|\sum_{\mathbf{I}\mathbf{J}}a_{\mathbf{I}\mathbf{J}}(u)\left[\hat{\xi}'_{\mathbf{I}\mathbf{J}}(1+\epsilon_{B})^{\sum_{j=1}^{n}I_{j}(\sum_{m=j}^{n}J_{m})}+\hat{\xi}_{\mathbf{I}\mathbf{J}}\left(\sum_{k=0}^{\sum_{j=1}^{n}I_{j}(\sum_{m=j}^{n}J_{m})-1}(1+\epsilon_{B})^{k}\right) \right]\braket{X^{I_{1}}P^{J_{1}}\cdots X^{I_{n}}P^{J_{n}}}\right|\\
        &\leq \sum_{\mathbf{I}\mathbf{J}}|a_{\mathbf{I}\mathbf{J}}(u)|\left[(1+\epsilon_{B})^{(\mathbf{I}+\mathbf{J})^2}+\sum_{k=0}^{{(\mathbf{I}+\mathbf{J})^2}-1}(1+\epsilon_{B})^{k} \right]|\braket{X^{I_{1}}P^{J_{1}}\cdots X^{I_{n}}P^{J_{n}}}|\\
        &\leq (2+\epsilon_{B})\sum_{\mathbf{I}\mathbf{J}}|a_{\mathbf{I}\mathbf{J}}(u)|(1+\epsilon_{B})^{(\mathbf{I}+\mathbf{J})^2} |\braket{X^{I_{1}}P^{J_{1}}\cdots X^{I_{n}}P^{J_{n}}}|
        \ea
    \end{equation}
    where we considered the notation $\mathbf{I}+\mathbf{J}=\sum_{j=1}^{n} I_j+J_{j}$, the fact that $\sum_{j=1}^{n}I_{j}(\sum_{m=j}^{n}J_{m})\leq (\mathbf{I}+\mathbf{J})^2$, and
    \begin{equation}
        \sum_{k=0}^{{(\mathbf{I}+\mathbf{J})^2}-1}(1+\epsilon_{B})^{k} =\frac{(1+\epsilon_{B})^{(\mathbf{I}+\mathbf{J})^2}-1}{\epsilon_{B}}\leq (1+\epsilon_{B})^{(\mathbf{I}+\mathbf{J})^2+1}.
    \end{equation}
    Substituting  $b_{\mathbf{I}\mathbf{J}}\to (1+\epsilon_{B})^{\mathbf{I}+\mathbf{J}}a_{\mathbf{I}\mathbf{J}}(u)$ and $s_{\mathbf{I}\mathbf{J}}\to \braket{X^{I_1}P^{J_{1}}\cdots X^{I_n}P^{J_{n}}}(1+\epsilon_{B})^{\mathbf{I}+\mathbf{J}}$ in the Proposition \ref{lemaconvergent}, then it follows that the series
    \begin{equation}
        \sum_{\mathbf{I}\mathbf{J}}|a_{\mathbf{I}\mathbf{J}}(u)|(1+\epsilon_{B})^{(\mathbf{I}+\mathbf{J})^2} |\braket{X^{I_{1}}P^{J_{1}}\cdots X^{I_{n}}P^{J_{n}}}|\label{guaux}
    \end{equation}
    will absolutely converge if there is a set of bounded real non-negative scalars $\{R_{\mathbf{I}\mathbf{J}}''\}$ such that
    \begin{equation}
    R_{\mathbf{I}\mathbf{J}}^{''\mathbf{I}+\mathbf{J}}>(1+\epsilon_{B})^{\mathbf{I}+\mathbf{J}}|\braket{X^{I_1}P^{J_{1}}\cdots X^{I_n}P^{J_{n}}}|\qquad\text{and}\qquad\limsup_{\mathbf{I}+\mathbf{J}\to \infty} \left[(1+\epsilon_{B})|a_{\mathbf{I}\mathbf{J}}|^{1/(\mathbf{I}+\mathbf{J})}R_{\mathbf{I}\mathbf{J}}''\right]\to 1\label{limpowers4}
    \end{equation}
    Now, from the convergence of the OBS characteristic function, it follows that there exists scalars $\{R_{\mathbf{I}\mathbf{J}}\}$ satisfying Eq.~\eqref{limpowers2}. Let us thus define $R_{\mathbf{I}\mathbf{J}}''=R_{\mathbf{I}\mathbf{J}}/(1+\epsilon_{B})$. Considering $\epsilon_{B}\ll 1$, it follows that, because Eq.~\eqref{limpowers2} holds,  $|\braket{X^{I_1}P^{J_{1}}\cdots X^{I_n}P^{J_{n}}}|\leq R_{\mathbf{I}\mathbf{J}}\approx R_{\mathbf{I}\mathbf{J}}''$ and the limit in Eq. \eqref{limpowers4} is satisfied. Therefore, from Proposition \ref{lemaconvergent}, the series in Eq. \eqref{guaux} absolutely converges and $g(u,\epsilon_{A},\epsilon_{B})$ is bounded. Hence, there exists a real non-negative bounded scalar $C'$ such that $|g(u,\epsilon_{A},\epsilon_{B})|\leq C'$.
\end{proof}
A crucial consequence of the Proposition \ref{lemagutau} is that whenever $\epsilon_{A},\epsilon_{B}\to 0$, then $\chi_{\text{\tiny OBS}}(u)\to \chi_{\text{\tiny CL}}(u)$. Therefore, in the classical limit, the OBS statistics converge to the classical one. In this sense, we deduced the main result of this paper.
\begin{result}
    If conditions (A) and (B) hold, then
    \begin{equation}
        \lim_{\epsilon_{\max}\to 0}\int_{-\infty}^{\infty}dw|P_{\text{\tiny OBS}}(w)- P_{\text{\tiny CL}}(w)|= 0\label{Res210}
    \end{equation}
    and 
    \begin{equation}
        \int_{-\infty}^{\infty}dw|P_{\text{\tiny OBS}}(w)-P_{\text{\tiny CL}}(w)|\leq K\epsilon_{\max},\label{main1}
    \end{equation}
    where $K$ is a bounded non-negative real scalar that is independent of $\epsilon_A$ and $\epsilon_B$ and $\epsilon_{\max}=\max\{\epsilon_{A},\epsilon_{B}\}>0$. 
\end{result}
\begin{proof}
    Notice that, from Proposition \ref{lemagutau}, $\lim_{\epsilon_{\max}\to 0}\chi_{\text{\tiny OBS}}(u)= \chi_{\text{\tiny CL}}(u)$. Therefore, $P_{\text{\tiny OBS}}(w)\to P_{\text{\tiny CL}}(w)$ (see the Continuity Theorem at section 10.3 in \cite{Athreya2006} or Theorem 15 at section 14.7 together with Proposition 3 in section 14.3 in \cite{Fristedt2013}). As a result, by Scheffe's Theorem (Theorem 2.5.4 in \cite{Athreya2006}),
    \begin{equation}
        \lim_{\epsilon_{\max}\to 0}\int_{-\infty}^{\infty}dw|P_{\text{\tiny OBS}}(w)- P_{\text{\tiny CL}}(w)|= 0.\label{Res21}
    \end{equation}
    To prove Eq. \eqref{main1}, we consider Eqs. \eqref{xdifference} and \eqref{gu2} and the definition of the characteristic function, to deduce that 
    \begin{equation}
        \int_{-\infty}^{\infty}dw|P_{\text{\tiny OBS}}(w)-P_{\text{\tiny CL}}(w)|=\frac{1}{2\pi}\int_{-\infty}^{\infty}dw\left|\int_{-\infty}^{\infty} du \mathrm{e}^{iuw}(\chi_{\text{\tiny OBS}}(u)- \chi_{\text{\tiny CL}}(u))\right|=g'(\epsilon_{A},\epsilon_{B})\epsilon_{\max}
    \end{equation}
    where
    \begin{equation}
        g'(\epsilon_{A},\epsilon_{B})=\frac{1}{2\pi}\int_{-\infty}^{\infty} dw \left|\int_{-\infty}^{\infty} du \mathrm{e}^{iuw}g(u,\epsilon_{A},\epsilon_{B})\right|.\label{glinha}
    \end{equation}
    Using Proposition 2, we prove below in subsection ``\emph{Boundness of $g'(\epsilon_{A},\epsilon_{B})$}''  that $g'(\epsilon_{A},\epsilon_{B})$ must be bounded for any $\epsilon_{\max}>0$, so that we can define
    \begin{equation}
        K=\max_{\epsilon_{A},\epsilon_{B}}|g'(\epsilon_{A},\epsilon_{B})|
    \end{equation}
    where $K<\infty$ is a positive bounded real scalar. Therefore, it follows that
    \begin{equation}
        \int_{-\infty}^{\infty}dw|P_{\text{\tiny OBS}}(w)-P_{\text{\tiny CL}}(w)|\leq K\epsilon_{\max}
    \end{equation}
    and the result is proved.
\end{proof}

It is important to remark that the results carried here can all be naturally extended to the scenario of many systems described by Schr{\"o}dinger observables $X$ and $P$ satisfying $[X,P]=i\hbar \mathbbm{1}$. In this sense, by suitably adapting conditions (A) and (B), similar lines of reasoning and deductions as the ones done here can be used to deduce Result 2.

\subsection{Boundness of $g'(\epsilon_{A},\epsilon_{B})$}
To deduce that $g'(\epsilon_{A},\epsilon_{B})$ is bounded for any $\epsilon_B$, we first consider the definition
\begin{equation}
    g'(\epsilon_{A},\epsilon_{B})=\frac{1}{2\pi}\int_{-\infty}^{\infty} dw \left|\int_{-\infty}^{\infty} du \mathrm{e}^{-iuw}g(u,\epsilon_{A},\epsilon_{B})\right|.\label{glinhaboundeddef}
\end{equation}
where
    \begin{equation}
        g(u,\epsilon_{A},\epsilon_{B})=\sum_{\mathbf{I}\mathbf{J}}a_{\mathbf{I}\mathbf{J}}(u)\left[\hat{\xi}'_{\mathbf{I}\mathbf{J}}(1+\epsilon_{B})^{\sum_{j=1}^{n}I_{j}(\sum_{m=j}^{n}J_{m})}+\hat{\xi}_{\mathbf{I}\mathbf{J}}\left(\sum_{k=0}^{\sum_{j=1}^{n}I_{j}(\sum_{m=j}^{n}J_{m})-1}(1+\epsilon_{B})^{k}\right) \right]\braket{X^{I_{1}}P^{J_{1}}\cdots X^{I_{n}}P^{J_{n}}}.\label{gu2proposition}
    \end{equation}
satisfying, respectively, 
\begin{equation}
    \int_{-\infty}^{\infty}\,dw\,|P_{\text{\tiny OBS}}(w)-P_{\text{\tiny CL}}(w)|=g'(\epsilon_{A},\epsilon_{B})\epsilon_{\max}\qquad\text{and} \qquad P_{\text{\tiny OBS}}(w)-P_{\text{\tiny CL}}(w)=\frac{\epsilon_{\max}}{2\pi}\int_{-\infty}^{\infty}du\,\mathrm{e}^{-iuw}g(u,\epsilon_{A},\epsilon_{B}).\label{relobvia}
\end{equation}
Considering the definitions in Eqs. \eqref{CIJdef}, \eqref{chidef}, \eqref{chilinhadef}, and \eqref{chilinhahatdef}, we can explicitly write
    \begin{equation}
        \ba{l}
        \displaystyle\hat{\xi}_{\mathbf{I}\mathbf{J}}\left(\sum_{k=0}^{(\sum_{j=1}^{n}I_{j}\sum_{m=j}^{n}J_{m})-1}(1+\epsilon_{B})^{k}\right)\braket{X^{I_{1}}P^{J_{1}}\cdots  X^{I_{n}}P^{J_{n}}}=\frac{\braket{X^{I_{1}}P^{J_{1}}\cdots X^{I_{n}}P^{J_{n}}}-\braket{P^{\text{\tiny $\sum_{i=1}^{n}J_i'$}}X^{\text{\tiny $\sum_{i=1}^{n}I_i'$}}}}{\epsilon_{\max}},\\
        \displaystyle\hat{\xi}_{\mathbf{I}\mathbf{J}}'(1+\epsilon_{B})^{\sum_{j=1}^{n}I_{j}(\sum_{m=j}^{n}J_{m})}\braket{X^{I_{1}}P^{J_{1}}\cdots X^{I_{n}}P^{J_{n}}}=\frac{\braket{X^{I_{1}}P^{J_{1}}\cdots X^{I_{n}}P^{J_{n}}}-\braket{p^{\sum_{j=1}^{n} J_{j}} x^{\sum_{i=1}^{n}I_i}}_{\text{\tiny CL}}}{\epsilon_{\max}},
        \ea\label{chidefprop0}
    \end{equation}
    where, for any state satisfying conditions A and B, $|\hat{\xi}_{\mathbf{I}\mathbf{J}}|,|\hat{\xi}_{\mathbf{I}\mathbf{J}}'|\leq 1$. As a result, we have that
\begin{equation}
        g(u,\epsilon_{A},\epsilon_{B})=\sum_{\mathbf{I}\mathbf{J}}a_{\mathbf{I}\mathbf{J}}(u)\frac{\braket{X^{I_{1}}P^{J_{1}}\cdots X^{I_{n}}P^{J_{n}}}-\braket{p^{\sum_{j=1}^{n} J_{j}} x^{\sum_{i=1}^{n}I_i}}_{\text{\tiny CL}}}{\epsilon_{\max}}.\label{gu2proposition2}
    \end{equation}

Taken into account that $P_{\text{\tiny OBS}}(w)-P_{\text{\tiny CL}}(w)$ is purely real, then it follows from Eq. \eqref{relobvia} that $\int_{-\infty}^{\infty}du\,\mathrm{e}^{-iuw}g(u,\epsilon_{A},\epsilon_{B})$ must also be a purely real number. Expanding $g(u,\epsilon_{A},\epsilon_{B})$ considering Eq. \eqref{gu2proposition2}, we get
    \begin{equation}
    \ba{rl}
        \displaystyle\frac{\int_{-\infty}^{\infty}du\,\mathrm{e}^{iuw}g(u,\epsilon_{A},\epsilon_{B})}{2\pi}&\displaystyle=\sum_{\mathbf{I}\mathbf{J}}\frac{a_{\mathbf{I}\mathbf{J}}''(w)}{2\pi}\frac{\braket{X^{I_{1}}P^{J_{1}}\cdots X^{I_{n}}P^{J_{n}}}-\braket{p^{\sum_{j=1}^{n} J_{j}} x^{\sum_{i=1}^{n}I_i}}_{\text{\tiny CL}}}{\epsilon_{\max}},\label{alleq0}
        \ea
    \end{equation}
    where 
    \begin{equation}
        a_{\mathbf{I}\mathbf{J}}''(w)=\int_{-\infty}^{\infty}du\,\mathrm{e}^{iuw}a_{\mathbf{I}\mathbf{J}}(u).\label{all}
    \end{equation}
    Since the right-hand side of Eq. \eqref{alleq0} is real, its absolute value can only be
    \begin{equation}
        \ba{rl}
        \displaystyle\frac{\left|\int_{-\infty}^{\infty}du\,\mathrm{e}^{iuw}g(u,\epsilon_{A},\epsilon_{B})\right|}{2\pi}\displaystyle=\pm\sum_{\mathbf{I}\mathbf{J}}\frac{a_{\mathbf{I}\mathbf{J}}''(w)}{2\pi}\frac{\braket{X^{I_{1}}P^{J_{1}}\cdots X^{I_{n}}P^{J_{n}}}-\braket{p^{\sum_{j=1}^{n} J_{j}} x^{\sum_{i=1}^{n}I_i}}_{\text{\tiny CL}}}{\epsilon_{\max}},
        \ea\label{absgw}
    \end{equation}
    considering the case in which it is either positive or negative, respectively. Denoting $\int_{\geq(<)}\,dw$ as the integral over all intervals of $w$ in which $\int_{-\infty}^{\infty}du\,\mathrm{e}^{iuw}g(u,\epsilon_{A},\epsilon_{B})$ is non-negative (negative), we thus obtain
    \begin{equation}
        \ba{rl}
        g'(\epsilon_{A},\epsilon_{B})&\displaystyle=\sum_{\mathbf{I}\mathbf{J}}\frac{a_{\mathbf{I}\mathbf{J}}^{\geq}}{2\pi}\frac{\braket{X^{I_{1}}P^{J_{1}}\cdots X^{I_{n}}P^{J_{n}}}-\braket{p^{\sum_{j=1}^{n} J_{j}} x^{\sum_{i=1}^{n}I_i}}_{\text{\tiny CL}}}{\epsilon_{\max}}-\sum_{\mathbf{I}\mathbf{J}}\frac{a_{\mathbf{I}\mathbf{J}}^{<}}{2\pi}\frac{\braket{X^{I_{1}}P^{J_{1}}\cdots X^{I_{n}}P^{J_{n}}}-\braket{p^{\sum_{j=1}^{n} J_{j}} x^{\sum_{i=1}^{n}I_i}}_{\text{\tiny CL}}}{\epsilon_{\max}}
        \ea
    \end{equation}
    where
    \begin{equation}
        a_{\mathbf{I}\mathbf{J}}^{\geq (<)}=\int_{\geq(<)}\,dw\, a_{\mathbf{I}\mathbf{J}}''(w).
    \end{equation}
    Moreover, since $\int_{-\infty}^{\infty}dw (P_{\text{\tiny OBS}}(w)-P_{\text{\tiny CL}}(w))=0$, it follows that
    \begin{equation}
        \ba{rl}
        0&\displaystyle=\int_{-\infty}^{\infty}dw\frac{\int_{-\infty}^{\infty}du\,\mathrm{e}^{iuw}g(u,\epsilon_{A},\epsilon_{B})}{2\pi}\\
        &\displaystyle=\sum_{\mathbf{I}\mathbf{J}}\frac{a_{\mathbf{I}\mathbf{J}}^{\geq}}{2\pi}\left[\hat{\xi}'_{\mathbf{I}\mathbf{J}}(1+\epsilon_{B})^{\sum_{j=1}^{n}I_{j}(\sum_{m=j}^{n}J_{m})}+\hat{\xi}_{\mathbf{I}\mathbf{J}}\left(\sum_{k=0}^{\sum_{j=1}^{n}I_{j}(\sum_{m=j}^{n}J_{m})-1}(1+\epsilon_{B})^{k}\right) \right]\braket{X^{I_{1}}P^{J_{1}}\cdots X^{I_{n}}P^{J_{n}}}+\\
        &\displaystyle+\sum_{\mathbf{I}\mathbf{J}}\frac{a_{\mathbf{I}\mathbf{J}}^{<}}{2\pi}\left[\hat{\xi}'_{\mathbf{I}\mathbf{J}}(1+\epsilon_{B})^{\sum_{j=1}^{n}I_{j}(\sum_{m=j}^{n}J_{m})}+\hat{\xi}_{\mathbf{I}\mathbf{J}}\left(\sum_{k=0}^{\sum_{j=1}^{n}I_{j}(\sum_{m=j}^{n}J_{m})-1}(1+\epsilon_{B})^{k}\right) \right]\braket{X^{I_{1}}P^{J_{1}}\cdots X^{I_{n}}P^{J_{n}}}\\
        \ea
    \end{equation}
    so that
    \begin{equation}
        \ba{rl}
        g'(\epsilon_{A},\epsilon_{B})&\displaystyle=\frac{2}{\epsilon_{\max}}\sum_{\mathbf{I}\mathbf{J}}\frac{a_{\mathbf{I}\mathbf{J}}^{\geq}}{2\pi}\braket{X^{I_{1}}P^{J_{1}}\cdots X^{I_{n}}P^{J_{n}}}-\braket{p^{\sum_{j=1}^{n} J_{j}} x^{\sum_{i=1}^{n}I_i}}_{\text{\tiny CL}}.\label{ultimaantesdofim}
        \ea
    \end{equation}
    Before proving that $g'(\epsilon_{A},\epsilon_{B})$ is bounded, it is important to demonstrate that $\frac{a_{\mathbf{I}\mathbf{J}}^{\geq}}{2\pi}$ are bounded. For that, we first notice that, from Eq. \eqref{relobvia},
    \begin{equation}
        g'(\epsilon_{A},\epsilon_{B})\epsilon_{\max}\leq 2.
    \end{equation}
    Therefore, 
    \begin{equation}
        \sum_{\mathbf{I}\mathbf{J}}a_{\mathbf{I}\mathbf{J}}^{\geq}[\braket{X^{I_{1}}P^{J_{1}}\cdots X^{I_{n}}P^{J_{n}}}-\braket{p^{\sum_{j=1}^{n} J_{j}} x^{\sum_{i=1}^{n}I_i}}_{\text{\tiny CL}}]\leq 2\pi.\label{boundedamais}
    \end{equation}
    As we demonstrated in Eqs. \eqref{limpowers2} and \eqref{limpowers3}, there exist real bounded non-negative scalars $\{R_{\mathbf{I}\mathbf{J}}\}$ and $\{R_{\mathbf{I}\mathbf{J}}'\}$ such that 
\begin{equation}
    R_{\mathbf{I}\mathbf{J}}^{\mathbf{I}+\mathbf{J}}>|\braket{X^{I_1}P^{J_{1}}\cdots X^{I_n}P^{J_{n}}}|\qquad\text{and}\qquad R_{\mathbf{I}\mathbf{J}}^{'\mathbf{I}+\mathbf{J}}>|\braket{p_0^{\sum_{j=1}^{n} J_{j}} x_0^{\sum_{i=1}^{n}I_i}}_{\text{\tiny CL}}|.\label{limpowers2bounded}
    \end{equation}
As a consequence, $|\braket{X^{I_{1}}P^{J_{1}}\cdots X^{I_{n}}P^{J_{n}}}-\braket{p^{\sum_{j=1}^{n} J_{j}} x^{\sum_{i=1}^{n}I_i}}_{\text{\tiny CL}}|\leq 2 \max\{R_{\mathbf{I}\mathbf{J}}^{\mathbf{I}+\mathbf{J}},R_{\mathbf{I}\mathbf{J}}^{'\mathbf{I}+\mathbf{J}}\}$ is also bounded. Therefore, we can consider Proposition 2, substituting  $b_{\mathbf{I}\mathbf{J}}\to |\braket{X^{I_{1}}P^{J_{1}}\cdots X^{I_{n}}P^{J_{n}}}-\braket{p^{\sum_{j=1}^{n} J_{j}} x^{\sum_{i=1}^{n}I_i}}_{\text{\tiny CL}}|$ and $s_{\mathbf{I}\mathbf{J}}\to a_{\mathbf{I}\mathbf{J}}^{\geq}$, to deduce that there exist real bounded non-negative scalars $\{A_{\mathbf{I}\mathbf{J}}\}$ such that 
\begin{equation}
    A_{\mathbf{I}\mathbf{J}}^{\mathbf{I}+\mathbf{J}}>|a_{\mathbf{I}\mathbf{J}}^{\geq}|\qquad\text{and}\qquad\limsup_{\mathbf{I}+\mathbf{J}\to \infty} \left(|\braket{X^{I_{1}}P^{J_{1}}\cdots X^{I_{n}}P^{J_{n}}}-\braket{p^{\sum_{j=1}^{n} J_{j}} x^{\sum_{i=1}^{n}I_i}}_{\text{\tiny CL}}|^{1/(\mathbf{I}+\mathbf{J})}A_{\mathbf{I}\mathbf{J}}\right)\to 1.\label{limpowersboundedfundamental}
    \end{equation}
Therefore, $|a_{\mathbf{I}\mathbf{J}}^{\geq}|$ is bounded for every indexes $\mathbf{I}\mathbf{J}$ in which $\braket{X^{I_{1}}P^{J_{1}}\cdots X^{I_{n}}P^{J_{n}}}-\braket{p^{\sum_{j=1}^{n} J_{j}} x^{\sum_{i=1}^{n}I_i}}_{\text{\tiny CL}}\neq 0$.

We are now in position to show that $g'(\epsilon_{A},\epsilon_{B})$ is bounded for any $\epsilon_{\max}>0$, even when $\epsilon_{\max}\to 0$. To check this, we consider $\epsilon_{A}>0$ or $\epsilon_{B}>0$, such that $\epsilon_{\max}=\max\{\epsilon_{A},\epsilon_{B}\}>0$, and Eq. \eqref{chidefprop0} to rewrite Eq. \eqref{ultimaantesdofim} as
\begin{equation}
        \ba{rl}
        g'(\epsilon_{A},\epsilon_{B})&\displaystyle=\frac{2}{\epsilon_{\max}}\sum_{\mathbf{I}\mathbf{J}}\frac{a_{\mathbf{I}\mathbf{J}}^{\geq}}{2\pi}\braket{X^{I_{1}}P^{J_{1}}\cdots X^{I_{n}}P^{J_{n}}}-\braket{p^{\sum_{j=1}^{n} J_{j}} x^{\sum_{i=1}^{n}I_i}}_{\text{\tiny CL}}\\
        &\displaystyle=\frac{1}{\pi}\sum_{\mathbf{I}\mathbf{J}}a_{\mathbf{I}\mathbf{J}}^{\geq}\left[\hat{\xi}'_{\mathbf{I}\mathbf{J}}(1+\epsilon_{B})^{\sum_{j=1}^{n}I_{j}(\sum_{m=j}^{n}J_{m})}+\hat{\xi}_{\mathbf{I}\mathbf{J}}\left(\sum_{k=0}^{\sum_{j=1}^{n}I_{j}(\sum_{m=j}^{n}J_{m})-1}(1+\epsilon_{B})^{k}\right) \right]\braket{X^{I_{1}}P^{J_{1}}\cdots X^{I_{n}}P^{J_{n}}}.\label{glinhaperto}
        \ea
    \end{equation}
If the initial quantum state $\rho$ and classical distribution $\varrho(\Gamma_0)$ are such as to satisfy conditions A and B, then, as we deduced in Proposition 1, $|\xi_{\mathbf{I}\mathbf{J}}|,|\hat{\xi}_{\mathbf{I}\mathbf{J}}|\leq 1$. Moreover, for $\epsilon_A$ and $\epsilon_B$ not tending to $0$, $g'(\epsilon_{A},\epsilon_{B})$ is surely bounded, since, from Eq. \eqref{relobvia},
\begin{equation}
    |g'(\epsilon_{A},\epsilon_{B})|=\frac{\int_{-\infty}^{\infty}\,dw\,|P_{\text{\tiny OBS}}(w)-P_{\text{\tiny CL}}(w)|}{\epsilon_{\max}}\leq \frac{2}{\epsilon_{\max}}<\infty.\label{relobvia2}
\end{equation}
Because $a_{\mathbf{I}\mathbf{J}}^{\geq}$ are bounded, then we can consider again Proposition 2 regarding the series in the last line of Eq. \eqref{glinhaperto} to show that there exist real bounded non-negative scalars $\{\hat{R}_{\mathbf{I}\mathbf{J}}\}$ such that 
\begin{equation}
    \hat{R}_{\mathbf{I}\mathbf{J}}^{\mathbf{I}+\mathbf{J}}>|\braket{X^{I_{1}}P^{J_{1}}\cdots X^{I_{n}}P^{J_{n}}}|\label{limpowersboundedfundamental1}
    \end{equation}
    and
    \begin{equation}
        \limsup_{\mathbf{I}+\mathbf{J}\to \infty} \left[\left|a_{\mathbf{I}\mathbf{J}}^{\geq}\left(\hat{\xi}'_{\mathbf{I}\mathbf{J}}(1+\epsilon_{B})^{\sum_{j=1}^{n}I_{j}(\sum_{m=j}^{n}J_{m})}+\hat{\xi}_{\mathbf{I}\mathbf{J}}\left(\sum_{k=0}^{\sum_{j=1}^{n}I_{j}(\sum_{m=j}^{n}J_{m})-1}(1+\epsilon_{B})^{k}\right)\right)\right|^{1/(\mathbf{I}+\mathbf{J})}\hat{R}_{\mathbf{I}\mathbf{J}}\right]\to 1.\label{limpowersboundedfundamental2}
    \end{equation}
   Considering that $\epsilon_B>0$, and the fact that $|\xi_{\mathbf{I}\mathbf{J}}|,|\hat{\xi}_{\mathbf{I}\mathbf{J}}|\leq 1$, $\mathbf{I}+\mathbf{J}=\sum_{j=1}^{n} I_j+J_{j}$, so that $\sum_{j=1}^{n}I_{j}(\sum_{m=j}^{n}J_{m})\leq (\mathbf{I}+\mathbf{J})^2$, we can deduce that
    \begin{equation}
        \sum_{k=0}^{{(\mathbf{I}+\mathbf{J})^2}-1}(1+\epsilon_{B})^{k} =\frac{(1+\epsilon_{B})^{(\mathbf{I}+\mathbf{J})^2}-1}{\epsilon_{B}}\leq (1+\epsilon_{B})^{(\mathbf{I}+\mathbf{J})^2+1},\label{expepsilons}
    \end{equation}
    and 
    \begin{equation}
        \left|\hat{\xi}'_{\mathbf{I}\mathbf{J}}(1+\epsilon_{B})^{\sum_{j=1}^{n}I_{j}(\sum_{m=j}^{n}J_{m})}+\hat{\xi}_{\mathbf{I}\mathbf{J}}\left(\sum_{k=0}^{\sum_{j=1}^{n}I_{j}(\sum_{m=j}^{n}J_{m})-1}(1+\epsilon_{B})^{k}\right)\right|\leq (2+\epsilon_B)\left|(1+\epsilon_{B})^{(\mathbf{I}+\mathbf{J})^2}\right|.\label{auxKK}
    \end{equation}
    As a result, it follows that
    \begin{equation}
    \ba{l}
        1=\limsup_{\mathbf{I}+\mathbf{J}\to \infty} \left[\left|a_{\mathbf{I}\mathbf{J}}^{\geq}\left(\hat{\xi}'_{\mathbf{I}\mathbf{J}}(1+\epsilon_{B})^{\sum_{j=1}^{n}I_{j}(\sum_{m=j}^{n}J_{m})}+\hat{\xi}_{\mathbf{I}\mathbf{J}}\left(\sum_{k=0}^{\sum_{j=1}^{n}I_{j}(\sum_{m=j}^{n}J_{m})-1}(1+\epsilon_{B})^{k}\right)\right)\right|^{1/(\mathbf{I}+\mathbf{J})}\hat{R}_{\mathbf{I}\mathbf{J}}\right]\\
        =\limsup_{\mathbf{I}+\mathbf{J}\to \infty} \left[\left|a_{\mathbf{I}\mathbf{J}}^{\geq}\frac{\left(\hat{\xi}'_{\mathbf{I}\mathbf{J}}(1+\epsilon_{B})^{\sum_{j=1}^{n}I_{j}(\sum_{m=j}^{n}J_{m})}+\hat{\xi}_{\mathbf{I}\mathbf{J}}\left(\sum_{k=0}^{\sum_{j=1}^{n}I_{j}(\sum_{m=j}^{n}J_{m})-1}(1+\epsilon_{B})^{k}\right)\right)}{(2+\epsilon_B)\left|(1+\epsilon_{B})^{(\mathbf{I}+\mathbf{J})^2}\right|}(2+\epsilon_B)\left|(1+\epsilon_{B})^{(\mathbf{I}+\mathbf{J})^2}\right|\right|^{1/(\mathbf{I}+\mathbf{J})}\hat{R}_{\mathbf{I}\mathbf{J}}\right]\\
        =\limsup_{\mathbf{I}+\mathbf{J}\to \infty} \left[\left(|a_{\mathbf{I}\mathbf{J}}^{\geq}|\left|(1+\epsilon_{B})^{(\mathbf{I}+\mathbf{J})^2}\right|\right)^{1/(\mathbf{I}+\mathbf{J})}\hat{R}_{\mathbf{I}\mathbf{J}}\right].
        \ea\label{limpowersboundedfundamental3}
    \end{equation}
    where we considered the fact that
    \begin{equation}
        \frac{\left|\hat{\xi}'_{\mathbf{I}\mathbf{J}}(1+\epsilon_{B})^{\sum_{j=1}^{n}I_{j}(\sum_{m=j}^{n}J_{m})}+\hat{\xi}_{\mathbf{I}\mathbf{J}}\left(\sum_{k=0}^{\sum_{j=1}^{n}I_{j}(\sum_{m=j}^{n}J_{m})-1}(1+\epsilon_{B})^{k}\right)\right|}{(2+\epsilon_B)\left|(1+\epsilon_{B})^{(\mathbf{I}+\mathbf{J})^2}\right|}(2+\epsilon_B)\leq 2+\epsilon_B 
    \end{equation}
    is bounded for any indexes $\mathbf{I}\mathbf{J}$. From Eq. \eqref{limpowersboundedfundamental3} and Proposition 2, it thus follows that
    \begin{equation}
        \sum_{\mathbf{I}\mathbf{J}}|a_{\mathbf{I}\mathbf{J}}^{\geq}|\left|(1+\epsilon_{B})^{(\mathbf{I}+\mathbf{J})^2}\right||\braket{X^{I_{1}}P^{J_{1}}\cdots X^{I_{n}}P^{J_{n}}}|<\infty,\label{menorinfty}
    \end{equation}
    is bounded, since $\hat{R}_{\mathbf{I}\mathbf{J}}^{\mathbf{I}+\mathbf{J}}>|\braket{X^{I_{1}}P^{J_{1}}\cdots X^{I_{n}}P^{J_{n}}}|$. Considering Eq. \eqref{glinhaperto} together with Eqs. \eqref{auxKK} and \eqref{menorinfty}, it follows that
    \begin{equation}
        \ba{rl}
        g'(\epsilon_{A},\epsilon_{B})&\leq |g'(\epsilon_{A},\epsilon_{B})|\leq \frac{(2+\epsilon_{B})}{\pi}\sum_{\mathbf{I}\mathbf{J}}|a_{\mathbf{I}\mathbf{J}}^{\geq}|(1+\epsilon_{B})^{(\mathbf{I}+\mathbf{J})^2}|\braket{X^{I_{1}}P^{J_{1}}\cdots X^{I_{n}}P^{J_{n}}}|<\infty.\label{boundglinhaij}
        \ea
    \end{equation}
    A similar result can be deduced for either $\epsilon_B=0$ and $\epsilon_A>0$ or $\epsilon_A=0$ and $\epsilon_B>0$. As a result, $g'(\epsilon_{A},\epsilon_{B})$ is bounded for any case in which $\epsilon_A>0$ or $\epsilon_B>0$. Even when $\epsilon_{\max}\to 0$, $g'(\epsilon_{A},\epsilon_{B})$ is bounded, since, considering Eq. \eqref{boundglinhaij}, it follows that
    \begin{equation}
        \ba{rl}
        \lim_{\epsilon_{\max}\to 0} g'(\epsilon_{A},\epsilon_{B})&\leq \lim_{\epsilon_{\max}\to 0} \frac{(2+\epsilon_{B})}{\pi}\sum_{\mathbf{I}\mathbf{J}}|a_{\mathbf{I}\mathbf{J}}^{\geq}|(1+\epsilon_{B})^{(\mathbf{I}+\mathbf{J})^2}|\braket{X^{I_{1}}P^{J_{1}}\cdots X^{I_{n}}P^{J_{n}}}|<\infty.
        \ea
    \end{equation}
     Consequently, we can consider the bounded real positive constant $K$ such that
\begin{equation}
    K=\max_{\epsilon_{A},\epsilon_{B}}g'(\epsilon_{A},\epsilon_{B}).
\end{equation}

\section{Example: non-autonomous harmonic oscillator}
In this section, we analyze in detail the model presented in Eqs. (4) and (5) in the main text. We consider the following Hamiltonian operator and function:
\begin{equation}
    H(t) = \frac{P^2}{2m} + m\omega^2(t)X^2, \quad H_{\text{\tiny CL}}^{t}(\Gamma_t,t) = \frac{p^2(t)}{2m} + m\omega^2(t)x^2(t), \label{hclhq}
\end{equation}
where $\omega^2(t) = \omega_0^2 + (\omega_1^2 - \omega_0^2)\frac{t}{\tau}$. This operators and function model the motion and energy of a particle within a time-dependent trap in the quantum and classical scenarios, respectively. Solving the Heisenberg and Hamilton equations, we obtain $\ddot{X}_{h}(t) + \omega^{2}(t)X_{h}(t) = 0$ and $\ddot{x}(t) + \omega^{2}(t)x(t) = 0$, with solutions given by
\begin{equation}
    X_{h}(t) = A(t)X + B(t)P, \quad P_h(t) = C(t)X + D(t)P, \quad \text{and} \quad x(t) = A(t)x_0 + B(t)p_0, \quad p(t) = C(t)x_0 + D(t)p_0. \label{xtpt}
\end{equation}
It is worth noting that $(X,P)$ represent the quantum Schr{\"o}dinger operators of position and momentum, with their Heisenberg versions $(X_{h}(t),P_{h}(t))$. Additionally, $\Gamma_0=(x_0,p_0)$ denotes the classical initial phase point, evolving to the phase point $\Gamma_t(\Gamma_0,t) = [x_t(\Gamma_0,t),p_t(\Gamma_0,t)]$ at time $t$. The functions $A(t)$, $B(t)$, $C(t)$, and $D(t)$ are defined as
\begin{align}
    \displaystyle A(t)&\displaystyle=\frac{\text{Ai}'\left[-\omega_0^2\tau^{'2}\right]
   \text{Bi}\left[-\omega^2(t)\tau^{'2}\right]-\text{Bi}'\left[-\omega_0^2\tau^{'2}\right] \text{Ai}\left[-\omega^2(t)\tau^{'2}\right]}{\text{Ai}'\left[-\omega_0^2\tau^{'2}\right] \text{Bi}\left[-\omega_0^2\tau^{'2}\right]-\text{Ai}\left[-\omega_0^2\tau^{'2}\right] \text{Bi}'\left[-\omega_0^2\tau^{'2}\right]},\\
    \displaystyle B(t)&\displaystyle=\frac{\tau'}{m}\frac{\text{Ai}\left[-\omega_0^2\tau^{'2}\right] \text{Bi}\left[-\omega^2(t)\tau^{'2}\right]-\text{Bi}\left[-\omega_0^2\tau^{'2}\right]
   \text{Ai}\left[-\omega^2(t)\tau^{'2}\right]}{ \text{Ai}'\left[-\omega_0^2\tau^{'2}\right] \text{Bi}\left[-\omega_0^2\tau^{'2}\right]-\text{Ai}\left[-\omega_0^2\tau^{'2}\right] \text{Bi}'\left[-\omega_0^2\tau^{'2}\right]},\\
   \displaystyle  C(t)&\displaystyle=\frac{m \tau' \left(\text{Ai}'\left[-\omega^2(t)\tau^{'2}\right]\text{Bi}'\left[-\omega_0^2\tau^{'2}\right]-\text{Ai}'\left[-\omega_0^2\tau^{'2}\right] \text{Bi}'\left[-\omega^2(t)\tau^{'2}\right] \right)}{ \text{Ai}'\left[-\omega_0^2\tau^{'2}\right] \text{Bi}\left[-\omega_0^2\tau^{'2}\right]-\text{Ai}\left[-\omega_0^2\tau^{'2}\right] \text{Bi}'\left[-\omega_0^2\tau^{'2}\right]},\\
   \displaystyle  D(t)&\displaystyle=\frac{\text{Bi}\left[-\omega_0^2\tau^{'2}\right]
   \text{Ai}'\left[-\omega^2(t)\tau^{'2}\right]-\text{Ai}\left[-\omega^2(t)\tau^{'2}\right] \text{Bi}'\left[-\omega^2(t)\tau^{'2}\right]}{ \text{Ai}'\left[-\omega_0^2\tau^{'2}\right] \text{Bi}\left[-\omega_0^2\tau^{'2}\right]-\text{Ai}\left[-\omega_0^2\tau^{'2}\right] \text{Bi}'\left[-\omega_0^2\tau^{'2}\right]}\label{ABCD}
\end{align}
where $\tau'=\sqrt[3]{\tau(\omega_1^2-\omega_0^2)^{-1}}$. $\text{Ai}(z)$ and $\text{Bi}(z)$ are the Airy functions, defined as the two linearly independent solutions of equations of the type $y''(z)-y(z)z=0$, with explicit forms
\begin{equation}
   \text{Ai}(z)=\frac{F_{1}^{0}(\frac{2}{3};\frac{1}{9}z^3)}{3^{2/3}Ga(2/3)}-\frac{F_{1}^{0}(\frac{4}{3};\frac{1}{9}z^3)z}{3^{1/3}Ga(1/3)},\quad \text{and}\quad \text{Bi}(z)=\frac{F_{1}^{0}(\frac{2}{3};\frac{1}{9}z^3)}{3^{1/6}Ga(2/3)}+\frac{3^{1/6}F_{1}^{0}(\frac{4}{3};\frac{1}{9}z^3)z}{Ga(1/3)},
\end{equation}
where 
\begin{equation}
    F_{1}^{0}(a;\theta)=\sum_{n}^{\infty}\frac{\theta^n}{(a)_n n!}\quad\text{and}\quad Ga(\theta)=\int_{0}^{\infty}t^{\theta-1}e^{-t}dt
\end{equation}
are respectively the confluent hypergeometric limit and Gamma functions and 
\begin{equation}
    (a)_{n}=\left\{\begin{array}{ll}
        1 &n=0  \\
         a(a+1)\cdots(a+n-1)& n>0.
    \end{array} \right.
\end{equation}
Moreover, $Ai'(z)=\partial_{z}Ai(z)$ and $Bi'(z)=\partial_{z}Bi(z)$. 

Given the solutions in Eq. \eqref{xtpt}, we can compare the classical work done during any interval $[t_1,t_2]$ with the statistics provided by OBS and TPM protocols in the quantum scenario. To begin, we compute the classical work. In this context, we define $H_{\text{\tiny CL}}(\Gamma_0,t)=H_{\text{\tiny CL}}^{t}(\Gamma_t,t)$ as the energy at time $t$ of the particle that was at $\Gamma_0$ at time $0$. Subsequently, by substituting the solutions from Eq. \eqref{xtpt} into Eq. \eqref{hclhq}, we obtain
\begin{equation}
    H_{\text{\tiny CL}}(\Gamma_0,t)=E(t)x_0^2+F(t)p_0^2+2G(t_2)x_0 p_0\label{Hclgamma0}
\end{equation}
where 
\begin{equation}
    E(t)=\frac{m\omega^2(t)}{2}A^2(t)+\frac{C^2(t)}{2m},\quad
    F(t)=\frac{m\omega^2(t)}{2}B^2(t)+\frac{D^2(t)}{2m},\quad
    G(t)=\frac{m\omega^2(t)}{2}A(t)B(t)+\frac{C(t)D(t)}{2m}.\label{EFG}
\end{equation}
Hence, the classical work performed on a particle, starting at the phase point $\Gamma_0=(x_0,p_0)$, during the interval $[t_1,t_2]$, can be determined from Eq. \eqref{Hclgamma0}, and it is given by 
\begin{equation}
    W_{\text{\tiny CL}}(\Gamma_0,t_2,t_1)=H_{\text{\tiny CL}}(\Gamma_0,t_2)-H(\Gamma_0,t_1)=(E(t_2)-E(t_1))x_0^2+(F(t_2)- F(t_1))p_0^2+2(G(t_2)-G(t_1))x_0 p_0.\label{Wgamma0}
\end{equation}
Therefore, the statistics of classical work can be raised using Eq. \eqref{Gucl} to obtain the characteristic function $\chi_{\text{\tiny CL}}$ for any initial phase distribution $\varrho(\Gamma_0)$.

\begin{figure}[h]

    \centering
    \includegraphics[width=0.4\linewidth]{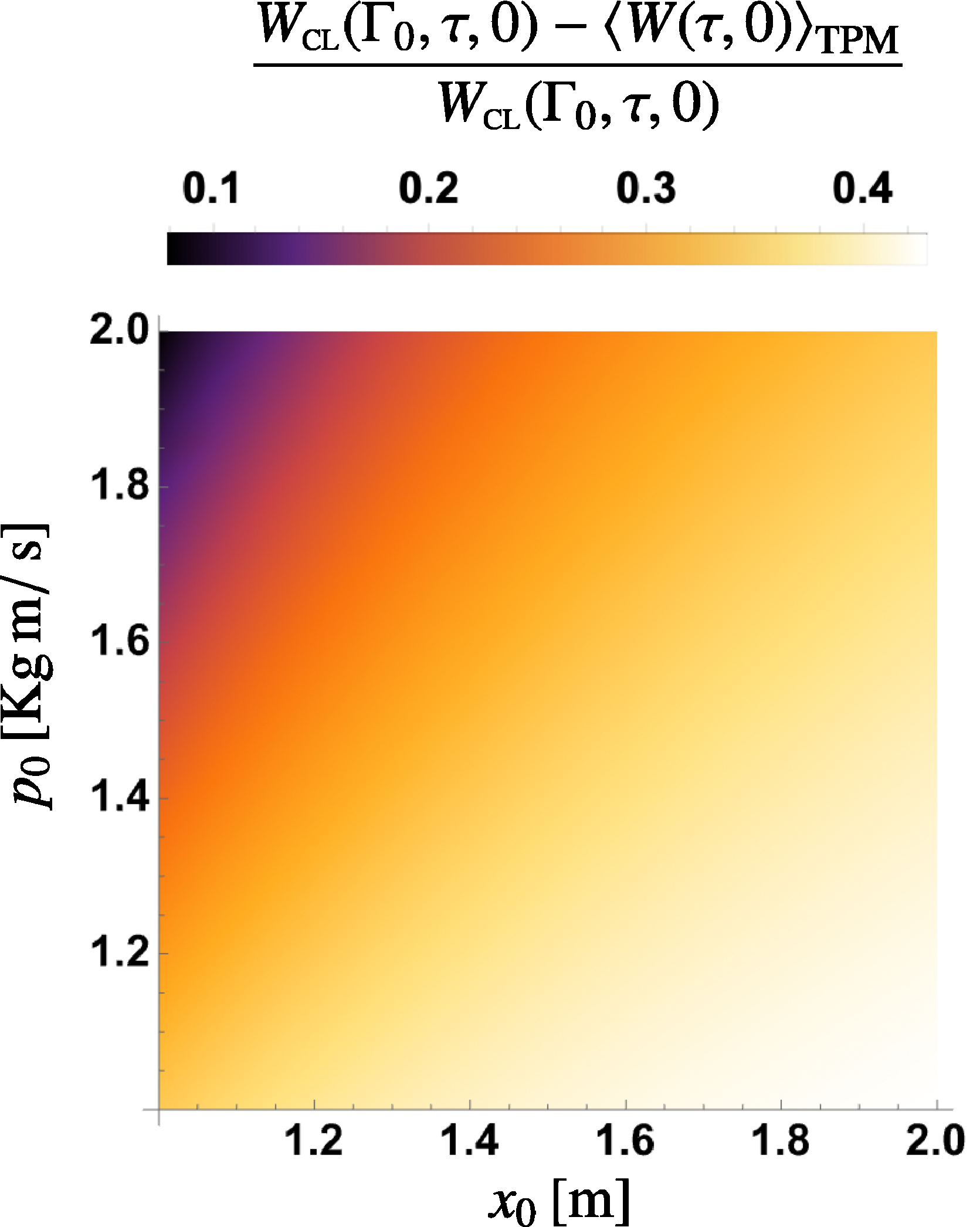}
    \caption{We present a plot illustrating the relative difference $(W_{\text{\tiny CL}}(\Gamma_0,\tau,0)-\braket{W(\tau,0)}_{\text{TPM}})/W_{\text{\tiny CL}}(\Gamma_0,\tau,0)$ in the classical limit, where $\hbar/|x_0p_0|\to 0$ and $|\alpha|\to \infty$. The axes represent the initial position and momentum of a particle in a time-dependent trap, and the color axis represents the ratio $(W_{\text{\tiny CL}}(\Gamma_0,\tau,0)-\braket{W(\tau,0)}_{\text{TPM}})/W_{\text{\tiny CL}}(\Gamma_0,\tau,0)$. The parameters used for this plot are $\omega_0=\omega_1/3=1\,\mathrm{s}^{-1}$, $m=1\,\mathrm{kg}$, and $\tau=1\,\mathrm{s}$. The significant differences between the TPM statistics and the classical results are evident, demonstrating that the TPM protocol cannot replicate the classical outcomes in this classical limit scenario.}
    \label{fig1}
\end{figure}

For the quantum scenario, we start computing the energy operator, in analogy to \eqref{Hclgamma0}. Substituting Eq. \eqref{xtpt} in Eq. \eqref{hclhq}, we thus derive
\begin{equation}
    H_{h}(t)=E(t)X^2+F(t)P^2+G(t)\{X,P\}\label{Hqht}
\end{equation}
where $\{X,P\}=XP+PX$ is the anticommutator of the Schr{\"o}dinger operators $X,P$. The work operator is thus given by
\begin{equation}
     W(t_2,t_1)=(E(t_2)-E(t_1))X^2+(F(t_2)-F(t_1))P^2+(G(t_2)-G(t_1))\{X,P\}\label{WqopHO}
\end{equation}
We now turn our attention to specifying the quantum state against which we will compare with the classical statistics. In this regard, we consider an initial coherent state defined as follows:
\begin{equation}
    \rho = \ket{\alpha}\bra{\alpha}, \quad 
    \ket{\alpha} = \mathrm{e}^{-|\alpha|^2/2}\sum_{n=0}^{\infty}\frac{\alpha^{n}}{\sqrt{n!}}\ket{e_{n}(0)}, \label{rhoinitial}
\end{equation}
where $\ket{e_{n}(0)}$ represents the eigenstates of $H(0)$, and $\alpha$ is a complex number. It is well-established \cite{Cohen2020} that coherent states satisfy the following relations:
\begin{equation}
p_{\alpha} = \braket{P} = \sqrt{2m\hbar\omega_0}\Im(\alpha), \quad x_{\alpha} = \braket{X} = \sqrt{\frac{2\hbar}{m\omega_0}}\Re(\alpha), \quad \braket{P^2} = p_{\alpha}^2 + \frac{m\hbar\omega_0}{2}, \quad \braket{X^2} = x_{\alpha}^2 + \frac{\hbar}{2m\omega_0}, \quad \braket{\{X,P\}} = 2x_{\alpha}p_{\alpha}. \label{alphadef}
\end{equation}
Here, $\Re(\alpha)$ and $\Im(\alpha)$ denote the real and imaginary parts of $\alpha$. Moreover, for sufficiently large $|\alpha|$, the center of the wave packet described by $\braket{X}$ and $\braket{P}$ exhibits classical particle-like behavior, following dynamics governed by the classical Hamiltonian $H_{\text{\tiny CL}}$. This represents the classical limit of this system. Indeed, by equating the quantum averages in the coherent scenario with the classical initial phase spaces, i.e., $\braket{X} = x_0$ and $\braket{P} = p_0$, and assuming $\hbar/|x_0p_0|\to 0$, it directly follows from Eqs. \eqref{xtpt} and \eqref{alphadef} that $|\alpha|\to \infty$, and the averages $\braket{X(t)}$, $\braket{P(t)}$, $\braket{X^2(t)}$, $\braket{P^2(t)}$, and $\braket{X(t)P(t)}$ exactly match their classical counterparts. Taking into account the explicit form in Eq. \eqref{Wnweirdtpm} for computing the statistical moments of work, for any initial state $\rho$, the $m$-th power of work using the OBS and TPM protocols are, respectively,
\begin{equation}
    \braket{W^{m}(t_2,t_1)}_{\text{OBS}} = \braket{W^{m}(t_2,t_1)} \quad \text{and} \quad \braket{W^{m}(t_2,t_1)}_{\text{TPM}} = \left\langle\Phi_{H(0)}\left[\mathcal{T}_{>}\left(W^{m}(t_2,t_1)\right)\right]\right\rangle. \label{Wnweirdt2}
\end{equation}
It follows directly from Eqs. \eqref{WqopHO} and \eqref{Wnweirdt2} assuming $\braket{X} = x_0$, $\braket{P} = p_0$, and $\hbar/|x_0p_0|\to 0$ that the OBS average of work results in
\begin{equation}
    \braket{W(t_2,t_1)}_{\text{OBS}}=\braket{W(t_2,t_1)} = (E(t_2) - E(t_1))x_0^2 + (F(t_2) - F(t_1))p_0^2 + 2(G(t_2) - G(t_1))x_0 p_0,
\end{equation}
which precisely matches the classical expression in Eq. \eqref{Wgamma0}. On the other hand, from Eq. \eqref{Wnweirdt2}, the TPM statistics result in the expression
\begin{equation}
    \braket{W(t_2,t_1)}_{\text{TPM}} = \left\langle\Phi_{H(0)}\left[\mathcal{T}_{>}\left(W(t_2,t_1)\right)\right]\right\rangle = \left\langle\Phi_{H(0)}\left(W(t_2,t_1)\right)\right\rangle = \Tr\left[W(t_2,t_1)\Phi_{H(0)}(\rho)\right],
\end{equation}
where we have utilized the cyclic property of the trace. To compute $\braket{W(t_2,t_1)}_{\text{TPM}}$, first notice that the incoherent part of the coherent state is given by
\begin{equation}
    \Phi_{H(0)}(\rho)=\sum_{n=0}^{\infty}\ket{n}\bra{n}\left(\ket{\alpha}\bra{\alpha}\right)\ket{n}\bra{n}=\mathrm{e}^{-|\alpha|^2}\sum_{n=0}^{\infty}\frac{|\alpha|^{2n}}{n!}\ket{n}\bra{n}\label{incopart}
\end{equation}
where $\ket{e_n(0)}\equiv \ket{n}$ is the eigenvector of both $H(0)$ and the number operator $N$, given by
\begin{equation}
    N=\frac{2}{\hbar\omega_0}H(0)-\frac{1}{2}=a^{\dagger}a,
\end{equation}
where $a$ and $a^{\dagger}$ are the annihilation and creation operators, related to position and momentum operator as follows
\begin{equation}
    X= \sqrt{\frac{\hbar}{2m\omega_0}}(a^\dagger +a)\quad \text{and}\quad P=i\sqrt{\frac{m\hbar\omega_0}{2}}(a^\dagger - a).
\end{equation}
Using the well-known algebra rules for $a$ and $a^\dagger$, it follows that
\begin{equation}
    \bra{n}\{X,P\}\ket{n}=0,\quad \bra{n}P^2\ket{n}=\frac{m\hbar\omega_0(2n+1)}{2},\quad \bra{n}X^2\ket{n}=\frac{\hbar(2n+1)}{2m\omega_0}.\label{n}
\end{equation}
Consequently, for the incoherent state it follows: 
\begin{equation}
    \Tr[\{X,P\}\Phi_{H(0)}(\rho)]=0,\quad \Tr[P^2\Phi_{H(0)}(\rho)] =\frac{m\hbar\omega_0}{2}(2|\alpha|^2+1),\quad \Tr[X^2\Phi_{H(0)}(\rho)] =\frac{\hbar}{2m\omega_0}(2|\alpha|^2+1),\label{XXPPinco}
\end{equation}
where we considered Eqs. \eqref{incopart} and \eqref{n} and 
\begin{equation}
    \sum_{n=0}^{\infty}\frac{|\alpha|^{2n}}{n!}=\mathrm{e}^{|\alpha|^2},\quad 
    \sum_{n=0}^{\infty}\frac{2n|\alpha|^{2n}}{n!}=2|\alpha|^2\mathrm{e}^{|\alpha|^2}.
\end{equation}
Considering Eq. \eqref{alphadef}, it follows that
\begin{equation}
    |\alpha|^2= \Re^2(\alpha)+ \Im^2(\alpha)= \frac{\frac{m\omega_0^2}{2}x_{\alpha}^2+\frac{p_{\alpha}^2}{2m}}{\hbar\omega_0}
\end{equation}
then we can rewrite Eq. \eqref{XXPPinco} as
\begin{equation}
    \Tr[\{X,P\}\Phi_{H(0)}(\rho)]=0,\quad \Tr[P^2\Phi_{H(0)}(\rho)] =\frac{m^2\omega_0^2}{2}x_{\alpha}^2+\frac{p_{\alpha}^2}{2}+\frac{m\hbar\omega_0}{2},\quad \Tr[X^2\Phi_{H(0)}(\rho)] =\frac{x_{\alpha}^2}{2}+\frac{p_{\alpha}^2}{2m^2\omega_0^2}+\frac{\hbar}{2m\omega_0},\label{XXPPinco2}
\end{equation}
Comparing TPM with the classical regime by assuming $\braket{X}=x_0$, $\braket{P}=p_0$, and $\hbar/|x_0p_0|\to 0$, it follows from Eqs. \eqref{WqopHO}, \eqref{Wnweirdt2}, and \eqref{XXPPinco2} that
\begin{equation}
    \braket{W(t_2,t_1)}_{\text{TPM}}=\Tr\left[W(t_2,t_1)\Phi_{H(0)}(\rho)\right]=(E(t_2)-E(t_1))\left(\frac{1}{2}x_0^{2}+\frac{p_0^2}{2m^2  \omega_0^2}\right)+(F(t_2)-F(t_1))\left(\frac{m^2 \omega_0^2}{2}x_0^{2}+\frac{p_0^2}{2}\right).
\end{equation}
which completely differs from the classical expression for work in Eq. \eqref{Wgamma0}. Indeed, to illustrate these disparities, we have depicted in Figure \ref{fig1} the quantity $(W_{\text{\tiny CL}}(\Gamma_0,\tau,0)-\braket{W(\tau,0)}_{\text{TPM}})/W_{\text{\tiny CL}}(\Gamma_0,\tau,0)$ as functions of $x_0$ and $p_0$. These results were obtained using the model parameters $\tau^{-1}=\omega_0=\omega_1/3=1\,\mathrm{s}^{-1}$, $m=1\,\mathrm{kg}$, and $\hbar/|x_0p_0|\to 0$. Notably, the difference can range between $10-40\,\%$, highlighting a significant and non-negligible divergence between the two approaches in an explicitly classical limit scenario. Consequently, we can conclude that TPM fails to provide the correct statistics in this classical limit. The observed divergence in this case is primarily attributable to the pronounced impact of the first measurement's vanishing due to the coherences, a contrast that becomes evident when compared with the OBS average. Indeed, if we take into account the same parameters $\{\omega_0,\omega_1,\tau,m\}$ above and take into account the value of the reduced Planck constant $\hbar=1.054\times 10^{-34}\, \mathrm{J}\, \mathrm{s}$ with the position and momentum within the range $x_0\in [1,2]\,\mathrm{m}$ and  $p_0\in [1,2]\,\mathrm{Kg}\,\mathrm{m}\,\mathrm{s}^{-1}$, the relative difference between the OBS and the classical statistics of work $(W_{\text{\tiny CL}}(\Gamma_0,\tau,0)-\braket{W(\tau,0)}_{\text{OBS}})/W_{\text{\tiny CL}}(\Gamma_0,\tau,0)$  is below the computer precision $\approx 10^{-16}$.

\subsection{OBS versus classical statistics}

In this subsection, our objective is to demonstrate that the OBS statistics can recover the classical limit for the trapped harmonic oscillator modeled in Eq. \eqref{hclhq}, whenever $|\alpha|\gg 1$. Referring to Result 2, it therefore suffices to establish that the oscillator satisfies all the classicality conditions, (A) and (B), as elaborated in the previous section and the main text. This path will guide our discussion throughout the remainder of this supplementary material.

Initially, it becomes immediately apparent from Eqs. \eqref{Hclgamma0} and \eqref{Hqht} that the second part of condition (A) is satisfied, i.e. Eq. \eqref{CondBsecondpart} holds. Consequently, our task is to establish that the first part of condition (A) and  the full condition (B) are satisfied. To this end, we employ the Wigner-Weyl formalism \cite{Schleich2001}. In this case, the expectation value of any momentum or position-dependent observable can be expressed as
\begin{equation}
    \Tr \left[G(X,P) \rho\right]=\int_{-\infty}^{\infty}\int_{-\infty}^{\infty} dq\,dp\, g(x,p)W(x,p),
\end{equation}
where 
\begin{equation}
    \mathcal{W}(q,p)=\frac{1}{2\pi\hbar}\int_{-\infty}^{\infty}dy\, \mathrm{e}^{ipy/\hbar}\bra{q-y/2}\rho\ket{q+y/2} 
\end{equation}
is the so-called Wigner quasi-probability distribution and
\begin{equation}
    g(q,p)=\int_{-\infty}^{\infty}dy \mathrm{e}^{ipy/\hbar}\bra{q-y/2}G(X,P)\ket{q+y/2} \label{weyl1}
\end{equation}
is the Weyl transform of $G(X,P)$. In the special case in which we consider the initial coherent state $\rho=\ket{\alpha}\bra{\alpha}$ described in Eq. \eqref{rhoinitial}, the Wigner function assumes a Gaussian form
\begin{equation}
    \mathcal{W}_{\alpha}(q,p)=\mathcal{G}\left[q,x_{\alpha},\sigma_x^{\alpha}\right]\mathcal{G}[p,p_\alpha,\sigma_p^{\alpha}],\label{Wigneralpha}
\end{equation}
where
\begin{equation}
    \mathcal{G}\left[r,\bar{r},\sigma_r\right]=\frac{1}{\sqrt{2\pi\sigma_r^2}}\exp\left[\frac{-(r-\bar{r})^2}{2\sigma_r^2}\right]
\end{equation}
refer to the Gaussian distribution centered around $\bar{r}$ with uncertainty $\sigma_r$. The position and momentum uncertainties of the coherent state (see Eq. \eqref{alphadef}) are
\begin{equation}
    \sigma_x^{\alpha}=\sqrt{\frac{\hbar}{2m\omega_0}},\quad \sigma_p^{\alpha}=\sqrt{\frac{m\hbar\omega_0}{2}}.\label{sigmaalpha}
\end{equation}
With these definitions and the commutation relation $[X,P]=i\hbar$, we now show that condition (A) is satisfied for the coherent state.

Let us first compute the Weyl transformation of $P X^m$. Let us call it $g_{1,m}(q,p)$. From the very definition of the Weyl transformation (Eq. \eqref{weyl1}), we get:
\begin{equation}
    g_{1,m}(q,p)=\int_{-\infty}^{\infty}dy \mathrm{e}^{ipy/\hbar}\bra{q-y/2} P X^m\ket{q+y/2}=(-i\hbar)\left[\partial_{y}(\mathrm{e}^{ipy/\hbar}(q+y/2)^m)\right]_{y=0}=pq^m-\frac{i\hbar}{2} mq^{m-1}.\label{inductionW0}
\end{equation}
Similarly, for $P^{n}X^{m}$, it follows: 
\begin{equation}
    g_{n,m}(q,p)=\int_{-\infty}^{\infty}dy \mathrm{e}^{ipy/\hbar}\bra{q-y/2} P^{n} X^m\ket{q+y/2}=\left[(-i\hbar\partial_{y})^{n}\mathrm{e}^{ipy/\hbar}(q+y/2)^m\right]_{y=0}
\end{equation}
where we used the identity $\int dx f(x)\partial_{x}^{m}\delta(y-a)=(-1)^{m}[\partial_{x}^{m}f(x)]_{x=a}$ \cite{Lighthill1959}. To express this relation in a more suitable form, we first prove the following identity:
\begin{equation}
    \ba{rl}
    (-i\hbar\partial_{y})^{k}\left[\mathrm{e}^{ipy/\hbar}(q+y/2)^m\right]&=\sum_{j=0}^{k} \frac{k!}{(k-j)!j!}p^{k-j}\mathrm{e}^{ipy/\hbar}(-i\hbar\partial_{y})^{j}(q+y/2)^m
    \ea\label{inductionW1}
\end{equation}
Indeed, Eq. \eqref{inductionW1} can be derived by induction. First, if Eq. \eqref{inductionW1} holds for $k$, then for $k+1$, we obtain that
\begin{equation}
    \ba{rl}
    \displaystyle(-i\hbar\partial_{y})^{k+1}\left[\mathrm{e}^{ipy/\hbar}(q+y/2)^m\right]&\displaystyle=(-i\hbar\partial_{y})\sum_{j=0}^{k} \frac{k!}{(k-j)!j!}p^{k-j}\mathrm{e}^{ipy/\hbar}(-i\hbar\partial_{y})^{j}(q+y/2)^m\\
    &\displaystyle=\sum_{j=0}^{k} \frac{k!}{(k-j)!j!}p^{k+1-j}\mathrm{e}^{ipy/\hbar}(-i\hbar\partial_{y})^{j}(q+y/2)^m+\\
    &\displaystyle+\sum_{j=1}^{k+1}\frac{k!}{(k-j+1)!(j-1)!}p^{k+1-j}\mathrm{e}^{ipy/\hbar}(-i\hbar\partial_{y})^{j}(q+y/2)^m\\
    &\displaystyle=p^{k+1}\mathrm{e}^{ipy/\hbar}(-i\hbar\partial_{y})^{0}(q+y/2)^m+\mathrm{e}^{ipy/\hbar}(-i\hbar\partial_{y})^{k+1}(q+y/2)^m+\\
    &\displaystyle+\sum_{j=1}^{k}\left(\frac{k!}{(k-j)!j!}+\frac{k!}{(k-j+1)!(j-1)!}\right)p^{k+1-j}\mathrm{e}^{ipy/\hbar}(-i\hbar\partial_{y})^{j}(q+y/2)^m.
    \ea
\end{equation}
Recognizing that
\begin{equation}
    \displaystyle\frac{k!}{(k-j)!j!}+\frac{k!}{(k-j+1)!(j-1)!}=k!\left(\frac{k+1-j}{(k+1-j)!j!}+\frac{j}{(k+1-j)!j!}\right)=\frac{(k+1)!}{(k+1-j)!j!}
\end{equation}
and
\begin{equation}    
    \displaystyle\frac{(k+1)!}{(k+1-j)!j!}p^{k+1-j}\mathrm{e}^{ipy/\hbar}(-i\hbar\partial_{y})^{j}(q+y/2)^m=\left\{\ba{ll} \displaystyle p^{k+1}\mathrm{e}^{ipy/\hbar}(-i\hbar\partial_{y})^{0}(q+y/2)^m, &\displaystyle\text{for}\quad j=0,\\
    \displaystyle\mathrm{e}^{ipy/\hbar}(-i\hbar\partial_{y})^{k+1}(q+y/2)^m, &\displaystyle\text{for}\quad j=k+1,\ea \right.
\end{equation}
then it follows that
\begin{equation}
    (-i\hbar\partial_{y})^{k+1}\left[\mathrm{e}^{ipy/\hbar}(q+y/2)^m\right]=\sum_{j=0}^{k+1}\frac{(k+1)!}{(k+1-j)!j!}p^{k+1-j}\mathrm{e}^{ipy/\hbar}(-i\hbar\partial_{y})^{j}(q+y/2)^m
\end{equation}
Therefore, if Eq. \eqref{inductionW1} is valid for $k$, then it is also valid for $k+1$. Since it can be directly checked that the expression Eq. \eqref{inductionW1} holds for $k=1$, then, by induction, it is valid for any natural $k\geq 1$. We thus have the general formula:
\begin{equation}
    g_{n,m}(q,p)=\int_{-\infty}^{\infty}dy \mathrm{e}^{ipy/\hbar}\bra{q-y/2} P^{n} X^m\ket{q+y/2}=
    \left[\sum_{j=0}^{n}\frac{(n)!}{(n-j)!j!}p^{n-j}\mathrm{e}^{ipy/\hbar}(-i\hbar\partial_{y})^{j}(q+y/2)^m\right]_{y=0}.
\end{equation}
Since $(-i\hbar\partial_{y})^{j}(q+y/2)^m=(-i\hbar)^{j}j!(1/2)^{j}(q+y/2)^{m-j}$ for $m\geq j$ and $0$ for $j>m$, then
\begin{equation}
    \displaystyle g_{n,m}(q,p)=\left\{\ba{ll}\displaystyle\sum_{j=0}^{n}\frac{n!}{(n-j)!}(-i\hbar/2)^{j}p^{n-j}q^{m-j},&\displaystyle \text{for} \quad n\leq m,\\
    \displaystyle\sum_{j=0}^{m}\frac{n!}{(n-j)!}(-i\hbar/2)^{j}p^{n-j}q^{m-j},& \displaystyle\text{for}\quad n>m.\ea\right.\label{g1}
\end{equation}
It then follows that
\begin{equation}
    (-i\hbar/2)g_{n-1,m-1}(q,p) =\left\{\ba{ll}\displaystyle=\frac{1}{n}\sum_{j=1}^{n}\frac{n!}{(n-j)!}(-i\hbar/2)^{j}p^{n-j}q^{m-j},&\displaystyle \text{for} \quad n\leq m,\\
    \displaystyle\frac{1}{n}\sum_{j=1}^{m}\frac{n!}{(n-j)!}(-i\hbar/2)^{j}p^{n-j}q^{m-1},& \displaystyle\text{for}\quad n>m.\ea\right.\label{f1}
\end{equation}
Comparing Eq. \eqref{g1} with \eqref{f1}, we obtain
\begin{equation}
    g_{n,m}(q,p)=p^{n}q^{m}-n (i\hbar/2)g_{n-1,m-1}(q,p).
\end{equation}
Multiplying both sides by the Winger function and integrating over $q$ and $p$, it follows:
\begin{equation}
    \braket{P^n X^m}-\iint dq\,dp\, p^{n}q^{m}\mathcal{W}(q,p)=-\frac{ni\hbar}{2}\braket{P^{n-1}X^{m-1}}.
\end{equation}
Using iteratively this equation, we deduce
\begin{equation}
    \braket{P^n X^m}-\iint dq\,dp\, p^{n}q^{m}\mathcal{W}(q,p)=-\frac{ni\hbar}{2}\left[\iint dq\,dp\, p^{n-1}q^{m-1}\mathcal{W}(q,p)-\frac{(n-1)i\hbar}{2}\braket{P^{n-2}X^{m-2}}\right],
\end{equation}
which results in
\begin{equation}
    \braket{P^n X^m}=\left\{\ba{ll}\displaystyle\sum_{k=0}^{n-1}\frac{n!}{(n-k)!}\left(-\frac{i\hbar}{2}\right)^{k}\iint dq\,dp\, p^{n-k}q^{m-k}\mathcal{W}(q,p)+n!\left(-\frac{i\hbar}{2}\right)^{n}\braket{X^{m-n}},&\displaystyle \text{for} \quad n\leq m,\\
    \displaystyle\sum_{k=0}^{m-1}\frac{n!}{(n-k)!}\left(-\frac{i\hbar}{2}\right)^{k}\iint dq\,dp\, p^{n-k}q^{m-k}\mathcal{W}(q,p)+\frac{n!}{(n-m)!}\left(-\frac{i\hbar}{2}\right)^{m}\braket{P^{n-m}},& \displaystyle\text{for}\quad n>m.\ea\right. \label{finallemaaux}
\end{equation}
Considering the limit in which the quantum system approach the classical regime, then $x_\alpha\to x_0$, $p_\alpha\to p_0$, and $\frac{\hbar}{|x_0 p_0|}\to 0$. As a result from Eq. \eqref{sigmaalpha}, the Wigner distribution then approximates Dirac's delta distribution
\begin{equation}
    \mathcal{W}_{\alpha}(q,p)=\mathcal{G}\left[q,x_{\alpha},\sigma_x^{\alpha}\right]\mathcal{G}[p,p_\alpha,\sigma_p^{\alpha}]\approx\delta(q-x_0)\delta(p-p_0).\label{WigDelta}
\end{equation}
As a result, the quantum scenario approximates a classical distribution of a particle with well-defined position and momentum. In this classical limit, we can deduce directly from Eq. \eqref{finallemaaux} that
\begin{equation}
    \frac{\braket{P^n X^m}- \iint dq\,dp\, p^{n}q^{m}\mathcal{W}(q,p)}{\left|\iint dq\,dp\, p^{n}q^{m}\mathcal{W}(q,p)\right|}\to \sum_{k=1}^{n_{\min}}\frac{n!}{(n-k)!}\left(-\frac{i\hbar}{2 |x_0 p_0|}\right)^{k}.\label{CondA}
\end{equation}
where $n_{\min}=\min\{n,m\}$.
Therefore, 
\begin{equation}
    \ba{rl}
    \displaystyle\frac{\left|\braket{P^n X^m}- \iint dq\,dp\, p^{n}q^{m}\mathcal{W}(q,p)\right|}{|\braket{P^n X^m}|}&\displaystyle=\frac{\braket{P^n X^m}- \iint dq\,dp\, p^{n}q^{m}\mathcal{W}(q,p)}{\left|\iint dq\,dp\, p^{n}q^{m}\mathcal{W}(q,p)+\braket{P^n X^m}- \iint dq\,dp\, p^{n}q^{m}\mathcal{W}(q,p)\right|}\\
    &\displaystyle=\,\frac{\left|\sum_{k=1}^{n_{\min}}\frac{1}{(n-k)!}\left(-\frac{i\hbar}{2|x_0 p_0|}\right)^{k}\right|}{\left|\sum_{k=0}^{n_{\min}}\frac{1}{(n-k)!}\left(-\frac{i\hbar}{2|x_0 p_0|}\right)^{k}\right|}\label{CondA2}
    \ea
\end{equation}
Notice that
 since for all $k\leq n_{\min}$, $(n-k)/((n-k)!)=1/((n-k-1)!)\leq 1$, then it follows that
\begin{equation}
    \left|\sum_{k=1}^{n_{\min}}\frac{1}{(n-k)!}\left(-\frac{i\hbar}{2|x_0 p_0|}\right)^{k}\right|=\frac{\hbar}{2|x_0 p_0|}\left|\sum_{k=0}^{n_{\min}-1}\frac{1}{(n-k-1)!}\left(-\frac{i\hbar}{2|x_0 p_0|}\right)^{k}\right|\leq \frac{\hbar}{2|x_0 p_0|}\sum_{k=0}^{n_{\min}-1}\left(\frac{\hbar}{2|x_0 p_0|}\right)^{k}\leq \frac{\hbar}{2|x_0 p_0|} \frac{1-\left(\frac{\hbar}{2|x_0 p_0|}\right)^{n_{\min}}}{1-\frac{\hbar}{2|x_0 p_0|}},\label{CondA3}
\end{equation}
where we considered $\frac{\hbar}{2|x_0 p_0|}\leq 1$ and
\begin{equation}
    \sum_{k=0}^{n_{\min}-1}\left(\frac{\hbar}{2|x_0 p_0|}\right)^{k}=\frac{1-\left(\frac{\hbar}{2|x_0 p_0|}\right)^{n_{\min}}}{1-\frac{\hbar}{2|x_0 p_0|}}.
\end{equation}
Also, we can deduce
\begin{equation}
    \left|\sum_{k=0}^{n_{\min}}\frac{1}{(n-k)!}\left(-\frac{i\hbar}{2|x_0 p_0|}\right)^{k}\right|=\left|1+\sum_{k=1}^{n_{\min}}\frac{1}{(n-k)!}\left(-\frac{i\hbar}{2|x_0 p_0|}\right)^{k}\right|=\left|1+\frac{z\hbar}{2|x_0 p_0|} \frac{1-\left(\frac{\hbar}{2|x_0 p_0|}\right)^{n_{\min}}}{1-\frac{\hbar}{2|x_0 p_0|}}\right|,\label{CondA4}
\end{equation}
where we defined 
\begin{equation}
    z=\frac{\sum_{k=1}^{n_{\min}}\frac{1}{(n-k)!}\left(-\frac{i\hbar}{2|x_0 p_0|}\right)^{k}}{\frac{\hbar}{2|x_0 p_0|} \frac{1-\left(\frac{\hbar}{2|x_0 p_0|}\right)^{n_{\min}}}{1-\frac{\hbar}{2|x_0 p_0|}}},
\end{equation}
which, from  Eq. \eqref{CondA3}, is a complex number such that $|z|\leq 1$ (notice that $|z|$ is just the ratio between the first term on the left-hand side over the last right-hand expression of Eq. \eqref{CondA3}). Taking into account Eqs. \eqref{CondA3} and \eqref{CondA4}, Eq. \eqref{CondA2} lead us to 
\begin{equation}
    \lim_{\frac{\hbar}{2|x_0 p_0|}\to 0}\frac{\left|\braket{P^n X^m}- \iint dq\,dp\, p^{n}q^{m}\mathcal{W}(q,p)\right|}{|\braket{P^n X^m}|}=0,\label{CondAfinal}
\end{equation}
for any $n_{\min}$. Notice that this result holds even if $n_{\min}\to \infty$. Therefore, taking into account the fact that Eq. \eqref{CondBsecondpart} holds, condition (A) is fully satisfied when $x_\alpha\to x_0$, $p_\alpha\to p_0$, and $\hbar/|x_0 p_0|\to 0$. 

Now we prove that condition (B) is satisfied for $x_\alpha\to x_0$, $p_\alpha\to p_0$, and $\hbar/|x_0 p_0|\to 0$. Let us first notice that, by using a similar induction procedure as used in the Results 1 and 2 above or considering the results in the reference \cite{Transtrum2005}, we can deduce that for any arbitrary set of powers $\mathbf{I}\mathbf{J}\equiv \{I_1,I_2,\cdots I_n,J_1,J_2,\cdots J_n\}\subset \mathbbm{Z}$, the following relations can be written:
\begin{equation}
\ba{rl}
    \braket{X^{I_1}P^{J_{1}}\cdots X^{I_k}P^{J_{k}}\cdots X^{I_n}P^{J_{n}}}&=\braket{P^{\sum_{\mu}I_{\mu}}X^{\sum_{\mu}J_{\mu}}}+\sum_{k=1}^{N}c_k\hbar^k\braket{P^{\sum_{\mu}I_{\mu}-k}X^{\sum_{\mu}J_{\mu-k}}} \\
    \braket{X^{I_1}P^{J_{1}}\cdots X^{I_k-1}P^{J_{k}-1}\cdots X^{I_n}P^{J_{n}}}&=\braket{P^{(\sum_{\mu}I_{\mu})-1}X^{(\sum_{\mu}J_{\mu})-1}}+\sum_{k=1}^{N'}c_k'\hbar^k\braket{P^{\sum_{\mu}I_{\mu}-k}X^{\sum_{\mu}J_{\mu-k}}}
    \ea\label{induccondiC}
\end{equation}
where $c_k$ and $c_k'$ are coefficients dependent on $\mathbf{I}\mathbf{J}$ and $N=\max\{ \sum_{\mu}I_{\mu},\sum_{\mu}J_{\mu}\}=N'+1$. To show that the condition (B) is satisfied for our example with $\hbar/2|x_0 p_0|\to 0$, we analyze the ratio
\begin{equation}
    \frac{|\braket{X^{I_1}P^{J_{1}}\cdots X^{I_k-1}[X,P]P^{J_{k}-1}\cdots X^{I_n}P^{J_{n}}}|}{|\braket{X^{I_1}P^{J_{1}}\cdots X^{I_k}P^{J_{k}}\cdots X^{I_n}P^{J_{n}}}|}=\frac{\hbar|\braket{X^{I_1}P^{J_{1}}\cdots X^{I_k-1}P^{J_{k}-1}\cdots X^{I_n}P^{J_{n}}}|}{|\braket{X^{I_1}P^{J_{1}}\cdots X^{I_k}P^{J_{k}}\cdots X^{I_n}P^{J_{n}}}|}.
\end{equation}
Denoting $c_{\max}=\max\{\{c_k\}_{1}^{N},\{c_k'\}_{1}^{N'}\}$ and taking into account Eq. \eqref{induccondiC}, it follows that
\begin{equation}
     \frac{\hbar|\braket{X^{I_1}P^{J_{1}}\cdots X^{I_k-1}P^{J_{k}-1}\cdots X^{I_n}P^{J_{n}}}|/c_{\max}}{|\braket{X^{I_1}P^{J_{1}}\cdots X^{I_k}P^{J_{k}}\cdots X^{I_n}P^{J_{n}}}|/c_{\max}}=\frac{\hbar|\braket{P^{\sum_{\mu}I_{\mu}-1}X^{\sum_{\mu}J_{\mu}-1}}+\sum_{k=1}^{N'}c_k'\hbar^k\braket{P^{\sum_{\mu}I_{\mu}-k}X^{\sum_{\mu}J_{\mu-k}}}|/c_{\max}}{|\braket{P^{\sum_{\mu}I_{\mu}}X^{\sum_{\mu}J_{\mu}}}+\sum_{k=1}^{N}c_k\hbar^k\braket{P^{\sum_{\mu}I_{\mu}-k}X^{\sum_{\mu}J_{\mu-k}}}|/c_{\max}}\label{auxCC}
\end{equation}
Assuming $\hbar/2|x_0 p_0|\to 0$ and taking into account Eqs. \eqref{WigDelta} and \eqref{CondAfinal}, then the right-hand side of \eqref{auxCC} satisfies the following relations
\begin{equation}
     \frac{\hbar|p_0^{\sum_{\mu}I_{\mu}-1}x_0^{\sum_{\mu}J_{\mu}-1}+\sum_{k=1}^{N'}c_k'\hbar^k p_0^{\sum_{\mu}I_{\mu}-k}x_0^{\sum_{\mu}J_{\mu-k}}|/c_{\max}}{|p_0^{\sum_{\mu}I_{\mu}}x_0^{\sum_{\mu}J_{\mu}}+\sum_{k=1}^{N}c_k\hbar^k p_0^{\sum_{\mu}I_{\mu}-k}x_0^{\sum_{\mu}J_{\mu-k}}|/c_{\max}}=\frac{\hbar}{|x_0 p_0|}\frac{|1/c_{\max}+\sum_{k=1}^{N'}(c_k'/c_{\max})\left(\frac{\hbar}{|x_0 p_0|}\right)^k|}{|1/c_{\max}+\sum_{k=1}^{N}(c_k/c_{\max})\left(\frac{\hbar}{|x_0 p_0|}\right)^k|}\to\frac{\hbar}{|x_0 p_0|}\frac{1/c_{\max}}{1/c_{\max}}\to 0.
\end{equation}
Therefore, whenever $\hbar/|x_0 p_0|\to 0$, then
\begin{equation}
    \frac{|\braket{X^{I_1}P^{J_{1}}\cdots X^{I_k-1}[X,P]P^{J_{k}-1}\cdots X^{I_n}P^{J_{n}}}|}{|\braket{X^{I_1}P^{J_{1}}\cdots X^{I_k}P^{J_{k}}\cdots X^{I_n}P^{J_{n}}}|}\to 0
\end{equation}
and a similar deduction can be made for  $|\braket{X^{I_1}P^{J_{1}}\cdots X^{I_k-1}[X,P]P^{J_{k}-1}\cdots X^{I_n}P^{J_{n}}}|/|\braket{X^{I_1}P^{J_{1}}\cdots X^{I_k-1}PXP^{J_{k}-1}\cdots X^{I_n}P^{J_{n}}}|$. As a consequence, condition (B) is satisfied. With all the conditions (A) and (B) fulfilled, it follows from Result 2 that the work statistics for both the classical and quantum scenarios will be identical in the limit where $x_\alpha\to x_0$, $p_\alpha\to p_0$, and $\hbar/|x_0 p_0|\to 0$. Consequently,
\begin{equation}
    P_{\text{\tiny OBS}}(w)\approx P_{\text{\tiny CL}}(w)=\delta[w-(W_{\text{\tiny CL}}(\Gamma_0,\tau,0))].
\end{equation} 
